\documentclass[fleqn,usenatbib]{mnras}

\usepackage{newtxtext,newtxmath}
\usepackage[normalem]{ulem}

\usepackage[T1]{fontenc}

\DeclareRobustCommand{\VAN}[3]{#2}
\let\VANthebibliography\thebibliography
\def\thebibliography{\DeclareRobustCommand{\VAN}[3]{##3}\VANthebibliography}

\newcommand{\hi}{\textrm{H\textsc{i}}}
\newcommand{\secref}[1]{\hyperref[#1]{Section~\ref*{#1}}}
\newcommand{\appref}[1]{\hyperref[#1]{Appendix~\ref*{#1}}}
\newcommand{\txteq}[1]{\,{#1}\,}
\newcommand{\Nfg}{N_\text{fg}}
\newcommand{\hMpc}{\,h\,\text{Mpc}^{-1}}

\newcommand{\orcid}[1]{\hspace{1pt}\raisebox{-1pt}{\href{https://orcid.org/#1}{\includegraphics[height=10pt]{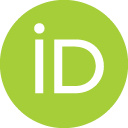}}}\hspace{-0.4ex}
}

\usepackage{graphicx}	
\usepackage{amsmath}	

\usepackage{calrsfs}
\DeclareMathAlphabet{\pazocal}{OMS}{zplm}{m}{n}
\SetMathAlphabet\pazocal{bold}{OMS}{zplm}{bx}{n}

\title[MeerKLASS L-band deep-field intensity maps]
{MeerKLASS L-band deep-field intensity maps: entering the \hi\ dominated regime}

\author[MeerKLASS Collaboration]{MeerKLASS Collaboration\thanks{Corresponding authors:}:
Matilde Barberi-Squarotti$^{1,2,3,4}$,
Jos\'e L. Bernal$^{5}$,
Philip Bull$^{6,7}$,
\newauthor
Stefano Camera$^{1,8,9,7}$,
Isabella P. Carucci$^{10,11}$,
Zhaoting Chen$^{12}$,
Steven Cunnington
$^{6}$\thanks{ \href{mailto:steven.cunnington@manchester.ac.uk}{\texttt{steven.cunnington@manchester.ac.uk}}},
\newauthor
Brandon N. Engelbrecht$^{7}$,
Jos\'e Fonseca$^{7,13,14}$,
Keith Grainge$^{6}$, 
Melis O. Irfan$^{15,7}$,
Yichao Li$^{16,7}$,
\newauthor
Aishrila Mazumder$^6$,
Sourabh Paul$^6$,
Alkistis Pourtsidou$^{12,17,7}$,
Mario G. Santos$^{7,18}$,
Marta Spinelli$^{19,7}$,
\newauthor
Jingying Wang$^{20,7}$\thanks{ \href{mailto:jywang@shao.ac.cn}{\texttt{jywang@shao.ac.cn}}},
Amadeus Witzemann$^{7,6}$,
Laura Wolz$^{6}$
\\
$^1$Dipartimento di Fisica, Universit\`a degli Studi di Torino, via P. Giuria 1, 10125, Torino, Italy\\
$^{2}$Dipartimento di Fisica, Universit\`a degli Studi di Milano, via G. Celoria 16, 20133, Milano, Italy\\
$^{3}$INFN – Istituto Nazionale di Fisica Nucleare, Sezione di Milano, via G. Celoria 16, 20133, Milano, Italy\\
$^{4}$INAF -- Istituto Nazionale di Astrofisica, Osservatorio Astrofisico di Brera-Merate, via Brera 28
20121 Milano, Italy\\
$^{5}$ Instituto de Física de Cantabria (IFCA), CSIC-Univ. de Cantabria, Avda. de los Castros s/n, E-39005 Santander, Spain\\
$^{6}$Jodrell Bank Centre for Astrophysics, Department of Physics \& Astronomy, The University of Manchester, Manchester M13 9PL, UK\\
$^7$Department of Physics and Astronomy, University of the Western
Cape, Robert Sobukwe Road, Cape Town 7535, South Africa\\
$^8$INFN – Istituto Nazionale di Fisica Nucleare, Sezione di Torino, via P. Giuria 1, 10125, Torino, Italy\\
$^9$INAF – Istituto Nazionale di Astrofisica, Osservatorio Astrofisico di Torino, 10025 Pino
Torinese, Italy\\
$^{10}$INAF -- Istituto Nazionale di Astrofisica, Osservatorio Astronomico di Trieste, Via G.B.\ Tiepolo 11, 34131 Trieste, Italy\\
$^{11}$IFPU -- Institute for Fundamental Physics of the Universe, Via Beirut 2, 34151 Trieste, Italy\\
$^{12}$Institute for Astronomy, The University of Edinburgh, Royal Observatory, Edinburgh EH9 3HJ, UK\\
$^{13}$Instituto de Astrof\'isica e Ci\^encias do Espa\c{c}o, Universidade do Porto CAUP, 4150-762 Porto, Portugal\\
$^{14}$Departamento de F\'isica e Astronomia, Faculdade de Ci\^{e}ncias, Universidade do Porto, Rua do Campo Alegre 687, PT4169-007 Porto, Portugal\\
$^{15}$Institute of Astrophysics, University of Cambridge, Madingley Road, CB3 0HA\\
$^{16}$Department of Physics, College of Sciences, Northeastern University, Wenhua Road, Shenyang, 11089, China\\
$^{17}$Higgs Centre for Theoretical Physics, School of Physics and Astronomy, The University of Edinburgh, Edinburgh EH9 3FD, UK\\
$^{18}$South African Radio Astronomy Observatory (SARAO), 2 Fir Street, Cape Town, 7925, South Africa\\
$^{19}$Observatoire de la Côte d’Azur, Laboratoire Lagrange, Bd de l’Observatoire, CS 34229, 06304 Nice cedex 4, France\\
$^{20}$Shanghai Astronomical Observatory, Chinese Academy of Sciences, 80 Nandan Road, Shanghai, 200030, China
}

\date{Accepted XXX. Received YYY; in original form ZZZ}

\pubyear{2024}

\begin{document}

\label{firstpage}
\pagerange{\pageref{firstpage}--\pageref{lastpage}}
\maketitle

\begin{abstract}
We present results from MeerKAT single-dish \hi\ intensity maps, the final observations to be performed in L-band in the MeerKAT Large Area Synoptic Survey (MeerKLASS) campaign. The observations represent the deepest single-dish \hi\ intensity maps to date, produced from 41 repeated scans over $236\,\deg^2$, providing 62 hours of observational data for each of the 64 dishes before flagging. By introducing an iterative \textit{self-calibration} process, the estimated thermal noise of the reconstructed maps is limited to ${\sim}\,1.21$\,mK ($1.2\,\times$ the theoretical noise level). This thermal noise will be sub-dominant relative to the \hi\ fluctuations on large scales ($k\,{\lesssim}\,0.15\,h\,\text{Mpc}^{-1}$), which demands upgrades to power spectrum analysis techniques, particularly for covariance estimation. In this work, we present the improved MeerKLASS analysis pipeline, validating it on both a suite of mock simulations and a small sample of overlapping spectroscopic galaxies from the Galaxy And Mass Assembly (GAMA) survey. Despite only overlapping with ${\sim}\,25\%$ of the MeerKLASS deep field, and a conservative approach to covariance estimation, we still obtain a ${>}\,4\,\sigma$ detection of the cross-power spectrum between the intensity maps and the 2269 galaxies at the narrow redshift range $0.39\,{<}\,z\,{<}\,0.46$. We briefly discuss the \hi\ auto-power spectrum from this data, the detection of which will be the focus of follow-up work. For the first time with MeerKAT single-dish intensity maps, we also present evidence of \hi\ emission from stacking the maps onto the positions of the GAMA galaxies. 
\end{abstract}




\section{Introduction}

Probing density fluctuations in large-scale cosmic structure is a cornerstone of precision cosmology. Constructing maps that span wide areas and reach deep redshifts provides a rich resource for testing and constraining the standard
$\Lambda$CDM cosmological model. Surveys at optical and near-infrared wavelengths have made impressive progress in this pursuit \citep{eBOSS:2020yzd,Heymans:2020gsg,DES:2021wwk,DESI:2024mwx}. However, extending experiments to a broader range of wavelengths boosts the quantity of available data. Furthermore, multi-wavelength observations provide robust checks of systematic influence due to the alternative instrumental approach required to survey other wavelengths. There can also be vastly different complex astrophysics for tracers at different wavelengths. Thus alternative tracers can validate the consistency of cosmological conclusions.

Of particular interest for surveying large-scale structure is mapping redshifted neutral hydrogen (\hi) gas at 21cm radio wavelengths. In the post-reionisation Universe, at redshifts $z\,{\lesssim}\,6$, the remaining \hi\ is overwhelmingly contained within dark matter halos \citep{Villaescusa-Navarro:2018vsg}, meaning the 21cm radiation spontaneously emitted from \hi\ acts as a tracer of the large-scale structure. However, at the redshifts of interest for large-scale structure cosmology ($z\,{\gtrsim}\,0.1$), the 21cm emission becomes faint and hard to detect with the high signal-to-noise required for a comprehensive mapping survey. This problem can be circumvented by abandoning the requirement to resolve individual galaxies and instead integrate the combined unresolved emission with low-aperture radio telescopes. This process is now commonly known as \hi\ intensity mapping \citep{Bharadwaj:2000av,Battye:2004re,Wyithe:2007rq,Chang:2007xk}.

The novel approach of intensity mapping presents some challenges. Since it records all radiation entering the telescope receiver within its frequency bandpass, the contribution to the signal will not come from cosmological \hi\ alone; in fact, only a small percentage is. There are multiple sources of foregrounds in the frequency ranges corresponding to redshifted 21cm emission. These mostly originate from our Galaxy and come from diffuse \textit{synchrotron} and \textit{free-free} emission, but there can also be strong extragalactic point sources \citep{2019MNRAS.483.4411M,Lian:2020jgd,2021MNRAS.505.1575S,Irfan:2021xuk,2022PASA...39...35H,Matshawule:2020fjz}. These sources dominate the cosmological \hi\ brightness temperature by three to five orders of magnitude. Furthermore, instrumental effects can distort the foreground spectra, making their removal additionally challenging \citep{Shaw:2014khi,Carucci:2020enz,Matshawule:2020fjz,2022MNRAS.509.2048S}. In addition to foregrounds, the telescopes will detect terrestrial, human-made radio frequency interference (RFI) \citep{Harper:2018ncl,Engelbrecht:2024eoc}. This can dominate even more than the foregrounds, but vary with time and sky position so can therefore be flagged and deleted from the final maps. Residual RFI will also average down if repeated scans are performed over the same patch of sky, as is the case with MeerKLASS observations. 

Despite foreground contamination and RFI challenges, the potential benefits from \hi\ intensity mapping are too large to ignore, thus justifying the ongoing endeavour to develop the technique into a tool for precision cosmology. Experiments have already delivered promising results, demonstrating how detections are possible even without purpose-built instruments \citep{Masui:2012zc,Wolz:2015lwa,Anderson:2017ert,eBOSS:2021ebm}. New experiments are starting to arrive \citep[e.g.][]{2011IJMPD..20..989N,Battye:2012fd,Chen:2012xu,Newburgh:2016mwi,PUMA:2019jwd,Vanderlinde:2019tjt,CHIME:2022dwe,Pal:2022xfm}, many designed specifically with \hi\ intensity mapping in mind, creating a bright future ahead.

The MeerKAT telescope,\footnote{\href{https://www.sarao.ac.za/science/meerkat/}{\texttt{www.sarao.ac.za/science/meerkat}}} an array of 64 dishes in the Upper Karoo region of South Africa, has already showcased its capability to perform \hi\ intensity mapping both as an \textit{interferometer}, claiming the first detection of the \hi\ power spectrum \citep{Paul:2023yrr}, and also in \textit{single-dish mode}, where the field-of-view can reach cosmological scales. Using the single-dish pilot survey \citep{Wang:2020lkn}, a detection of the cross-correlation power spectrum was achieved with overlapping WiggleZ galaxies \citep{Cunnington:2022uzo,Drinkwater:2009sd}.

The MeerKAT Large Area Synoptic Survey (MeerKLASS) \citep{MeerKLASS:2017vgf} will continue adding to these observations until it is merged with the SKA Observatory (SKAO) \citep{Bacon:2018dui} on the same site.\footnote{\href{https://www.skao.int/en}{\texttt{www.skao.int/en}}} MeerKAT can observe in two frequency bands relevant for 21cm-cosmology: L-band ($900\,\text{MHz}\,{<}\,{\nu}\,{<}\,1420\,\text{MHz}$, reaching redshifts $z\,{<}\,0.58$) and UHF-band ($580\,\text{MHz}\,{<}\,{\nu}\,{<}\,1000\,\text{MHz}$, $1.45\,{>}\,z\,{>}\,0.4$).

In this work, we present results from MeerKAT single-dish observations in the MeerKLASS L-band deep field, a set of observations performed over a single $236\,\deg^2$ patch with 41 repeated scans for each of the 64 dishes. Integrating over the sky in this way will average down \textit{time-varying} contributions from thermal noise and RFI, thus boosting the signal-to-noise of \textit{time-constant} sky signals, which includes the foregrounds and the cosmological \hi. Previous analysis \citep[e.g.][]{eBOSS:2021ebm,Cunnington:2022uzo} exploited the fact that the foreground cleaned intensity maps were thermal noise dominated when estimating error bars on the power spectrum measurements. For this MeerKLASS deep-field and future MeerKLASS observations, it can no longer be assumed that the thermal noise will be dominant, therefore demanding more emphasis on covariance estimation. We therefore explore using the scatter in the foreground transfer function \citep{Cunnington:2023jpq}, primarily used to correct for signal loss from foreground cleaning, as a measure of the covariance in measurements. The transfer function is constructed using mock signal injection into the real data, hence its scatter should encapsulate cosmic variance, thermal noise fluctuations, residual RFI contributions, foreground residuals and signal loss correction uncertainties.

We validate the precision and accuracy of the analysis pipeline to ensure estimator accuracy is within the statistical error associated with these MeerKLASS observations. We formalise the main steps in our pipeline, such as beam reconvolution, foreground cleaning and sampling to a Cartesian grid, presenting validation steps on a suite of simulations which emulate the MeerKLASS L-band deep-field observations. We also use a small sample of overlapping spectroscopic galaxies from the Galaxy And Mass Assembly (GAMA) \citep{Drinkwater:2009sd} survey to study the cross-power spectrum and its covariance under different analysis choices.

The paper is outlined as follows: \secref{sec:ObsData} outlines the observations for the MeerKLASS deep field, our new calibration strategy (with an iterative \textit{self-calibration} process) and overlapping GAMA galaxies. \secref{sec:Mocks} presents the methods for generating our suite of mock data which emulate the observations. \secref{sec:FGcleaning} provides details of our foreground cleaning technique. \secref{sec:PkEst} focuses on the power spectra, formalising our estimation and modelling process, and presents the results. \secref{sec:stack} shows the results from stacking the MeerKLASS deep-field maps onto the locations of GAMA galaxies and discusses the process and complications that come with this. We then finally conclude in \secref{sec:conclusion}.

\section{Observational data}\label{sec:ObsData}

\subsection{MeerKLASS L-band deep-field}
Using the same scan strategy as the MeerKLASS L-band pilot survey in \citet{Wang:2020lkn}, a series of observations were conducted using the MeerKAT telescope between September 1st and December 29th 2021. 
Observations with single dishes typically require a scanning strategy where the dishes are rapidly moved across the sky. The MeerKAT antennas were set to scan in azimuth at constant elevation to minimise fluctuations of ground spill and airmass. To make sure the projected speed on the sky is $5\,\mathrm{arcmin\,s^{-1}}$, the telescope scan speed was set to $5/\cos(el)\,\mathrm{arcmin\,s^{-1}}$ (where $el$ is the elevation angle) along the azimuth. This speed also ensures that the telescope pointing does not move significantly compared to the width of the primary beam (${\sim}\,1\,$deg) during a single time dump ($2\,\mathrm s$).

With a time resolution of $2\,\mathrm s$, this gives a scan speed of no more than $10\,\mathrm{arcmin}$ per time sample, which is well within the beam size. Noise diodes attached to each receiver were fired for $0.585\,\mathrm s$ once every $19.5\,\mathrm s$ during the observation to provide a relative time-ordered data (TOD) calibration reference. The dishes are moved back and forth with a slew of about 10\,deg sky projection in each direction, corresponding to an observing time of about $120\,\mathrm s$ per stripe. The duration of each set of scans with 48 strips is about 100 minutes, and before and after each scan we spent  ${\sim}\,8$ minutes tracking a nearby celestial point source to use as a bandpass calibrator and absolute flux calibrator.

At fixed elevation, two scans can be performed per night, corresponding to when the field is rising and setting respectively. The two scans will cross each other to achieve good sky coverage in the region of overlap.
There were 41 observations in total (each one referred as a \textit{block}), the details of each are listed in the Appendix in \autoref{tab:block}. The observations repeated scans covering a patch of sky approximately $236\,\deg^2$ in the southern sky, spanning in R.A.\ $(330^{\circ},360^{\circ})$ and Dec.\ $(-36^{\circ}, -25^{\circ})$. The 62 hours obtained for each of the 64 dishes before flagging, provide a cumulative 3968 hours of observations,\footnote{This number is reduced for the final useable data since some observation blocks were entirely flagged and some dishes failed in observation. We provide details of this later in this section.} likely to be the \textit{deepest} single-dish intensity maps, in terms of observational hours per square degree, in the MeerKLASS campaign.

\subsection{Standard MeerKLASS calibration}\label{sec:StandardCalibration}

We employ the MeerKAT single-dish calibration pipeline {\tt KATcali}
to calibrate the deep-field observations and create the combined 3D data cubes. {\tt KATcali} includes several cycles of RFI flagging, calibration, and map-making. The details of the calibration are thoroughly described in \citet{Wang:2020lkn}. Here, we provide a brief overview of the main points. We take a two-step approach to the calibration process. The first step is to use the flux model of a strong point source during tracking observations to calibrate the bandpass and absolute flux, as well as the injection power of the noise diode. The second step then uses the noise diode injections to calibrate the TOD during the constant-elevation scanning phase.

To calibrate the scanning data, we build the model as 
\begin{gather}\label{equ:model_scan}
  T_{\rm model}(t, \nu)= T_{\rm diffuse} (t, \nu)+ T_{\rm el} (t, \nu)
    + T_{\rm diode}(t, \nu)  + T_{\rm rec}(t, \nu)\,,
\end{gather}
where $T_{\rm el}$ is the antenna temperature model of elevation-dependent terrestrial emission (atmosphere and ground spill),  $T_{\rm diode}$ is the noise diode contribution, $T_{\rm rec}$ is the receiver temperature, while 
the diffuse celestial components $T_{\rm diffuse}$ includes Galactic emission and the cosmic microwave background (CMB), 
\begin{gather}
  T_{\rm diffuse}(t,\nu)= T_{\rm Gal}(t, \nu) +T_{\rm CMB} \,,
\end{gather}
where the CMB emission is set to be a uniform $T_{\rm CMB}\txteq{=}2.725\,$K \citep{2009ApJ...707..916F}, and the Galactic emission $T_{\rm Gal} (t, \nu)$ is derived from {\tt PySM}
(\citealt{2017MNRAS.469.2821T}) that is based on the reprocessed Haslam 408\,MHz total intensity map (\citealt{2015MNRAS.451.4311R}). 

To compare with the raw data from the correlator, we multiply the signal model by the gain, $g(t, \nu)$, so that
\begin{gather}\label{equ:model2}
  \hat T_{\rm model}(t, \nu)= g(t, \nu)\, T_{\rm model}(t, \nu)\,, 
\end{gather}
where the hat indicates a temperature in the (arbitrary) correlator units. 

The signal and gain models are then fitted (with a Bayesian framework; see more details in Section 3.3 of \citealt{Wang:2020lkn}) to the TOD for each polarisation, frequency channel, dish, and observation scan, all of which are treated independently. \textcolor{black}{The function of gain over time is a 4th order polynomial (Equation 17 in \citealt{Wang:2020lkn}). We checked that the order of the polynomial was sufficient to represent the data; choosing higher orders produced noise-dominated coefficients.} Using the gain solution, we can then obtain the calibrated temperature,
\begin{gather} \label{eq:T}
    T_{\rm cal}(t,\nu)\equiv \frac{\hat T_{\rm raw}(t, \nu)}{g(t, \nu)}\,,
\end{gather}
where ${\hat T_{\rm raw}}$ is the raw data collected by MeerKAT.
The residual temperature difference between the signal model and calibrated data is then
\begin{gather} \label{eq:T_resi}
    T_{\rm res}(t,\nu)\equiv T_{\rm cal}(t,\nu) - T_{\rm model}(t, \nu)\,.
\end{gather}

In standard calibration, we assumed that $T_{\rm res}$ is noise-dominated during the fitting, ignoring the faint point sources in the target field.  This was a fair assumption when the noise level was relatively high \citep{Wang:2020lkn}. In this work, we seek a better calibration approach by implementing a self-calibration method. 

\begin{figure}
\centering
\includegraphics[width=\columnwidth]{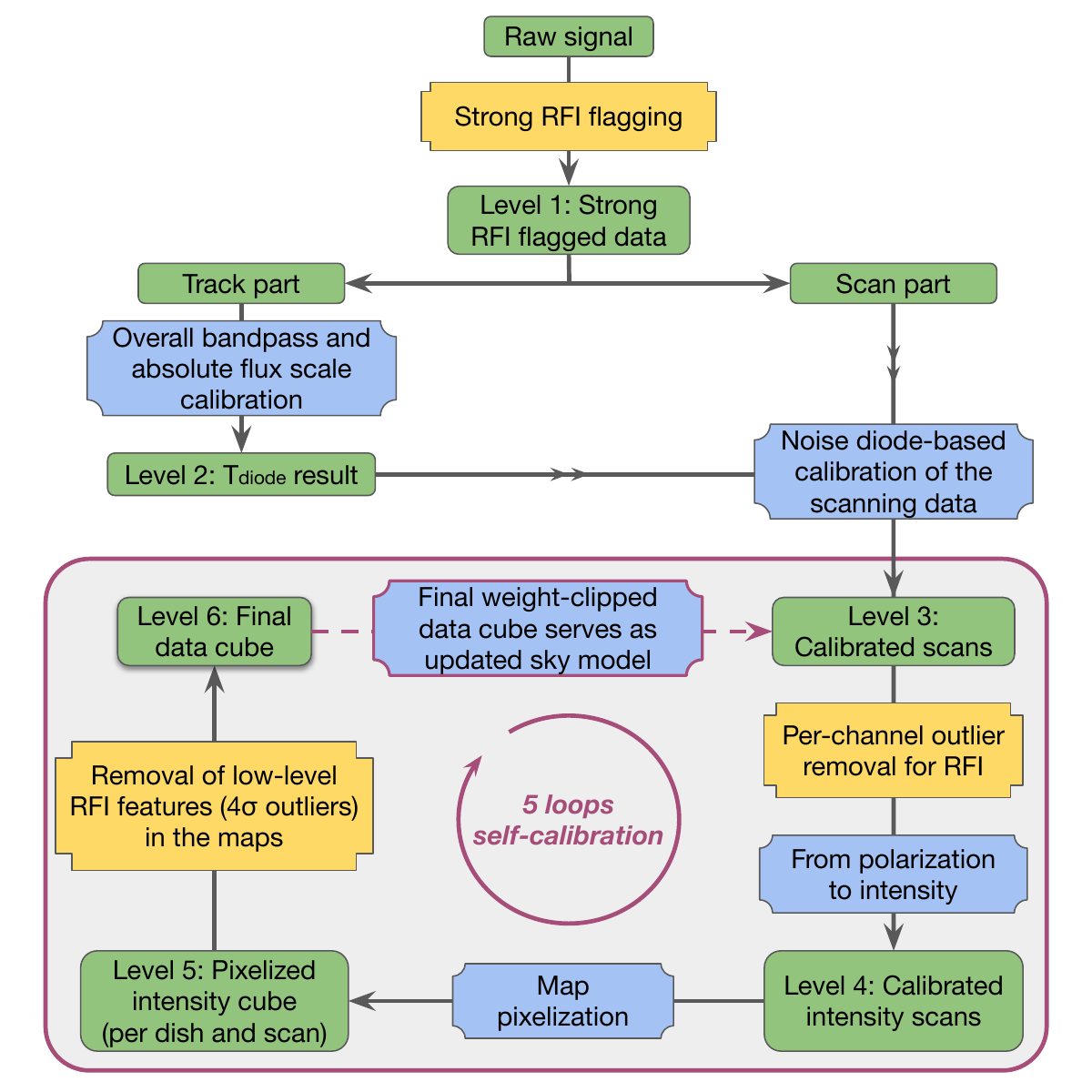}
\caption{Flowchart showing each step in the calibration pipeline. This is an extension beyond the \citet{Wang:2020lkn} pipeline since it includes the new self-calibration strategy.}
\label{fig:pip}
\end{figure} 

\begin{figure}
\centering
\includegraphics[width=\columnwidth]{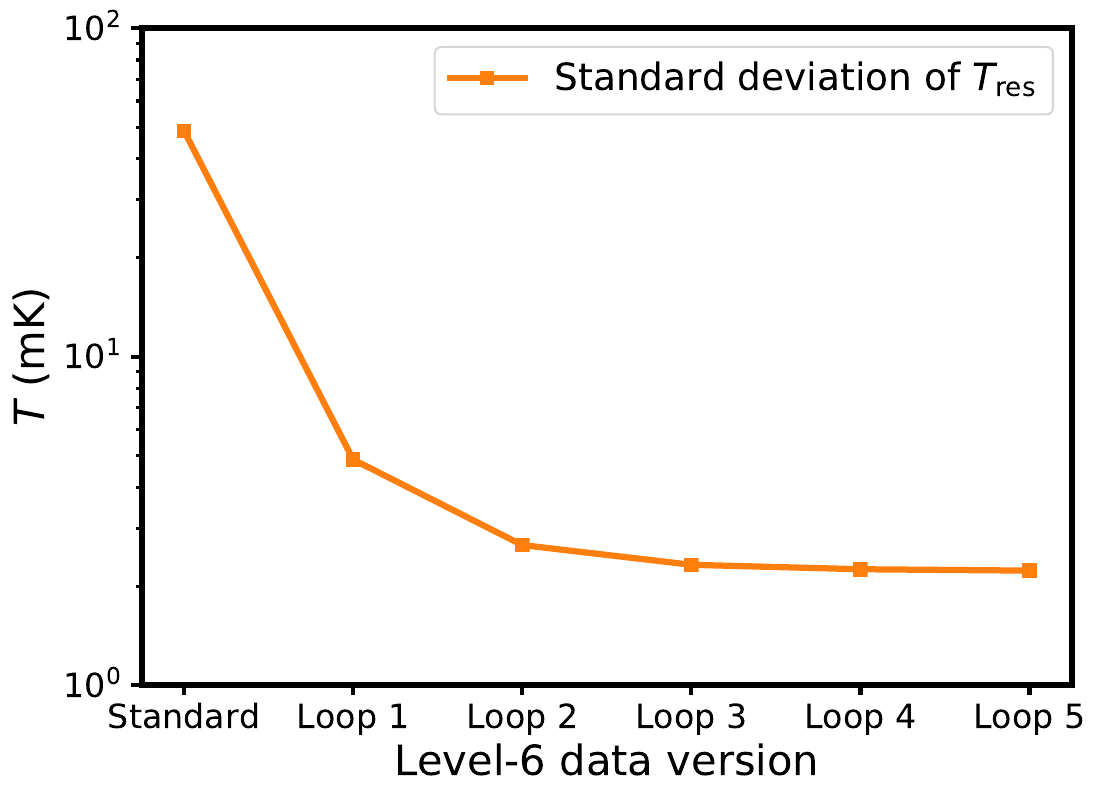}

\caption{The standard deviations of $T_\text{res}$ for different self-calibration cycles (Loops 1-5) in the 971–1023 MHz region. The standard calibration value is also shown for reference (first point).}
\label{fig:loop_statis}
\end{figure}

\subsection{Self-calibration based on the MeerKLASS sky map}\label{sec:SelfCal}


The higher signal-to-noise from the MeerKLASS deep-field data may allow us to deal more carefully with the point sources, which were ignored in the standard calibration, to achieve a better calibration result. Additionally, we aim to use a more accurate sky model to replace the $T_{\rm Gal}$ model that is based on the Haslam map. 
The 408\,MHz radio continuum all-sky map, which has been widely used to study diffuse synchrotron radiation from our Galaxy and works as a Galactic foreground model for cosmological research, combines data from four different surveys that happened more than 40 years ago (\citealt{1981A&A...100..209H}; \citealt{1982A&AS...47....1H}). The instrument performance in the 1980s limited the data quality, especially for the Galactic substructures and the zero level of the maps \citep{2015MNRAS.451.4311R}. The selection of the spectral index map, used to generate Galactic maps at MeerKAT frequencies, is also a matter of debate \citep{2022MNRAS.509.2048S}. 

We therefore extend the calibration process described in \secref{sec:StandardCalibration} and test the behaviour of the results by introducing an iterative \textit{self-calibration} process with several loops.
This includes \\
(1) replacing $T_{\rm diffuse} (t, \nu)$ with  $T_{\rm sky} (t, \nu)$, obtained from the previous Level-6 $T_{\rm sky}$ result, when the corresponding Level-6 pixel has a hit-count$\txteq{>}40$ (which ensures that $T_{\rm sky}$ values are obtained from high signal-to-noise data).  \textcolor{black}{Here Level-6 $T_{\rm sky}$ are the combined calibrated temperature maps for all scans and all dishes at all frequencies, after several rounds of RFI flagging, as shown in \autoref{fig:pip}.}\\
(2) setting the time-order weight $w({t_i})\txteq{=}1$ for all the scan time samples; instead of setting a higher weight for the diode (main calibrator during the standard calibration) injection samples (see more details in Section 3.3 of \citealt{Wang:2020lkn}).
Thus \autoref{equ:model_scan} becomes 
\begin{equation}\label{equ:model2_scan}
  T^{\rm upd}_{\rm model}(t, \nu)= T_{\rm sky} (t, \nu)+ T_{\rm el} (t, \nu)
    + T_{\rm diode}(t, \nu)  + T_{\rm rec}(t, \nu)\,.
\end{equation}
In this way, we have all components (especially point sources and Galactic substructures) included in the input ``model''. The difference between the data and our calibrated sky data is
\begin{equation} \label{eq:T_resi2}
    T^{\rm upd}_{\rm res}(t,\nu)\equiv T_{\rm cal}(t,\nu) - T^{\rm upd}_{\rm model}(t, \nu)\,,
\end{equation}
which should be noise-dominated compared to that in \autoref{eq:T_resi}. The self-calibration process is iterative for five loops, \textcolor{black}{as shown in \autoref{fig:pip}}, in which the $T_{\rm sky} (t, \nu)$ always uses the Level-6 result of the previous calibration loop. Each iteration refines the calibration, reducing the discrepancies between the observed data and the model. We stopped the self-calibration iteration after five loops because the standard deviation of $T_{\rm res}$ converged to a stable value, as shown in \autoref{fig:loop_statis}.

\begin{figure*}
\centering
\includegraphics[width=1.9\columnwidth]{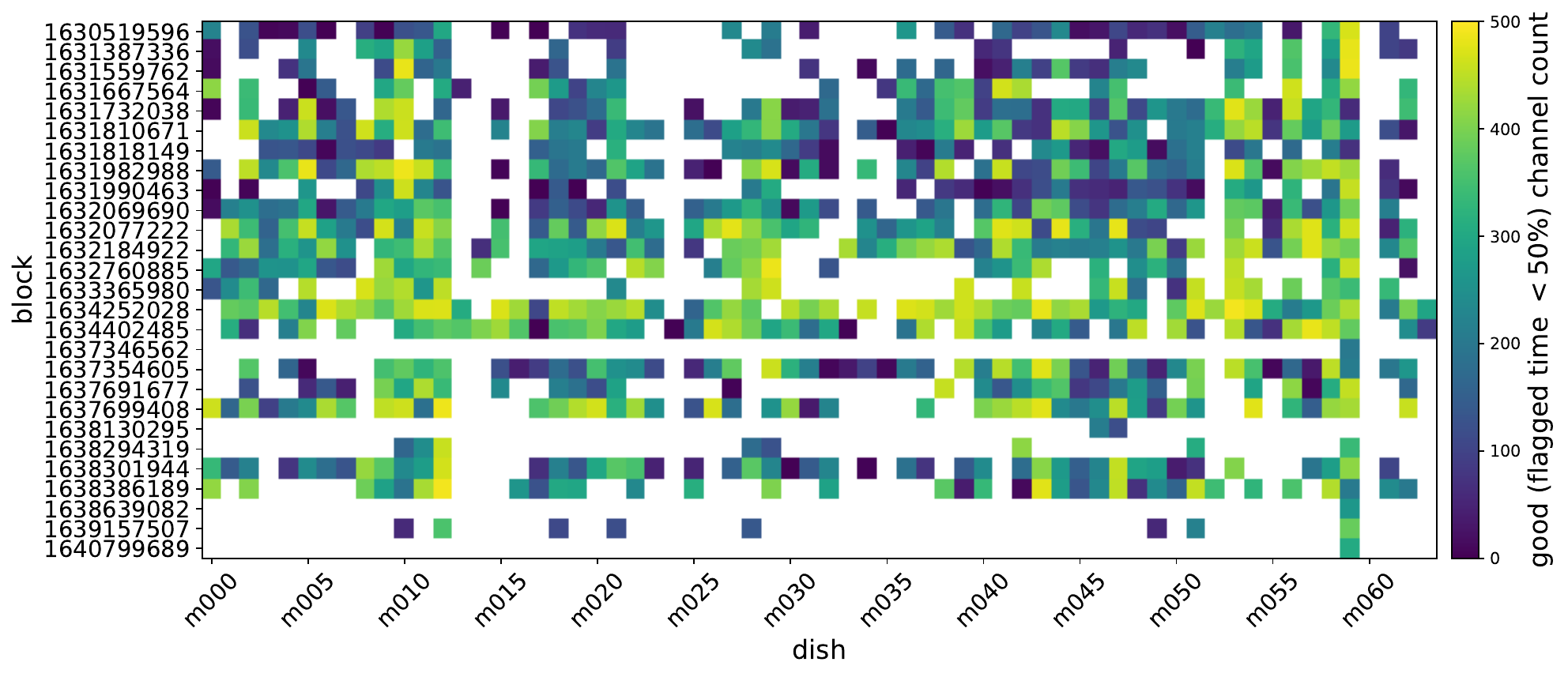}
\caption{Number count of good (flagged time $<50\%$) channels within the 971–1075\,MHz band (L-band channels 550–1050) for each dish ($x$-axis) and observation block ($y$-axis). We only show the 27 observation blocks used, out of the possible 41. The other blocks were entirely removed mainly due to RFI contamination, and one failed observation run. The details of \textit{all} 41 blocks are listed in \autoref{tab:block}.}
\label{fig:block_statis}
\end{figure*}

\begin{figure}
    \centering
    \includegraphics[width=1\linewidth]{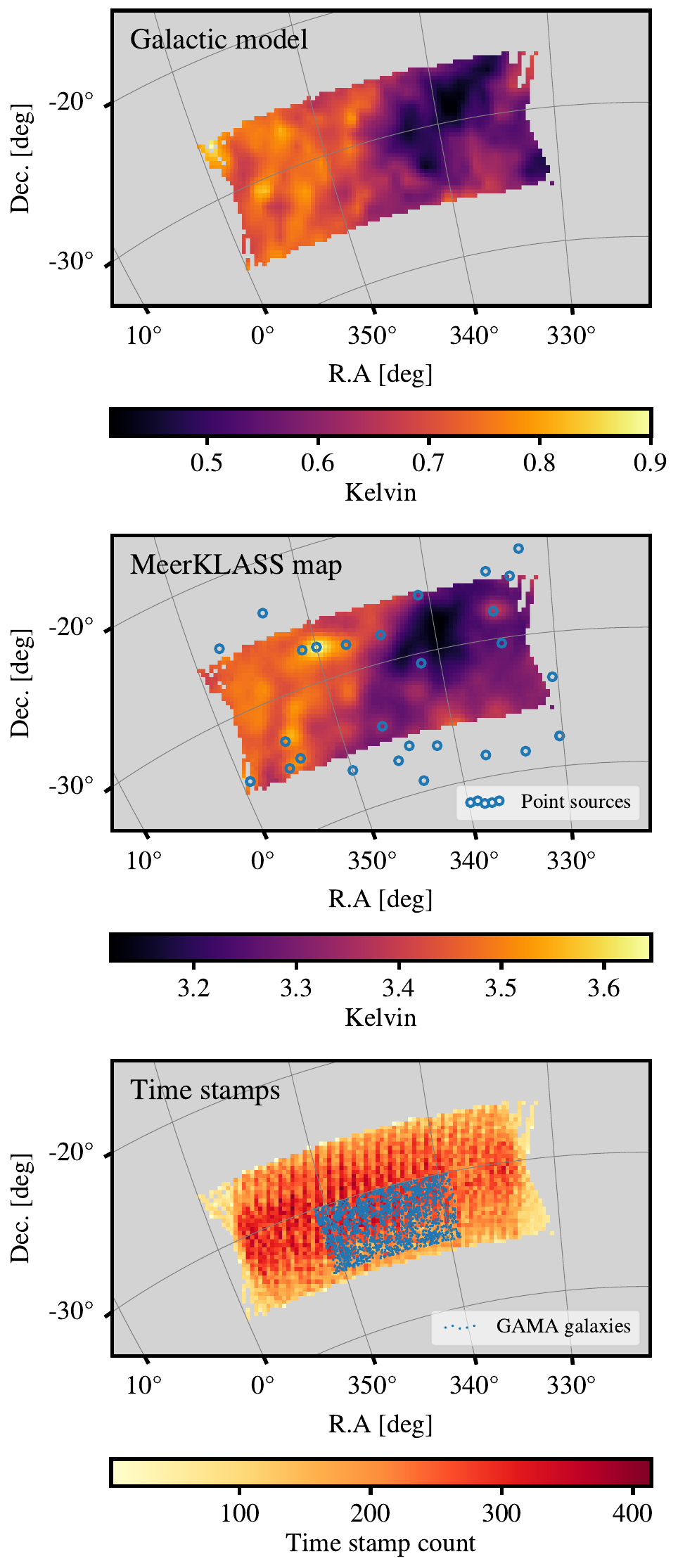}
    \caption{MeerKLASS L-band deep-field map at $1023\,$MHz (middle panel) in comparison with the Galactic emission (top panel) derived from PySM. Blue circles mark the positions of known point sources with flux ${>}\,1$\,Jy at 1.4\,GHz. The bottom panel shows the number of observational recordings i.e.\ \textit{timestamps} in each pixel at 1023\,MHz, from the surviving 27 blocks. Each timestamp is 2 seconds and we use this for noise estimation and weighting in the power spectrum analysis. Blue scatter points are the positions of all overlapping GAMA galaxies.}
    \label{fig:FGmap_and_counts}
\end{figure}

\begin{figure}
    \centering
    \includegraphics[width=1\linewidth]{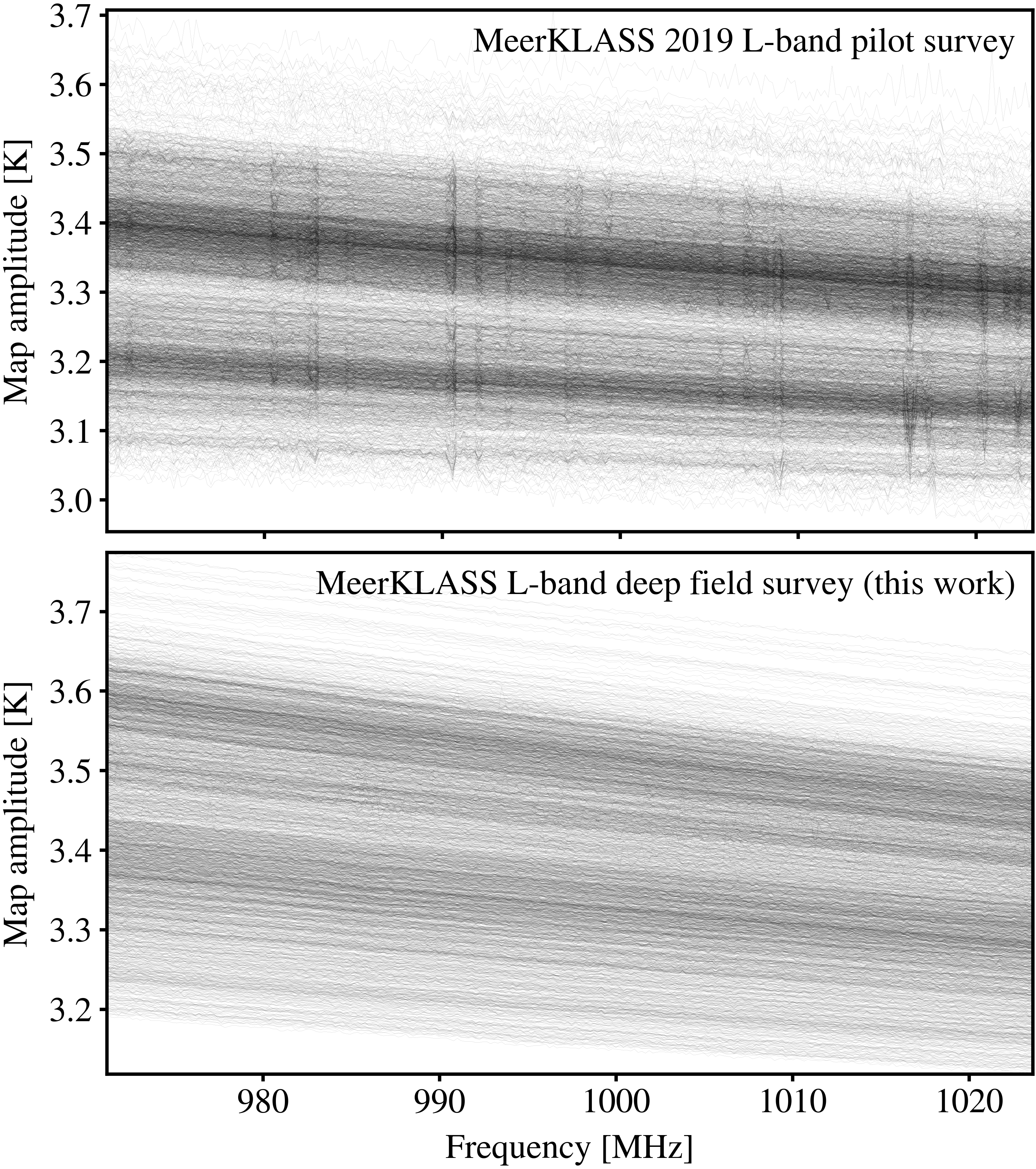}
    \caption{Amplitude of each map pixel along the line-of-sight for the previous MeerKLASS pilot survey (top panel) and the deep field used in this work (bottom panel. We only show the ${\sim}\,250$ channels most clean of RFI used in later cosmological analysis. This has the frequency range $971.2\,{<}\,\nu\,{<}\,1023.6\,\text{MHz}$ corresponding to $0.46\,{>}\,z\,{>}\,0.39$ for redshifted 21cm emission.}
    \label{fig:LoS_spectra}
\end{figure}

\subsection{Characteristics of final MeerKLASS data cubes}

\subsubsection{Final data cubes}

The combined calibrated temperature cubes (with a pixel size of 0.3\,deg and frequency resolution of 0.2\,MHz) are derived from all available scans and dishes. However, predominantly due to RFI, an identified repeated source of which we discuss in the following sub-section, only 27 of the 41 observation blocks yielded well-calibrated results. We mark them in the last column of \autoref{tab:block}. In \autoref{fig:block_statis} we present the number count of good (flagged time $<50\%$) channels within the 971–1075 MHz band (channels 550–1050) for each block and each dish. In particular, we found that antenna {\tt m059} has a higher chance of surviving calibration even when all other dishes fail.\footnote{For consistency with other MeerKAT work, we indicate MeerKAT's 64 antennae with this notation spanning {\tt m000}-{\tt m063}.} This may be because {\tt m059} is the easternmost dish of MeerKAT, located far from the array core, providing it with a more isolated environment.

We show the final sky temperature map from the combination of the remaining 27 scans after 5 loops of self-calibration at 1023\,MHz in the middle panel of \autoref{fig:FGmap_and_counts}. For comparison, the top panel shows the PySM diffuse Galactic emission model with the same sky coverage. We also show in the bottom panel the total number of time samples (across all dishes and observations) that were combined in each pixel. \textcolor{black}{The temperature offset between the top and middle panels is mainly attributed to the CMB, with additional contributions from point sources.}

To demonstrate the improvement in data quality from the deep-field intensity maps, \autoref{fig:LoS_spectra} plots the amplitude of each pixel along the line-of-sight as a function of frequency. We compare this to the previous MeerKLASS L-band pilot survey (top-panel) relative to the deep field used for this work. The maps should be dominated by the continuum synchrotron spectra, thus their smoothness is a rough proxy for how dominant any systematics are. There are numerous perturbations in the pilot survey spectra that have consequences for the effectiveness of foreground cleaning, required to access the cosmological \hi. However, for the deep field, the spectra appear smoother and well-calibrated. In addition to the improvements shown in \autoref{fig:loop_statis}, we also find a marginal performance improvement on the line-of-sight rms when using the self-calibration technique relative to the previous standard calibration in the deep-field. The average variance along the line-of-sight of each pixel decreases to 977.2\,mK$^2$ for the self-calibration, down from 984.3\,mK$^2$ for the standard calibration. 


\subsubsection{ RFI contaminations}\label{sec:zebra}
From the 41 observation blocks, we had to remove 14 blocks due to excessive, low-level RFI contaminations. In affected blocks, these contaminations were sometimes recognisable on the map level of individual dish data as horizontal stripes. After careful investigations, we have now identified the source of the RFI contamination as the downlink of the mobile communication tower (${\sim}\,54\,\text{km}$ away) that enters through the sidelobes and causes a non-linear gain effect in the entire band. Our aggressive flagging approach in this study, with the entire removal of 14 blocks, was to ensure high data quality for our analysis. We are currently developing a method to correct for this effect, which should allow us to use the currently flagged data for future projects.

\subsubsection{Noise level estimation}

Here we estimate the noise levels from the final maps to check if they are consistent with theoretical expectations. We use the final data cube, $T_{\rm sky}$, to estimate the noise level using the difference between four neighbouring channels (the so-called ABBA technique). By adopting the ABBA pattern, this difference statistic removes both constant and linearly varying signals to cancel signals detected by the receiver, leaving only the noise.  The noise level is estimated by
\begin{gather}
  \Delta T_i(\nu^{\star})= \frac{1}{2} \left( T^i_{\rm sky}(\nu) + T^i_{\rm sky}(\nu+\delta\nu) \right ) \nonumber\\
  ~~~~~~~~~~~~~~~~~-\frac{1}{2}\left ( T^i_{\rm sky}(\nu-\delta\nu)+T^i_{\rm sky}(\nu+2\delta\nu)\right),
\end{gather}
where the index $i$ goes over all pixels in the map and $^{\star}$ means the combined result from the four channels $[\nu{-}\delta\nu,\,\nu,\,\nu{+}\delta\nu,\,\nu{+}2\delta\nu]$. 
From the observation data, we take a weighted rms,
\begin{gather} \label{eq:rms_obs}
     \Delta T_{\rm RMS}^2(\nu^{\star}) = 
     \frac {N_\text{pix}} {N_\text{pix} - 1}
     \left (\frac {\sum_i w_i\Delta T^2_i} {\sum_i w_i} -
     \left[\frac {\sum_i w_i\Delta T_i} {\sum_i{w_i}}\right]^2 \right),
\end{gather}
where the sum is over the number of pixels in the map, $N_\text{pix}$, and we have suppressed the dependence of $w_i$ and $\Delta T_i$ on $\nu^\star$. For the weight, $w_i$, we used the theoretical expected variance itself, $w_i(\nu^{\star})=1/\sigma^2_{{\rm th},i}(\nu^{\star})$.
The corresponding variance in each pixel would be,
\begin{gather}
\sigma^2_{{\rm th},i}(\nu^{\star})=\frac{1}{4} \Big(\sigma^2_{{\rm th},i}(\nu)+\sigma^2_{{\rm th},i}(\nu{+}\delta\nu)+\sigma^2_{{\rm th},i}(\nu{-}\delta\nu)\nonumber\\
     ~~~~~~~~~~~~~~~~~~~ + \sigma^2_{{\rm th},i}(\nu{+}2\delta\nu) \Big).
\end{gather}
Here the noise in the initial polarised TOD is given by the radiometer equation 
\begin{equation}
    \sigma^2_{\rm th,pol}(t,\nu) = \frac{T^2_{\rm sys}(t,\nu)}{\delta\nu \, \delta t},
\end{equation}
where $\delta\nu\txteq{=}0.209\,\text{MHz}$ is the frequency width of the MeerKAT L-band channels and $\delta t\txteq{=}2$\,s is the length of each time stamp recording. 
For the system temperature we use the calibrated data itself, $T_{\rm sys}\,{=}\,T_{\rm cal,HH}(t,\nu)$ or $T_{\rm cal,VV}(t,\nu)$  for horizontal and vertical polarisation data, respectively. 
Thus the noise in the initial intensity TOD is
\begin{equation}
    \sigma^2_{\rm th}(t,\nu) = \frac{T^2_{\rm cal,HH}(t,\nu)+T^2_{\rm cal,VV}(t,\nu)}{4\, \delta\nu \, \delta t}.
\end{equation}
In order to get the final data cube, $T^i_{\rm sky}(\nu)$, the data goes through several stages of averaging. We propagate the variance taking into account this averaging in order to get to the final $\sigma^2_{{\rm th},i}(\nu)$.

The theoretical rms which can be compared to $\Delta T_{\rm RMS} (\nu^{\star})$ in \autoref{eq:rms_obs}, is then simply
\begin{gather}
     \sigma^2_{\rm th} (\nu^{\star}) = N_\text{pix} \left ( \sum_i \sigma^{-2}_{{\rm th},i}(\nu^{\star})  \right )^{-1}.
\end{gather}
More details can be found in Section 5.6 of \citet{Wang:2020lkn}.

\begin{figure}
\centering
\includegraphics[width=\columnwidth]{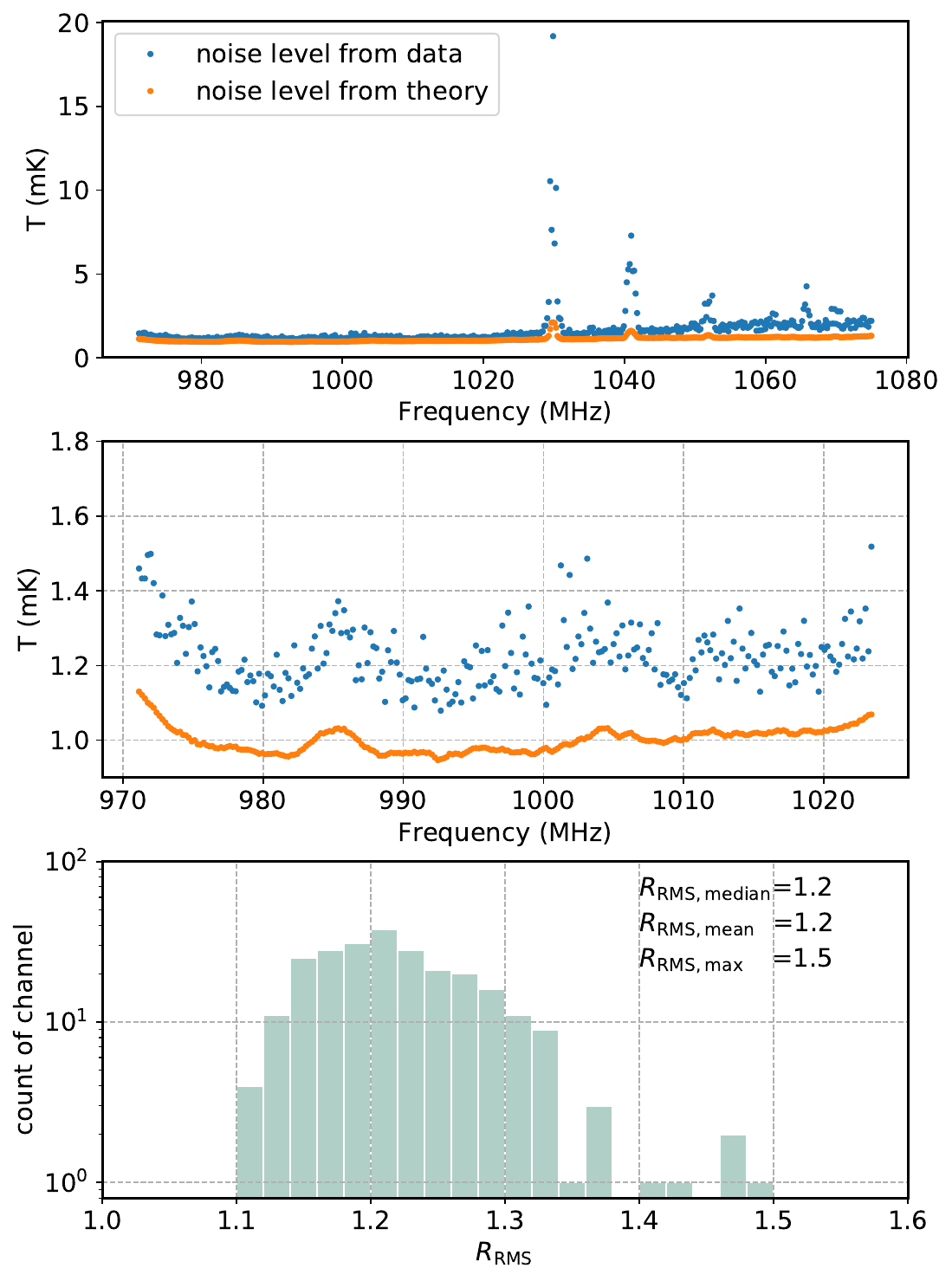}
\caption{Comparison between noise levels derived from calibrated observational data ($\Delta T_{\rm RMS} (\nu^{\star})$) and from theory equation ($\sigma_{\rm th} (\nu^{\star})$) in 971–1075 MHz (upper panel),  a zoom-in for above in 971–1023 MHz (middle panel), and the histogram of the ratio $R_{\rm RMS}=\Delta T_{\rm RMS} (\nu^{\star})/ \sigma_{\rm th} (\nu^{\star}) $ in 971–1023 MHz (lower panel).}
\label{fig:rms}
\end{figure} 

In the upper panel of \autoref{fig:rms}, we show the noise levels (from data and theory) in the 971–1075\,MHz frequency band (channels 550–1050). Note here both the real and theory values are related to the real $T_{\rm sys}$ and hit counts in corresponding map pixels. The count reduction due to RFI flagging, and unflagged low-level RFIs that are hiding in the data would cause wiggles. Four groups of spikes are present at higher frequencies, while below 1023\,MHz, there is much less scatter compared to the previous MeerKLASS pilot survey result. We therefore chose to restrain our cosmological analysis to the 971–1023\,MHz region, for which we show the noise level in the middle panel, finding a median of $\Delta T_{\rm RMS}\txteq{=}1.21\,$mK. In the lower panel of \autoref{fig:rms}, we also show the histogram of the ratio 
$R_{\rm RMS}\txteq{=}\Delta T_{\rm RMS} (\nu^{\star}) / \sigma_{\rm th} (\nu^{\star})$ in the 971–1023\,MHz band. We find a median value of $R_{\rm RMS}\txteq{=}1.2$, while some channels are noisy and $R_{\rm RMS}$ can reach $1.5$.

\subsection{GAMA spectroscopic galaxies}\label{sec:GAMAobs}

Overlapping with the MeerKLASS deep field, is a catalogue of spectroscopic galaxies recorded in the Galaxy and Mass Assembly (GAMA) survey. GAMA began observing in February 2008 \citep{Driver:2010zb,LiskeGAMA15}. The survey aimed to construct catalogues of galaxies to test the cold dark matter paradigm and probe galaxy evolution \citep{Driver:2009hm}. For our purposes, we exploit the fact that these galaxies will trace dark matter and should therefore correlate with the overlapping MeerKLASS intensity maps after foreground cleaning.  

We use the fourth and final data release from the Galaxy and Mass Assembly survey, GAMA DR4 \citep{Driver:2022vyh}. This includes galaxies from the GAMA II survey phase which enlarged the GAMA I equatorial survey, by including extra survey regions. One of these extra regions was the 23\,hr (G23) field, which completely overlaps in area with the MeerKLASS L-band deep field, shown by the blue scatter points in the bottom panel of \autoref{fig:FGmap_and_counts}, which represent the GAMA G23 galaxies relative to the observational time footprint map for the L-band deep field. The G23 field covered $339\,\deg\,{<}\,\text{R.A.}\,{<}\,351\,\deg$ and ${-}35\,\deg\,{<}\,\text{Dec.}\,{<}\,{-}30\,\deg$. The magnitude limit of the field was reduced from the planned 19.8 in $r$-band to 19.2 in $i$-band due to limited observational time, but this allowed the GAMA II survey to obtain a high $94.2\%$ redshift completion in the G23 region  \citep{LiskeGAMA15}.

\begin{figure}
    \centering
    \includegraphics[width=1\linewidth]{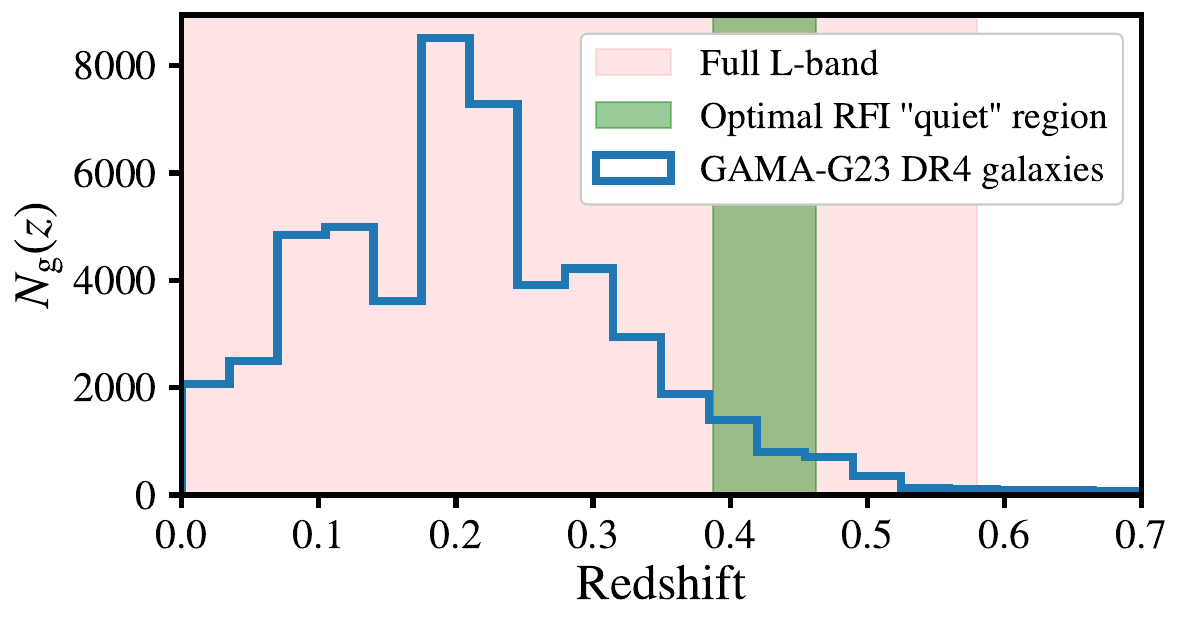}
    \caption{Redshift distribution of GAMA G23 galaxies in the 23hr field. The accessible redshifted 21cm emission for the full L-band range is shown by the light-pink shading. The region used in this work between $0.39\,{<}\,z\,{<}\,0.46$ ($1023.6\,{<}\,\nu\,{<}\,971.2\,\text{MHz}$) is shown by the darker green shading, in which 2269 GAMA galaxies exist.}
    \label{fig:GAMA_Nz}
\end{figure}

The GAMA DR2 only provides random catalogues that emulate the survey selection function of the real galaxies for the G09, G12, and G15 regions, (not the G23 used in this work). Ideally, these would be used to construct a survey completion window function, which would be used in the cross-power spectrum estimation. In the absence of this, we assume a uniform survey completion across the G23 area and a constant distribution with redshift, given the very flat distribution for the redshift range used in this analysis (green region of \autoref{fig:GAMA_Nz}). Given our signal-to-noise levels, survey masks will be a sub-dominant effect in the cross-power, thus this is an adequate approximation to make.

Due to our aggressive flagging strategy on the MeerKLASS L-band data, the overlap region in redshift is restricted to $0.39\,{<}\,z\,{<}\,0.46$. As shown by \autoref{fig:GAMA_Nz}, this catches the GAMA survey galaxy redshift distribution, $N_\text{g}(z)$, near the end of its optimal range, but still provides 2269 galaxies to cross-correlate with. At this redshift range, the comoving (volumetric) number density of the GAMA galaxies is $\bar{n}_\text{g}\,{=}\,4.8\,{\times}\,10^{-4}\,h^{3}\,\text{Mpc}^{-3}$.

\section{Simulation of mock data}\label{sec:Mocks}

To accompany the calibrated observations outlined in \secref{sec:ObsData}, we produce a suite of 500 mock simulations, used for validating the analysis pipeline, calculating signal loss from foreground cleaning and estimating covariance in the power spectrum measurements. To create this large number of mocks we generate lognormal density fields \citep{ColesLognormal1991} from which a map of \hi\ brightness temperatures can be cut, and galaxies Poisson sampled onto for the spectroscopic galaxy counterpart. 
\textcolor{black}{The lognormal distribution has been shown to be a more accurate representation of the PDF for the distribution of cosmic matter, relative to a more simple Gaussian approximation \citep{Shin:2017cwu,Agrawal:2017khv}. The consequences of only using a Gaussian approximation have not been fully investigated but could potentially impact the accuracy of covariance and signal loss estimation. We therefore adopt the lognormal approach since this adds no extra computational demand yet will provide a more realistic suite of mocks.
}

An important stage of the analysis pipeline is regridding the intensity mapping voxels with celestial sky coordinates (R.A., Dec., $\nu$) onto a regular 3D Cartesian grid with lengths $l_\text{x}$, $l_\text{y}$, $l_\text{z}$ in $h^{-1}\text{Mpc}$, on which a Fast Fourier Transform (FFT) can be applied for the power spectrum estimation \citep{Cunnington:2023aou}. We therefore require the mocks to begin in celestial coordinates so we can validate the regridding process. To achieve this, we begin by fitting a Cartesian cube around the full MeerKLASS deep-field footprint, creating a grid with dimensions $l_\text{x},l_\text{y},l_\text{z}\,{=}\,774.9,335.7,252.2\,h^{-1}\text{Mpc}$ over $n_\text{cell}\,{=}\,512^3$ cells. We use a large number of cells for this initial grid to approximately emulate a continuous sky, aiming to avoid the simulated maps resolving the finite input grid structure. Onto this grid, we simulate a lognormal density field. This is done by interpolating $\boldsymbol{k}$-modes from a model power spectrum onto the FFT grid. The \hi\ model power spectrum is given by
\begin{equation}\label{eq:PkHImodel}
    P_{\hi}(\boldsymbol{k})\equiv P_{\hi}(k,\mu) = \frac{\overline{T}_{\hi}^2\,b_{\hi}^2\,\left(1 + f\,\mu^2/b_\hi\right)^2\,P_\text{m}(k)}{1 + \left(k\,\mu\,\sigma_\text{v}/H_0\right)^2}\,.
\end{equation}
The anisotropy in this model makes it convenient to parameterise $\boldsymbol{k}$ using $\mu$, defined as the cosine of the angle between the line-of-sight and $\boldsymbol{k}$. The \hi\ parameters are the mean \hi\ gas temperature ($\overline{T}_\hi$) and the \hi\ bias ($b_\hi$). $f$ is the growth rate of structure and is included to model linear redshift space distortions (RSD) which enhance clustering across all scales. The denominator is included to approximately model the non-linear RSD; the so-called Fingers-of-God (FoG), which suppresses clustering at small scales. The FoG are modulated by $\sigma_\text{v}$, which is the velocity dispersion of the tracer, in this case, the \hi\ gas. All of these quantities have a redshift dependence, which we omit for brevity; they will also have little variation given our narrow redshift range. Lastly, $H_0$ is the Hubble parameter at $z\,{=}\,0$, known as the Hubble constant. An alternative process for generating brightness temperatures is pursued for our stacking analysis discussed later in the paper.
More realistic lognormal line intensity mapping mocks are available. For example, in future work, we intend to adopt the approach in \citet{Niemeyer:2023yeu} where luminosities from the spectral lines are painted given a luminosity function. This has been shown to reproduce better RSD for galaxy distributions and models the whole temperature distribution expected in the maps.

\textcolor{black}{To obtain the \hi\ correlation function, $\xi_\hi$, needed for the lognormal transform we first inverse Fourier transform $P_\hi/\overline{T}^{2}_\hi$, where $P_\hi$ is the model in \autoref{eq:PkHImodel} and we factor out the background average temperature $\overline{T}_\hi$ to remove the temperature units. The correlation function is 
}
%
then log-transformed via $\xi^\prime_\hi\,{=}\,\ln(1\,{+}\,\xi_\hi)$. Fourier transforming $\xi^\prime_\hi$ back gives the modified lognormal power spectrum $P^\prime_\hi$\textcolor{black}{, without temperature units}. Gaussian random numbers are then sampled into each cell of the Fourier grid $\delta_\text{G}(\boldsymbol{k})$ with variance $\sigma^2(\boldsymbol{k}) \propto P^\prime_\hi(\boldsymbol{k})$. This Gaussian random field is then inverse Fourier transformed to real space giving $\delta_\text{G}(\boldsymbol{x})$ from which the lognormal \hi\ 
\textcolor{black}{density field is given by}

%
\begin{equation}
    \textcolor{black}{\delta_\hi(\boldsymbol{x})} = \exp\left[\delta_\text{G}(\boldsymbol{x}) - \frac{\sigma^2_\text{G}}{2}\right] - 1\,,
\end{equation}
where $\sigma_\text{G}^2$ is the measured variance of the Gaussian random number field $\delta_\text{G}$.
\textcolor{black}{This is converted into the \hi\ brightness temperature field given by $\delta T_\hi\txteq{=}\overline{T}_\hi \delta_\hi$.}
Further details on this process can be found in the Appendix A of \citet{Beutler:2011hx}. For the galaxy over-density field $\delta_\text{g}(\boldsymbol{x})$, we follow the same procedure except the input model is for the galaxy power spectrum, given by
\begin{equation}
    P_\text{g}(\boldsymbol{k})\equiv P_\text{g}(k,\mu) = \frac{b_\text{g}^2\,\left(1 + f\,\mu^2/b_\text{g}\right)^2\,P_\text{m}(k)}{1 + \left(k\,\mu\,\sigma_\text{v}/H_0\right)^2}\,,
\end{equation}
where $b_\text{g}$ is the bias for the galaxies. We assume that the velocity dispersion $\sigma_\text{v}$ for the \hi\ gas and galaxies is the same for simplicity. This will be sufficient for this work where signal-to-noise is low at the small scales where $\sigma_\text{v}$ has the main effect.

The galaxy over-density $\delta_\text{g}$ is unit-less whereas the intensity maps $\delta T_\hi$ are brightness temperature fluctuations measured in mK. We use the same seed in the random number generation to ensure that both the $\delta T_\hi$ and $\delta_\text{g}$ correlate. The lognormal code we adopt utilises some functions from the publically available package \texttt{powerbox} \citep{Murray2018}.\footnote{\href{https://github.com/steven-murray/powerbox}{\texttt{github.com/steven-murray/powerbox}}}

\subsection{Transforming between Cartesian grids and sky maps}\label{sec:RegriddingMocks}

As already mentioned, the galaxies and the intensity mapping voxels from observations are in celestial coordinates at a certain frequency (or redshift) i.e.\ $(\boldsymbol{\theta},\nu)\,{\equiv}\,(\theta_\text{R.A.},\theta_\text{Dec.},\nu)$. However, for the final power spectrum estimation, we require the fields be in a Cartesian space with dimensions ($l_\text{x},l_\text{y},l_\text{z}$) in $h^{-1}\text{Mpc}$. For this we follow the recent work in \citet{Cunnington:2023aou} which we discuss further in \secref{sec:PkEst}.

Creating lognormal mocks in celestial coordinates that emulate observations requires the inverse of this step i.e.\ $\delta T_\hi(\boldsymbol{x})\,{\rightarrow}\,\delta T_\hi(\boldsymbol{\theta},\nu)$, since the lognormal fields are generated onto Cartesian grids measured in $h^{-1}\text{Mpc}$. Transforming mock galaxy coordinates (the Poisson sampling of which is discussed in \secref{sec:Galmocks}) from $(l_\text{x},l_\text{y},l_\text{z})\,[h^{-1}\text{Mpc}]\,{\rightarrow}\,(\theta_\text{R.A.},\theta_\text{Dec.},\nu)$ is trivial. We can use \texttt{astropy}\footnote{\href{https://docs.astropy.org/en/stable/index.html}{\texttt{docs.astropy.org}}} \citep{Astropy} transformation routines on each galaxy's coordinate. However, transforming the tessellating field of Cartesian cells, each with an associated brightness temperature, into a \textit{gap-less} map of voxels with identical celestial coordinates and frequency binning as the real MeerKAT maps, requires a resampling of the field. For this we also utilise the techniques and public code\footnote{\href{https://github.com/stevecunnington/gridimp}{\texttt{github.com/stevecunnington/gridimp}}} outlined \citet{Cunnington:2023aou}. 

This process involves a Monte-Carlo particle sampling technique where each Cartesian cell is filled with $N_\text{p}$ particles, randomly distributed within the cell. The particles are assigned the brightness temperature of the cell they fall within. The particles are then transformed to celestial coordinates ($\theta_\text{R.A.},\theta_\text{Dec.},z$) and added into the map voxels associated with their celestial coordinate and the frequency channel corresponding to their redshift. Each voxel is normalised by the number of particles falling within it. Providing enough sampling particles per cell are used (in this work we use $N_\text{p}\,{=}\,10$) and the Cartesian grid is of sufficient resolution (hence our choice of $n_\text{cell}\,{=}\,512^3$), it will provide an intensity map in celestial sky coordinates, that traces the fluctuations generated by the lognormal field. 

Throughout this work, we do not require the necessary extensions of higher-order particle assignment or field interlacing to suppress the effects of aliasing \citep[as trialled in][]{Cunnington:2023aou}, since these effects are below the current statistical noise from this data. We are however, still required to implement a correction to the power spectrum estimation from the Nearest-Grid-Point assignment which smooths the field, which we formalise in \secref{sec:PkEst}.

\subsection{\hi\ intensity map mocks}\label{sec:HIIMmocks}

Using the regridding process described in the previous section we are provided with \hi\ intensity mapping mocks $\delta T_\hi({\boldsymbol{\theta},\nu})$, with identical coordinates and dimensions as the observed MeerKAT intensity maps. From here we can add in observational effects and simulate various systematics. The most dominant effect comes from foreground contamination. Systematics will have a large influence on foreground contamination, and this can be extremely complex. Therefore, this is not something we attempt to directly simulate in this work since we risk under-estimating the problem. Instead, we use the observational data itself which has the relevant systematics contained within, albeit only one realisation of them. However, adding these directly onto different realisations of \hi\ mocks, still recreates stochasticity in the \hi\ signal's response to systematics. We discuss this further in \secref{sec:FGcleaning}.

Another important addition to the mocks will be a smoothing to the intensity maps corresponding to the angular pixelisation. This is necessary to emulate the radio telescope's beam, which for single-dish intensity mapping is quite broad giving a low angular resolution to the maps and damping small-scale fluctuations perpendicular to the line-of-sight. In this work, we assume the MeerKAT beam profile to be Gaussian, which is a simplification but still valid given the signal-to-noise at small scales where the beam suppresses signal below the noise. We will investigate implementing a more realistic MeerKAT beam profile in future work \citep[see e.g.][]{2021MNRAS.502.2970A,Matshawule:2020fjz,2022MNRAS.509.2048S}. The assumed Gaussian beam profile has a frequency ($\nu$) dependent standard deviation in radians given by
\begin{equation}\label{eq:sigbeamsize}
    \sigma_\text{beam}(\nu) = \frac{1}{2\sqrt{2\ln 2}}\frac{c}{\nu D_\text{dish}}\,,
\end{equation}
where we use $D_\text{dish}\,{=}\,13.5\,\text{m}$ for the diameter of the MeerKAT dishes and $c$ is the speed of light. To emulate the beam effects we smooth the mock maps by creating a 2D Gaussian array kernel centred on the median angular pixel $\theta_0$ given by
\begin{equation}\label{eq:GaussKernel}
    \pazocal{B}_\text{beam}(\boldsymbol{\theta},\nu_i) = \exp\left[-\frac{1}{2}\left(\frac{\boldsymbol{\theta} - \theta_0}{\sigma_\text{beam}(\nu_i)}\right)^2\right]\,.
\end{equation}
This is normalised such that $\sum_{\boldsymbol{\theta}}\pazocal{B}_\text{beam}\txteq{=}1$, so that convolution with this beam kernel does not change the total brightness temperatures within the maps. The \hi\ intensity map mocks are convolved with \autoref{eq:GaussKernel} at each frequency $\nu_i$ to give the beam-smoothed \hi\ map;
\begin{equation}\label{eq:beam_convolve}
    \delta T_{\circledast\hi}(\boldsymbol{\theta},\nu_i) = \delta T_\hi(\boldsymbol{\theta},\nu_i)\circledast\pazocal{B}_\text{beam}(\boldsymbol{\theta},\nu_i)\,,
\end{equation}
where the $\circledast$ symbol represents a convolution \textcolor{black}{over the angular $\boldsymbol{\theta}$ space}.

To realistically emulate additional components such as foregrounds, noise and their response to systematics, we add the \textit{real} observations to each mock iteration. This is to avoid the requirement of simulating the complex and poorly understood systematics. 

In some cases, mostly for studying the \hi\ auto-correlation, it will be useful to simulate \textit{just} the thermal noise originating from the thermal signature of the receivers.
We do this by generating Gaussian random noise the power of which can then be measured with the same pipeline as the rest of the analysis, providing a robust estimate of the thermal noise floor. We begin with an analytical description of the thermal noise fluctuations per pixel using the radiometer equation (as in \secref{sec:ObsData} but now at \textit{map-level})
\begin{equation}\label{eq:sigma_N}
    \sigma_\text{N}(\boldsymbol{\theta},\nu) = \frac{T_\text{sys}(\nu)}{\sqrt{2\,\delta\nu\,t_\text{p}(\boldsymbol{\theta},\nu)}}\,,
\end{equation}
where $t_\text{p}$ is the summed observation time contained in each voxel, which can be taken directly from the time stamp counts as shown in \autoref{fig:FGmap_and_counts}, with each count representing $\delta t \txteq{=}2\,$sec. This emulates the real data by creating a non-uniform noise map, generating higher noise around the edges where scan coverage is lower. For the system temperature, we follow \citet{Cunnington:2022ryj} and use a model tuned to coincide with the estimated system temperature of the MeerKAT telescope \citep{Wang:2020lkn}\footnote{See \href{https://skaafrica.atlassian.net/wiki/x/AYCHE}{\texttt{skaafrica.atlassian.net/wiki/x/AYCHE}}}, assuming 
\begin{equation}
    T_{\mathrm{sys}}(\nu)=T_{\mathrm{rx}}(\nu)+T_{\mathrm{spl}}+T_{\mathrm{CMB}}+T_{\mathrm{gal}}(\nu)\,,
\end{equation}
where we assume the contribution from our own galaxy is $T_{\mathrm{gal}}\txteq{=}25\text{K}\,(408 \mathrm{MHz} / \nu)^{2.75}$, the CMB temperature is $T_\text{CMB}\txteq{=}2.725\,\text{K}$, as in \secref{sec:ObsData}. The \textit{spill-over} temperature is assumed as $T_\text{spl}\txteq{=}3\,\text{K}$ and the receiver temperature we model as
\begin{equation}
    T_{\mathrm{rx}}(\nu) = 7.5\mathrm{K}+10\mathrm{K}\left(\frac{\nu}{\mathrm{GHz}}-0.75\right)^2\,.
\end{equation}
Due to the narrow frequency band we use, this gives a near-constant system temperature with the range $15.97\txteq{<}T_\text{sys}\txteq{<}16.02\,[\text{K}]$. Gaussian random numbers can then be generated across the MeerKLASS deep-field map with the voxel dependent rms of $\sigma_\text{N}(\boldsymbol{\theta},\nu)$.

\subsection{Galaxy mock catalogues}\label{sec:Galmocks}

To accompany the mock intensity maps, we also require corresponding galaxy mocks with the same correlated clustering.  Rather than transforming the Cartesian galaxy over-density field $\delta_\text{g}(\boldsymbol{x})$ into sky coordinates, it is more realistic to instead Poisson sample the field to generate a catalogue of galaxies with the same number density as the GAMA G23 survey. This also allows additional validity checks such as testing the stacking pipeline which cannot be done without a catalogue of galaxy coordinates. 

We generate the galaxy catalogue coordinates initially in the Cartesian domain to exploit the high $n_\text{cell}\,{=}\,512^3$ resolution. We first calculate the average galaxy counts in cells $\langle{n}_\text{g}\rangle$ for the real GAMA G23 region in the redshift range where there is overlap with the intensity map frequency range. We then take the galaxy survey selection window function, which for the GAMA G23 region, we have to assume is constant between $339\,{<}\,\text{R.A.}\,{<}\,351\,\deg$ and ${-}35\,{<}\,\text{Dec.}\,{<}\,{-}30\,\deg$ (see discussion in \secref{sec:GAMAobs}). We then transform this window function into the high resolution Cartesian space. A galaxy number counts field $n_\text{g}(\boldsymbol{x})$ is then given by
\begin{equation}
    n_\text{g}(\boldsymbol{x}) =  \langle n_\text{g}\rangle\left(\delta_\text{g}(\boldsymbol{x}) + 1\right) W_\text{g}(\boldsymbol{x})\,,  
\end{equation}
where we use the transformed window function $W_\text{g}(\boldsymbol{x})$ and the real GAMA average counts $\langle n_\text{g}\rangle$ to ensure the mock galaxy counts will share the same survey selection as the real galaxy survey, with the same number density. In this process we normalise the window function $W_\text{g}$ such that its maximum value is $1$, so for the GAMA G23 survey, this becomes a binary mask of 0 and 1. We then Poisson sample the $n_\text{g}$ field to introduce some stochasticity which we expect to be present in the real galaxy catalogues. This provides an integer value in each Cartesian cell for the galaxy counts. The last step is to assign random Cartesian coordinates within the cell boundaries for the $n_\text{g}(\boldsymbol{x})$ galaxies within each cell. At this stage we rely on the Cartesian field having sufficiently high resolution such that the random assignment of a coordinate within the cell does not shift the galaxies away from the peaks of the density fluctuations. The catalogue of galaxy coordinates (x, y, z [$h^{-1}$Mpc]) can then be transformed into celestial coordinates $(\theta_\text{R.A.},\theta_\text{Dec.},z)$. \newline
\\

\begin{figure}
    \centering
    \includegraphics[width=1\linewidth]{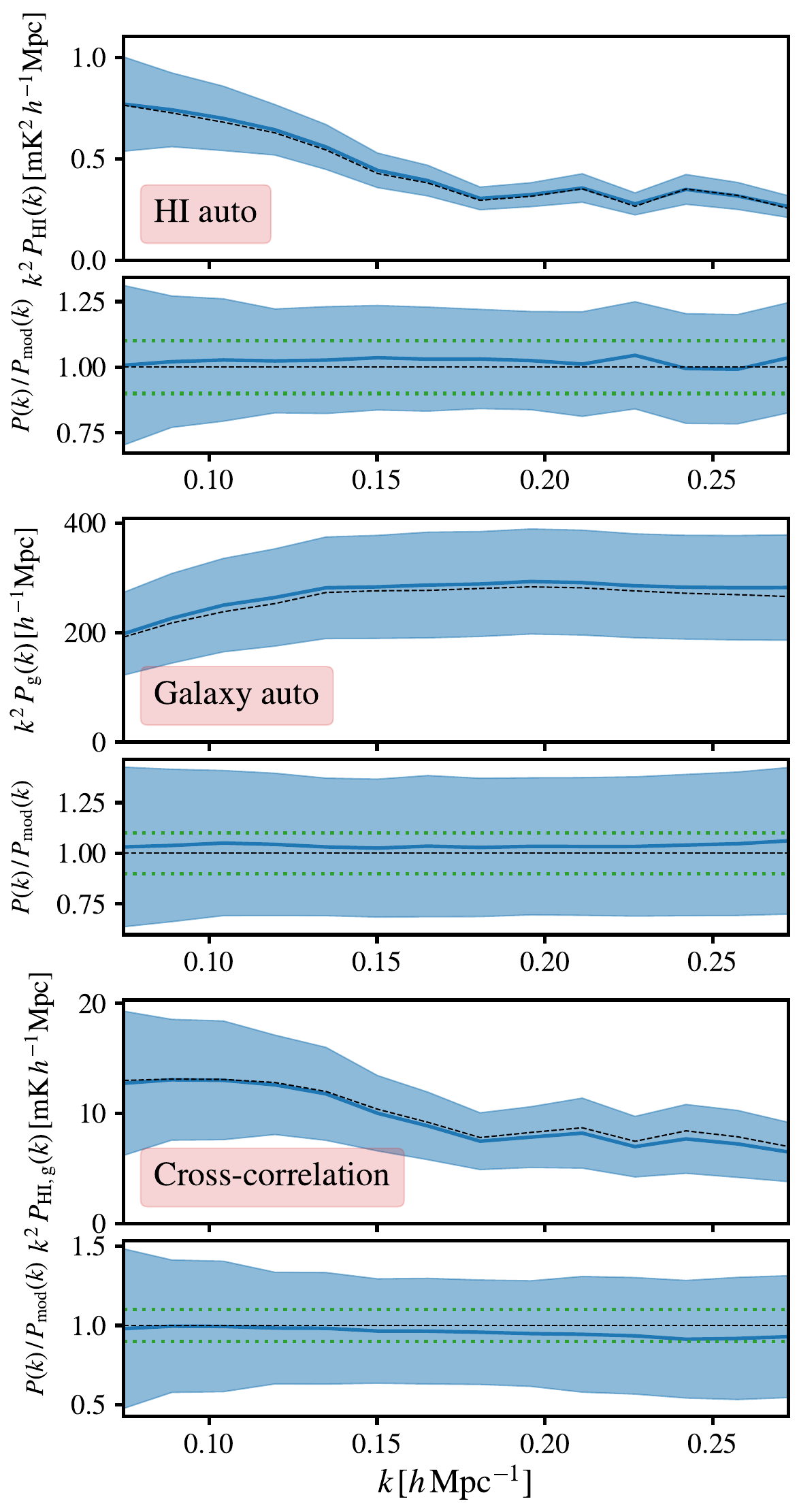}
    \caption{Averaged power spectra (blue solid line) of the 500 regridded mocks and agreement with the fiducial model (black dashed line). Mocks include the beam but no other systematics. Green dotted lines in the bottom panels mark the edges of the $10\%$ residual region, while the shaded regions mark the diagonal variance inferred from the simulations.}
    \label{fig:PksFromMocks}
\end{figure}

\noindent In this work, we generate 500 mock intensity maps with corresponding mock GAMA galaxy catalogues. As some early validation of these mocks, we show in \autoref{fig:PksFromMocks} how the power spectra for both the \hi\ intensity map and galaxy auto-correlations, along with their cross-correlation, agree with the fiducial model (black dashed line). This demonstrates that our estimator is unbiased and processes such as the regridding are performing to within an acceptable tolerance. The power spectrum estimation process used here on the mocks is identical to that used on the real data and is detailed later in \secref{sec:PkEst}, along with the modelling.

The mocks used in the test of \autoref{fig:PksFromMocks} do not include any additional systematics beyond the telescope beam in the intensity maps. We will investigate the inclusion of foregrounds and systematics in \secref{sec:FGcleaning}. The large scatter shown by the shaded blue region, which represents the $1\,\sigma$ range across the 500 mocks, is therefore driven by cosmic variance due to the small volume covered by the data. We see that the errors for the galaxy power spectrum appear larger than the \hi\ intensity maps. This is because we have not introduced any thermal noise (or other observational or systematic effects) into the intensity maps. However, the galaxy catalogues will have shot noise contributing to the uncertainties. This nicely showcases the potential for \hi\ intensity mapping if noise and systematics can be controlled, however, this is not the purpose of this exercise so is not a point we will labour. There are also additional caveats when considering high-density tracers such as \hi. As shown by \citet{Foreman:2024kzw}, despite the low shot noise, stochastic noise and non-linear bias in the \hi\ field can limit the potential of constraints inferred from the power spectrum.

The green dashed lines in the residual panels mark the regions where there is ${>}\,90\%$ accuracy relative to the model which we can see is achieved at all scales for all three cases. This is also well within the statistical uncertainty (blue-shaded regions). The largest residual comes in the cross-correlation where discrepancies seem to increase with $k$ nearly reaching the $-10\%$ boundary. We speculate that this may be caused by the Poisson sampling of mock galaxies followed by random assignment of coordinates within the cell. This extra stochasticity is not repeated in the \hi\ intensity map mocks and could therefore be generating a pseudo-cross-correlation coefficient, which would depart from $r\,{=}\,1$ at smaller scales (higher-$k$). However, this appears to be a minor effect. Furthermore, this could be seen as a feature, since we expect $r\,{<}\,1$ in the real observations, which will also have a scale dependence. For these reasons, we deem this to be a suitable result for the purposes of this work, and defer a higher precision investigation and validation to future work.  

\section{Foreground cleaning}\label{sec:FGcleaning}


Foreground cleaning \hi\ intensity maps is a well-studied problem \citep{Santos:2004ju,Ansari:2011bv,Wolz:2013wna,Alonso:2014dhk,Shaw:2014khi,Olivari:2015tka,Carucci:2020enz,Cunnington:2020njn,2022MNRAS.509.2048S,Soares:2021ohm,Irfan:2021bci}, and generally \textit{blind} cleaning techniques are the currently preferred method for single-dish experiments. As their name suggests, blind methods do not rely on prior knowledge of the foreground structure. This is advantageous for current experiments because the foregrounds at these frequencies lack observational constraints. Furthermore, blind techniques avoid placing overwhelmingly stringent calibration requirements on the data, since the dominant foregrounds require beyond sub-percent precision modelling for a successful subtraction. Blind techniques instead work by exploiting the foreground dominance and their frequency-correlated structure to assume that they can be reduced to a distinct set of correlated modes separated from the oscillating, sub-dominant fluctuations which will contain most of the \hi\ information.

We outline the formalism for our blind foreground cleaning approach in \secref{sec:PCAcleaning}, but first we discuss two \textit{pre-processing} steps before foreground cleaning. These are designed to improve the foreground clean by reducing the impact from residual RFI or systematics, plus any known frequency-dependent observational effects. 

\subsection{Reconvolution} \label{subsec:reconvolution}

All radio intensity mapping experiments will have a chromatic beam whose size evolves with frequency. This can add to the number of degrees of freedom in the radial foreground structure, inhibiting foreground removal since more modes need to be removed to perform a suitable clean. Smoothing the maps differently in each frequency channel can homogenise the angular resolution, giving them a consistent effective beam size at all frequencies. This reconvolution should also suppress residual contamination manifesting as localised amplitude spikes in the temperature maps, smoothing them out so there are fewer discontinuous fluctuations which will impact the PCA and Fourier transforms for power spectrum estimation.

The reconvolution step involves convolving the intensity maps, one frequency channel $\nu_i$ at a time with the kernel
%
\begin{equation}\label{eq:ResmoothKernel}
    \pazocal{K}(\boldsymbol{\theta},\nu_i)=
    \exp\left[-\frac{1}{2}\frac{\left(\boldsymbol{\theta} - \theta_0\right)^2}{\gamma \sigma_{\max}^{2}-\sigma_\text{beam}^{2}(\nu_i)}\right]\,,
\end{equation}
where $\sigma_\text{beam}$ is the beam size as previously defined in \autoref{eq:sigbeamsize}, $\sigma_\text{max}$ is the maximum $\sigma_\text{beam}(\nu_\text{min})$ value at the lowest frequency,
and $\gamma$ is a scaling factor which governs how much the final effective resolution decreases. The reconvolution kernel is normalised such that the sum over all pixels is unity i.e.\ $\Sigma_{\boldsymbol{\theta}}\pazocal{K}(\boldsymbol{\theta},\nu_i)\,{=}\,1$. 

As in previous work \citep{Cunnington:2022uzo}, we use a weighted reconvolution to mitigate the impact of edge pixels with low scan coverage. From hereon, the weights for the intensity maps derive directly from the time stamp count in each pixel, as shown in \autoref{fig:FGmap_and_counts}. The weighted reconvolved intensity maps are given by
\begin{equation}\label{eq:dT_resmooth}
    \delta T_\circledast(\boldsymbol{\theta},\nu_i) = \frac{\left[\delta T(\boldsymbol{\theta},\nu_i)w(\boldsymbol{\theta},\nu_i) \right] \circledast \pazocal{K}(\boldsymbol{\theta},\nu_i)}{w(\boldsymbol{\theta},\nu_i) \circledast \pazocal{K}(\boldsymbol{\theta},\nu_i)}
\end{equation}
\textcolor{black}{As in \autoref{eq:beam_convolve}, the $\circledast$ symbol represents a convolution}. The weights themselves must also be reconvolved to keep them consistent and optimal for the new $\delta T_\circledast$ maps. The reconvolved weights are given by 
\begin{equation}\label{eq:w_resmooth}
    w_\circledast(\boldsymbol{\theta},\nu_i) = \frac{\left[w(\boldsymbol{\theta},\nu_i) \circledast \pazocal{K}(\boldsymbol{\theta},\nu_i)\right]^2}{w(\boldsymbol{\theta},\nu_i) \circledast \pazocal{K}^2(\boldsymbol{\theta},\nu_i)}\,.
\end{equation}
In this work we use a value of $\gamma\,{=}\,1.4$ for the reconvolution kernel in \autoref{eq:ResmoothKernel}. We find that neglecting this resmoothing step entirely increases the influence of systematics in the results which is likely due to residual RFI or contaminants not being smoothed and depressing their dominance. We discuss this further in \appref{sec:ResmoothingAppendix} but also plan a dedicated study into this aspect of the analysis where a more robust model of the MeerKAT beam will also be implemented.

\begin{figure}
    \centering
    \includegraphics[width=1\linewidth]{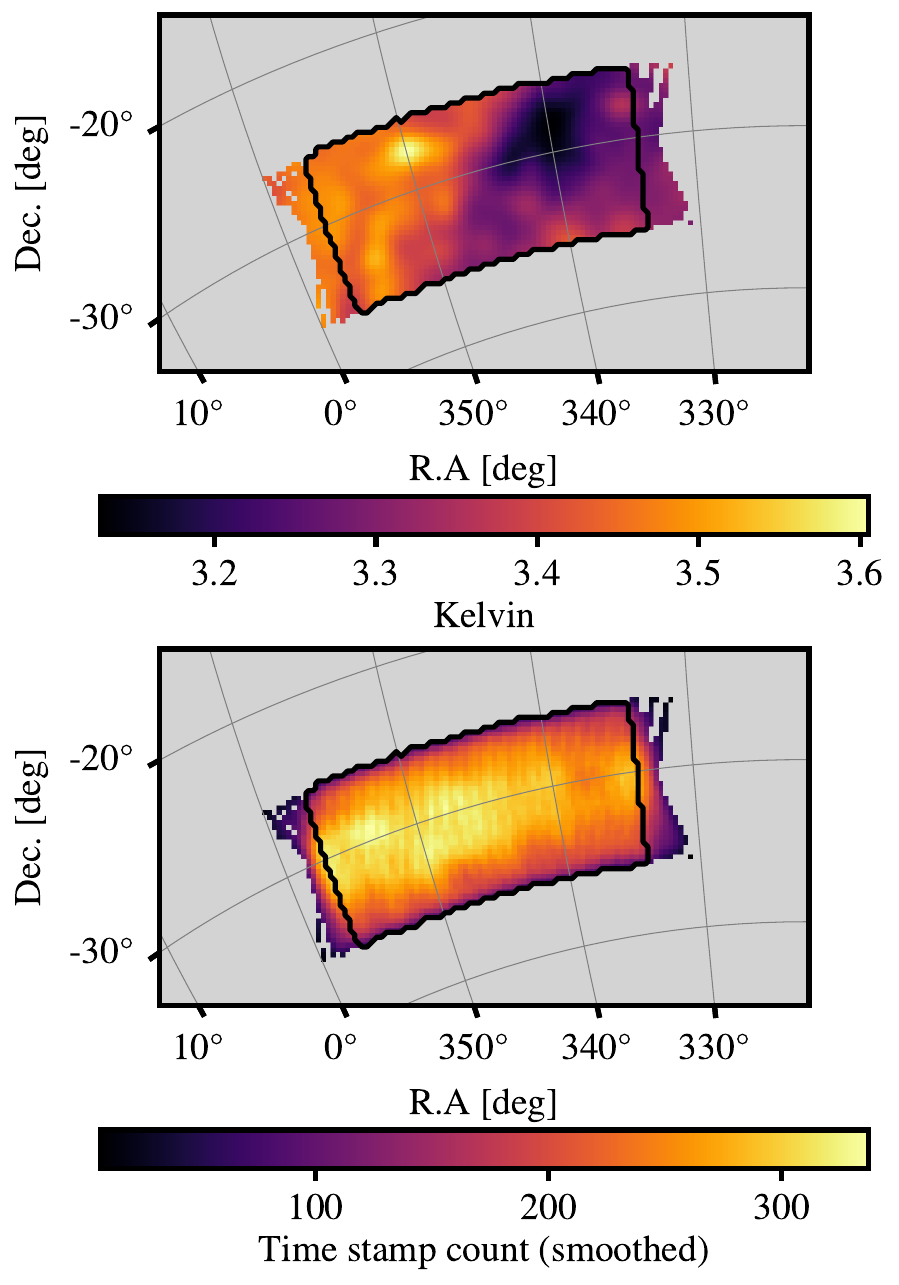}
    \caption{Reconvolved MeerKLASS deep field (top-panel) using $\gamma\txteq{=}1.4$ (see \autoref{eq:dT_resmooth}). The bottom panel shows the reconvolved time-stamp field which we use for weighting, thus also requires reconvolution (using \autoref{eq:w_resmooth}) for consistency with the intensity maps. Both panels 
    \textcolor{black}{are at 1023\,MHz to match \autoref{fig:FGmap_and_counts}.}}
    \label{fig:ResmoothMap_and_weights}
\end{figure}

The resmoothed, uncleaned map is shown in \autoref{fig:ResmoothMap_and_weights} (top panel). Careful comparison with \autoref{fig:FGmap_and_counts} reveals how some of the foreground structure has been smoothed, albeit subtly, which is the aim. Too much reconvolution will suppress power in high $k_\perp$ modes in the final power spectrum estimate. The reconvolution is more noticeable in the time stamp count field, which we use as the weights $w_\circledast$, shown by the bottom panel of \autoref{fig:ResmoothMap_and_weights}.

\subsection{Map trimming}\label{sec:maptrim}

Shown by the time stamp count field of \autoref{fig:FGmap_and_counts} (and the resmoothed version in \autoref{fig:ResmoothMap_and_weights}), the MeerKLASS scanning strategy of combining numerous rising and setting scans does not generate a uniform coverage. The edges of the patch result in relatively low recorded time stamps due to a lack of overlapping scans in these regions. 

We found that implementing an edge trim to the field improved the quality of the foreground clean and final power spectrum estimates. This is motivated and explored in greater detail in \citet{2024arXiv241206750C}. The improvement from map trimming is because pixels are removed where residual time-varying contamination has been averaged down less, due to fewer scans. 

In \autoref{fig:ResmoothMap_and_weights}, the black contour shows the cut to the patch we make, removing everything outside the trimmed lines. This amounts to all pixels at $334^\circ\,{<}\,\text{R.A.}\,{<}\,357^\circ$ being set to zero in both the weights and intensity map field.

\subsection{Blind foreground cleaning with PCA}\label{sec:PCAcleaning}

We represent the intensity maps with matrices $\textbf{\textsf{X}}$ which are equivalent to the $T(\boldsymbol{\theta},\nu)$ temperature fields, discretised into arrays on which matrix algebra can be performed.

Blind foreground cleaning assumes the matrix of the observational data cube $\textbf{\textsf{X}}_{\mathrm{obs}}$, which contains the intensity maps with $n_\theta$ pixels at each of the $n_\nu$ frequency channels, can be generalised by the linear equation
\begin{equation}\label{eq:linearsystem}
    \textbf{\textsf{X}}_{\mathrm{obs}} = \textbf{\textsf{A}} \textbf{\textsf{S}}+\textbf{\textsf{R}}\,.
\end{equation}
$\textbf{\textsf{A}}$ is referred to as the \textit{mixing matrix} and is the set of $N_\text{fg}$ spectral functions estimated to project out the foregrounds. $\textbf{\textsf{S}}$ are the $N_\text{fg}$ source maps obtained by projecting $\textbf{\textsf{A}}$ along the data;
\begin{equation}
    \textbf{\textsf{S}} = \textbf{\textsf{A}}^\text{T}\textbf{\textsf{X}}_{\mathrm{obs}}\,.
\end{equation}
The dot product $\textbf{\textsf{A}} \textbf{\textsf{S}}$ thus has shape $[n_\nu,n_\theta]$, along with the other terms in \autoref{eq:linearsystem}, and represents a subset of the observations estimated to contain the bulk of the foreground contamination. The residual term $\textbf{\textsf{R}}$ will contain the remaining \hi\ signal and thermal noise. In reality, the situation is more complex and there is cross-term leakage. Both \hi\ signal will leak into the subtracted $\textbf{\textsf{A}} \textbf{\textsf{S}}$ term causing \textit{signal loss}, and foreground will leak into the remaining $\textbf{\textsf{R}}$ causing \textit{foreground residuals}. The control of both these is crucial and we discuss them further throughout the remaining sections.

\begin{figure}
    \centering
    \includegraphics[width=1\linewidth]{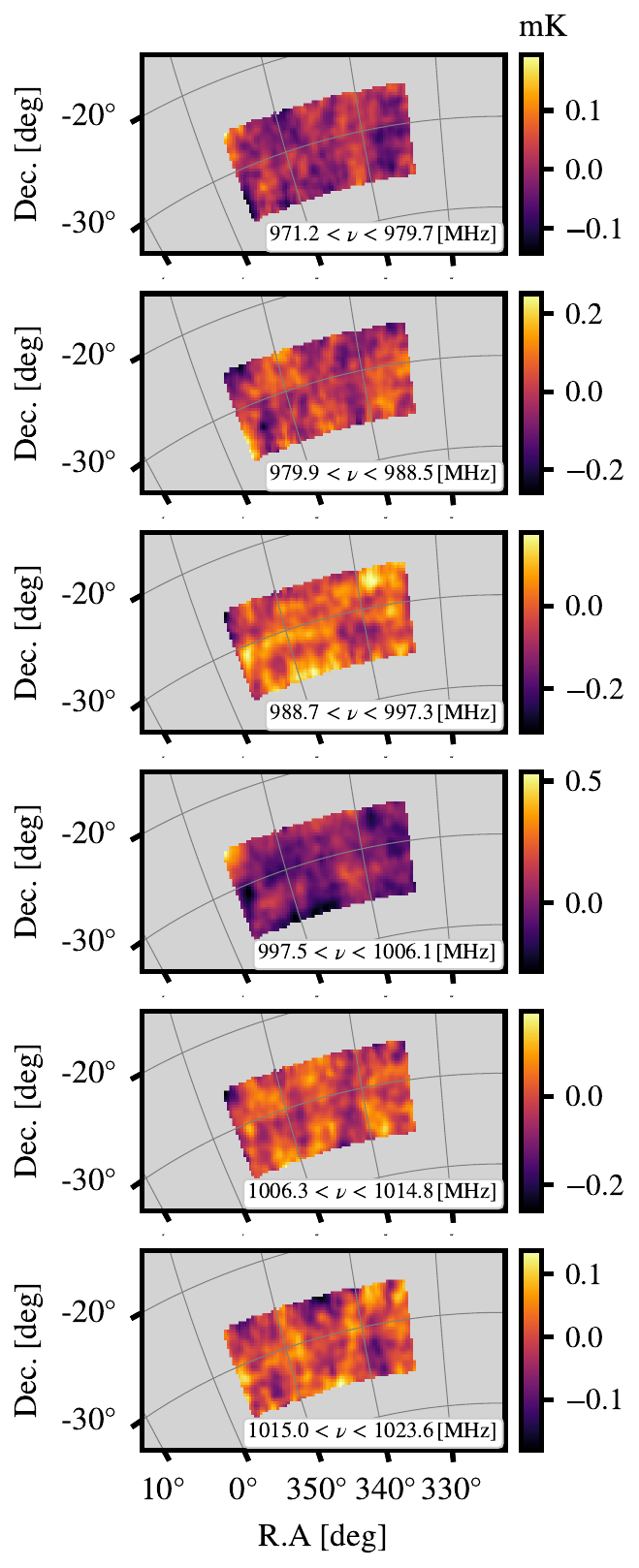}
    \caption{Foreground cleaned intensity maps for the MeerKLASS L-band deep field in frequency sub-bands (stated in bottom left). Maps have been cleaned by removing $\Nfg\txteq{=}10$ PCA modes.}
    \label{fig:CleanedMaps_nu}
\end{figure}

Different blind foreground techniques only vary in their approach to estimating the mixing matrix $\textbf{\textsf{A}}$.
In this work, we adopt the straightforward principal component analysis (PCA) approach, since it is well-tested and understood. Implementing more sophisticated blind techniques is planned in future work. PCA foreground cleaning begins by computing the frequency covariance matrix of the weighted observations
\begin{equation}\label{eq:weightedcov}
    \textcolor{black}{\textbf{\textsf{C}}=\frac{\left(\textbf{\textsf{w}} \circ \textbf{\textsf{X}}_{\mathrm{obs}}\right)\left(\textbf{\textsf{w}} \circ \textbf{\textsf{X}}_{\mathrm{obs}}\right)^{\mathrm{T}}}{n_\theta-1}\,,}
\end{equation}
where we have used the intensity mapping weights to optimise the estimation and $\textbf{\textsf{w}} \circ \textbf{\textsf{X}}_{\mathrm{obs}}$ represents the element-wise product between the weights and the observational data matrix. Since the weights are the stacked average of the counts along the line-of-sight, they are constant in frequency by design to avoid increasing the rank of $\textbf{\textsf{X}}_{\mathrm{obs}}$ i.e.\ not inject additional structure into the frequency spectra. Weighting in this way means $\textbf{\textsf{C}}$ is not a formally normalised covariance estimation, however, this is not a concern for us since we are only extracting unitless eigenvectors from the covariance and using these to project out modes in the real observed data so sensible amplitudes are recovered in the component separated pieces. We discuss this further, with some examples in \appref{sec:FGweights}. $\textbf{\textsf{C}}$ is then eigen-decomposed by
\begin{equation}
    \textbf{\textsf{C}} = \textbf{\textsf{U}}\boldsymbol{\Lambda}\textbf{\textsf{U}}^\text{T}\,,
\end{equation}
where $\textbf{\textsf{U}}$ is the matrix of eigenvectors with the corresponding eigenvalues contained in $\boldsymbol{\Lambda}$. The mixing matrix is then formed by taking the first $N_\text{fg}$ column vectors $\boldsymbol{u}_i$ from $\textbf{\textsf{U}}$ i.e.\ $\textbf{\textsf{A}}\,{=}\,[\boldsymbol{u}_1,\boldsymbol{u}_2,...,\boldsymbol{u}_{N_\text{fg}}]$. The first 9 eigenmodes are shown in the Appendix plot \autoref{fig:eigenmodes}. It is at this stage a choice is required for $N_\text{fg}$, which determines how many modes are removed and deemed as foregrounds. If the $N_\text{fg}$ is too high, the \hi\ signal loss increases; on the contrary, if $N_\text{fg}$ is too low, residual contamination will dominate more. The estimate of the foregrounds is provided by $\textbf{\textsf{X}}_{\mathrm{fg}}\,{\equiv}\,\textbf{\textsf{A}} \textbf{\textsf{S}}$ which is subtracted to provide the residuals, revealing the final cleaned data used in the analysis i.e.\ $\textbf{\textsf{X}}_\text{clean}\,{\equiv}\,\textbf{\textsf{R}}$ where
\begin{equation}
    \textbf{\textsf{X}}_\text{clean} = \textbf{\textsf{X}}_{\mathrm{obs}} -   \textbf{\textsf{X}}_{\mathrm{fg}}\,.  
\end{equation}
\autoref{fig:CleanedMaps_nu} shows the resulting cleaned maps by removing $N_\text{fg}\txteq{=}10$ PCA modes. We show the maps split into six frequency sub-bands (given in the bottom-left of each panel) to illustrate the changing intensity structures. Each panel is the frequency average of the 42 channels in each sub-band showing a reasonably consistent amplitude of temperature fluctuations across the sub-bands. The amplitudes in the maps are still likely driven by non-cosmological contributions, mostly thermal noise, but they are not far from the $0.1\,\text{mK}$ fluctuations we expect from \hi. As we will show statistically in \secref{sec:PkEst}, over large-scale fluctuations, the predominant contribution comes from the cosmological \hi.

\subsection{Foreground cleaning tests using mocks}\label{sec:FGcleanMocks}

In order to validate our foreground removal and estimate the signal loss it induces, we run the same PCA cleaning technique on mock intensity maps. As discussed in \secref{sec:HIIMmocks}, simulating foreground contamination that emulates the systematics and complex observational effects is beyond current capabilities. We therefore use the real observations themselves and inject mock \hi\ intensity maps into these. We then apply the foreground removal pipeline on the mock-injected data and estimate the signal loss from the injected mock whose original signal is known.

We follow the formalism outlined in \citep{Cunnington:2022uzo}, beginning with a \hi\ mock, whose matrices we notify by $\textbf{\textsf{M}}$ to distinguish from the real observation $\textbf{\textsf{X}}$. The mock is injected into the real observations i.e.\ $\textbf{\textsf{X}}_\text{obs} + \textbf{\textsf{M}}_\hi$. This is then cleaned by projecting out a subset of modes defined by the mixing matrix, as in the previous sections, such that
\begin{equation}\label{eq:Mclean}
    \textbf{\textsf{M}}_\text{clean} = \left(\textbf{\textsf{X}}_\text{obs} + \textbf{\textsf{M}}_\hi\right) - \textbf{\textsf{A}}\textbf{\textsf{A}}^\text{T}\left(\textbf{\textsf{X}}_\text{obs} + \textbf{\textsf{M}}_\hi\right)\,,
\end{equation}
The data in $\textbf{\textsf{M}}_\text{clean}$ will now contain signal loss emulating that of the real data clean, along with residual systematics and noise from the real data.

\begin{figure}
    \centering
    \includegraphics[width=1\linewidth]{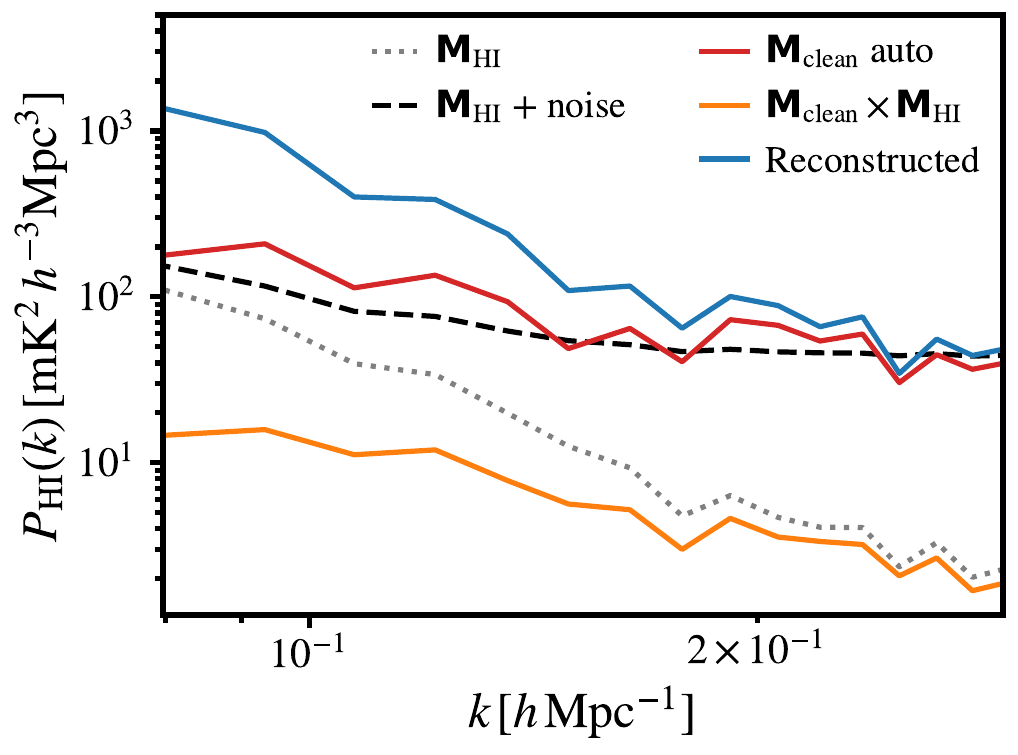}
    \caption{Power spectra for the \hi\ intensity map mocks under different scenarios, each averaged over 500 mock realisations. Including: \hi-only mock (grey dotted line). \hi\ plus thermal noise realisations (black dashed). The \hi\ mock with added foreground and systematics from the real MeerKLASS data, then cleaned as in \autoref{eq:Mclean} (red solid). The cross-correlation between \hi-only and cleaned mock (orange solid). Lastly, the blue solid line is the auto-correlation of the cleaned mock but with signal loss reconstructed.}
    \label{fig:PksFromMocks_wFG}
\end{figure}

We show the results of mocks contaminated by the real data in \autoref{fig:PksFromMocks_wFG}. We show measured power spectra averaged over all 500 mocks under different scenarios (details of power spectrum estimation are provided in the following \secref{sec:PkEst}). Firstly, the grey dotted line shows the \hi-only mocks, $\textbf{\textsf{M}}_\hi$, with no foreground cleaning effects, representing the \textit{true} power. The red solid line shows the result of adding the real observations to the mock, then PCA cleaning as in \autoref{eq:Mclean}. The power is higher than the grey line, despite the inevitable signal loss the clean will have caused. This is evidence that the $\textbf{\textsf{M}}_\text{clean}$ maps contain residual systematics and thermal noise, which are additively biasing the power\footnote{There will also be a small contribution from the true \hi\ signal in the observed data.}. We can demonstrate this by considering the orange solid line which is the cross-correlation of the cleaned mocks $\textbf{\textsf{M}}_\text{clean}$ with the \hi-only mocks $\textbf{\textsf{M}}_\hi$, a test only possible with simulations. Since the residual systematics and noise will be uncorrelated with the \hi-only mocks, the distortion in power comes purely from signal loss, hence why it is lower than the \hi-only power at all scales.

To isolate the contribution from \textit{unknown} systematics, we need to account for thermal noise and signal loss from foreground cleaning. The thermal noise contribution can be approximated using \autoref{eq:sigma_N} to generate 500 noise maps and add these to the \hi-only mocks. The black-dashed line shows this and demonstrates the boost in power relative to the grey-dotted line, more so at small scales where, due to the beam, the \hi\ signal is low relative to the scale-independent noise\footnote{The noise is not completely scale-independent because it has gone through the same regridding and reconvolution processes as the data which causes some subtle damping.}. The last result presented in \autoref{fig:PksFromMocks_wFG} is the reconstructed power (i.e.\ correcting for foreground cleaning signal loss) shown by the blue solid line. This utilises the signal-loss evaluated from the cross-power (orange line) to reconstruct the $\textbf{\textsf{M}}_\text{clean}$ auto-power. This is discussed more in the following sub-section. The discrepancy between the blue-solid and black-dashed line must be caused by systematics, since signal-loss and thermal noise have now been accounted for. We find a similar excess power with the real observations, which we discuss in the following \secref{sec:PkEst}. The results in \autoref{fig:PksFromMocks_wFG} are to validate the realism of the mock pipeline, and also to motivate the signal loss reconstruction process, which we formalise in the next sub-section.

\subsection{Signal loss correction with transfer functions}\label{sec:TFsignalloss}

We can utilise the method of mock signal injection from the previous \secref{sec:FGcleanMocks} for estimating the signal loss in the power spectrum of the real data. This is now a well-understood process \citep{Switzer:2015ria,Cunnington:2023jpq} and used in previous single-dish intensity mapping analyses \citep{Masui:2012zc,Anderson:2017ert,eBOSS:2021ebm,Cunnington:2022uzo}.

We begin with the $\textbf{\textsf{M}}_\text{clean}$ term in \autoref{eq:Mclean}, which as demonstrated, will contain signal loss that should emulate the real data. In some cases, the cleaned real data $\textbf{\textsf{X}}_\text{clean}$ can be subtracted from $\textbf{\textsf{M}}_\text{clean}$. This accelerates the convergence of the transfer function calculation by removing the additional noise caused by the real data but has no bearing on the signal loss calculation. Leaving this piece in the calculation of $\textbf{\textsf{M}}_\text{clean}$ is however crucial when using the transfer function for estimating errors in the power spectrum. We also find that due to the large number of mocks used in this work, an equivalent accuracy of the transfer function is obtained whether subtracting $\textbf{\textsf{X}}_\text{clean}$ or not. For simplicity, we therefore only use the case as defined by \autoref{eq:Mclean}, \textit{without} the $\textbf{\textsf{X}}_\text{clean}$ subtraction.

The transfer function is defined by the power spectrum, signified by the operator $\pazocal{P()}$, of the cleaned mock $\textbf{\textsf{M}}_\text{clean}$ with the foreground-free \hi-only mock $\textbf{\textsf{M}}_\hi$, and the ratio of this with the power spectrum of the \hi-only mock
\begin{equation}\label{eq:TF_HI}
    \pazocal{T}_\hi(\boldsymbol{k})=\left\langle\frac{\pazocal{P}\left(\textbf{\textsf{M}}_\text{clean},\textbf{\textsf{M}}_\hi\right)}{\pazocal{P}\left(\textbf{\textsf{M}}_\hi,\textbf{\textsf{M}}_\hi\right)}\right\rangle_{N_{\text {mock}}}\,.
\end{equation}
The angled brackets indicate an averaging over a large number of mocks such that a converged result is obtained. In this work, we use all mocks in our suite, thus $N_\text{mock}\txteq{=}500$. The operator $\pazocal{P()}$ can represent any power spectrum estimator i.e.\ we use this same recipe for estimating signal loss in the spherically averaged power and also in the cylindrical ($k_\perp$, $k_\parallel$) power spectrum, which we later utilise in some analysis.

In this work, we will mostly focus on the MeerKLASS deep field's cross-correlation with the overlapping GAMA galaxies. The signal loss from PCA cleaning in the cross-power spectrum should be identical to that in the auto-power. This was established empirically in \citet{Cunnington:2023jpq} which showed how removing a subset of PCA modes entirely but leaving their contribution in the other half of the cross-power product (e.g.\ in the galaxy maps which have no cleaning, thus no signal loss) generates no more power than the auto-\hi. We present further reasoning for why this is the case in \appref{sec:TF_signalloss_autoVcross}. To summarise, we believe it is explained by the PCA modes being orthogonal, thus maximally uncorrelated. Therefore, the additional remaining modes in the uncleaned field are linearly uncorrelated with what survives in the cleaned \hi, producing no extra power. 

Despite the approximately equal signal loss between auto and cross-correlations, we still define a separate transfer function for the cross-power spectrum purely for covariance estimation purposes whereby the only difference is we switch in the GAMA galaxy mocks $\textbf{\textsf{M}}_\text{g}$ instead which will emulate the signal strength and shot-noise of the GAMA galaxies, thus providing a more robust error estimate in the cross-power.
\begin{equation}\label{eq:TF_gHI}
    \pazocal{T}_{\hi,\text{g}}(\boldsymbol{k})=\left\langle\frac{\pazocal{P}\left(\textbf{\textsf{M}}_\text{clean},\textbf{\textsf{M}}_\text{g}\right)}{\pazocal{P}\left(\textbf{\textsf{M}}_\hi,\textbf{\textsf{M}}_\text{g}\right)}\right\rangle_{N_{\text {mock}}}\,.
\end{equation}
Any measured power spectra should be divided by $\pazocal{T}(\boldsymbol{k})$, thus the transfer function represents a reconstruction of signal at the power spectrum level. 

\section{Power spectrum estimation}\label{sec:PkEst}

In this section we present power spectra estimations from the foreground cleaned MeerKLASS L-band deep field. We first outline the pipeline for performing these measurements then present the results. By default, we use maps that have been foreground cleaned by removing $\Nfg\txteq{=}10$ PCA modes, as outlined in the previous section. However, we also examine the results from varying $\Nfg$. 

\subsection{Transforming to a Cartesian grid}

The first task to perform in order to measure a power spectrum is to transform the foreground cleaned intensity maps onto a Cartesian grid with comoving $h^{-1}\text{Mpc}$ coordinates. This allows for a Fast Fourier transform to be calculated on the Cartesian array, from which the power at different Fourier bandpasses $\boldsymbol{k}$, can be averaged and presented. 

We discussed the transformation between sky maps and Cartesian grids already in \secref{sec:RegriddingMocks} when we performed the inverse routine of sampling Cartesian cells from lognormal mocks into celestial sky map voxels that emulate the MeerKLASS intensity maps. We again utilise the recent work and code in \citet{Cunnington:2023aou} and first define a grid of size $l_\text{x},l_\text{y},l_\text{z}\,{=}\,632.2,341.5,231.5\,h^{-1}\text{Mpc}$ with $n_\text{x},n_\text{y},n_\text{z}\,{=}\,88,48,168$ cells along each dimension. The choice of cells is a factor of $1.5$ smaller than the number of map voxels in R.A., Dec., and $\nu$. Making the Cartesian grid slightly lower resolution than the map and using 10 sampling particles per map voxel ensures each Cartesian cell is well-sampled, avoiding empty cells.

The foreground cleaned maps $\delta T_\hi(\boldsymbol{\theta},\nu)$ are transformed to this Cartesian grid, giving $\delta T_\hi(\boldsymbol{x})$. The intensity mapping weights also follow an identical transformation process $w_\hi(\boldsymbol{\theta},\nu)\,{\rightarrow}\,w_\hi(\boldsymbol{x})$. The catalogue of GAMA galaxies can be directly gridded by transforming their celestial coordinates into Cartesian, then assigning them onto the Cartesian grid $n_\text{g}(\boldsymbol{x})$. For all sampling particles and galaxies we use a straightforward nearest grid point (NGP) assignment. \citet{Cunnington:2023aou} presented extensions to this method for avoiding aliased power and obtaining sub-percent accuracy. However, given the current signal-to-noise of our data at small scales, aliasing effects will be negligible, thus higher-order interpolations and interlacing are overkill and serve only to increase compute demand. However, we correct for the dominant effect which is the field's smoothing by the (NGP) particle assignment. This requires deconvolving the field, achieved by dividing through by the NGP mass assignment window function in Fourier space \citep{Jing:2004fq}
\begin{equation}
    \tilde{\delta T}_\hi(\boldsymbol{k}) \rightarrow \frac{\tilde{\delta T}_\hi(\boldsymbol{k})}{W_\text{ngp}(\boldsymbol{k})}\,,
\end{equation}
where the window function $W_\text{ngp}$ accounts for the sampling particles discretisation into cells of size $H$. These are separable functions along each Cartesian dimension, thus modelled by a Fourier transform of a top-hat function, which becomes a sinc function; $\operatorname{sinc}(x)\txteq{=}\sin(x)/x$ and therefore
\begin{equation}
    \tilde{W}_{\mathrm{ngp}}(\boldsymbol{k})=\operatorname{sinc}\left(\frac{k_{\mathrm{x}} H_{\mathrm{x}}}{2}\right) \operatorname{sinc}\left(\frac{k_{\mathrm{y}} H_{\mathrm{y}}}{2}\right) \operatorname{sinc}\left(\frac{k_{\mathrm{z}} H_{\mathrm{z}}}{2}\right)\,.
\end{equation}

\begin{figure}
    \centering
    \includegraphics[width=0.9\linewidth]{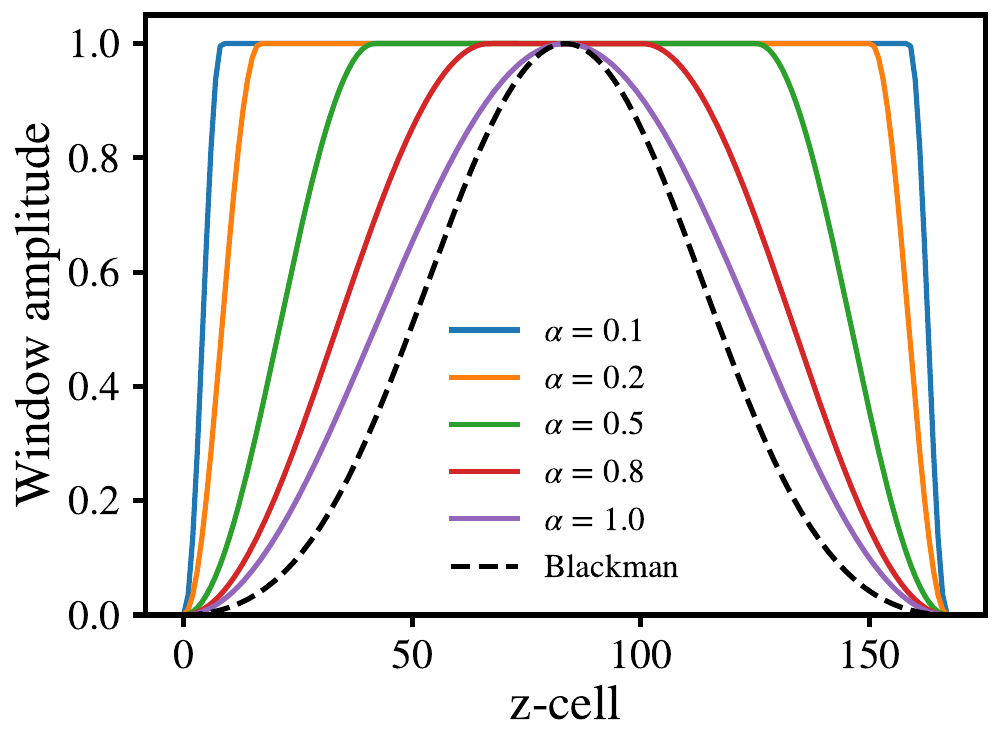}
    \caption{Tukey window functions with differing shape parameters $\alpha$. For comparison, we also show the Blackman window used in previous work.}
    \label{fig:tukey_windows}
\end{figure}

Previous work \citep[e.g.][]{Switzer:2013ewa,Cunnington:2022uzo} employed a tapering along the radial direction of the regridded field with the primary purpose of reducing edge effects in the Fourier transform. Our modelling approach should however account for these effects since the model is also convolved with the window function \citep{Blake:2019ddd}. We still found results were worse when omitting a radial tapering though, which we suspect is instead caused by edge-effects from the foreground clean \citep{Alonso:2014dhk} which are suppressed by the tapering. We experimented with a less aggressive taper by using a Tukey window, also known as a cosine window (this is also explored in \citet{2024arXiv241206750C}). This has a shape parameter $\alpha$ which is simply a top-hat with no apodisation when $\alpha\txteq{=}0$ and equivalent to a Hann window when $\alpha\txteq{=}1$. We show these window functions in \autoref{fig:tukey_windows}. By opting for less aggressive Tukey windows we still effectively suppress the extreme edges where foreground cleaning effects may be present but avoid unnecessary tapering on the central data. We will explore this further with future data in MeerKAT's UHF band where the larger useable frequency ranges may render this less of an issue. In this work we found optimal results when using $\alpha\txteq{=}0.8$ for the Tukey window which is applied to the intensity map weights $w_\hi(\boldsymbol{x})$. We do not apply this to the galaxy data as we found this increased the correlations between $k$-bins in the power spectrum. \textcolor{black}{A tapering in the angular directions could also be applied, but again as argued above, since we convolve our model with the survey window function, the bulk of these effects should be controlled. There is also no requirement for angular tapering to suppress any radial-only foreground-cleaning edge effects. We therefore chose to omit angular tapering from our analysis. 
}

\subsection{Power spectrum formalism}

With the fields assigned to a Cartesian space and appropriately tapered, the power spectra can then be measured. The Fourier transformed field of the \hi\ temperature fluctuation maps $\delta T_\hi$ is given by
\begin{equation}
    \tilde{F}_\hi(\boldsymbol{k})=\sum_{\boldsymbol{x}} \delta T_\hi(\boldsymbol{x}) w_\hi(\boldsymbol{x}) \exp (i \boldsymbol{k}{\cdot}\boldsymbol{x})\,,
\end{equation}
where the weights, now denoted by $w_\hi$ to distinguish them from the galaxy weights, still derive from the field of time-stamp counts, which are regridded following the same process as outlined in the previous sub-section. The Fourier transformed field of the the galaxy count field $n_\text{g}$ is given by
\begin{equation}
    \tilde{F}_\text{g}(\boldsymbol{k})=\sum_{\boldsymbol{x}} n_\text{g}(\boldsymbol{x}) w_\text{g}(\boldsymbol{x}) \exp (i \boldsymbol{k}{\cdot}\boldsymbol{x}) - N_\text{g}\tilde{W}_\text{g}(\boldsymbol{k})\,,
\end{equation}
where $N_\text{g}\,{=}\,\sum n_\text{g}$ is the total number of galaxies in the optical maps and $\tilde{W}_\text{g}$ is the weighted Fourier transform of the selection function
\begin{equation}
    \tilde{W}_\text{g}(\boldsymbol{k})=\sum_{\boldsymbol{x}} W_\text{g}(\boldsymbol{x}) w_\text{g}(\boldsymbol{x}) \exp (i \boldsymbol{k}{\cdot}\boldsymbol{x})\,.
\end{equation}
The galaxy window function $W_\text{g}$ is normally constructed by stacking mock galaxies that carefully follow the survey selection characteristics of the experiment to account for its incompleteness. As discussed previously, there are no GAMA G23 mocks available for this purpose so we therefore construct a simple window function which is 1 within the survey boundaries and 0 everywhere else. Some inclusion of the redshift evolution of $N(z)$ could have been included, but as \autoref{fig:GAMA_Nz} showed, the variation over our narrow redshift range is minimal. The window function is normalised so that $\sum_{\boldsymbol{x}}W_\text{g}(\boldsymbol{x})\,{=}\,1$. Since we lack this proper galaxy window function we do not implement the optimal weighting as per \citet{Feldman:1993ky}, but instead use uniform weighting, so equivalent of assuming $w_\text{g}(\boldsymbol{x})\txteq{\equiv}W_\text{g}(\boldsymbol{x})$. 

The cross-power spectrum estimator is then given by
\begin{equation}\label{eq:CrossPk}
    \hat{P}_{\hi,\text{g}}(\boldsymbol{k}) = \frac{V_\text{cell}}{\sum_{\boldsymbol{x}} w_\hi(\boldsymbol{x})w_\text{g}(\boldsymbol{x})W_\text{g}(\boldsymbol{x})}\operatorname{Re}\left\{\tilde{F}_\hi(\boldsymbol{k})\tilde{F}^{*}_\text{g}(\boldsymbol{k})\right\}\frac{1}{N_\text{g}}\,,
\end{equation}
where $V_\text{cell}\txteq{=}l_\text{x}l_\text{y}l_\text{z}/(n_\text{x}n_\text{y}n_\text{z})$. The hat notation $\left(\hat{P}\right)$ indicates that this is the \textit{estimated} power from the data, to distinguish from the \textit{modelled} power, which we discuss shortly. The \hi\ auto power spectrum estimator is more straightforwardly given by
\begin{equation}\label{eq:HIautoPk}
    \hat{P}_\hi(\boldsymbol{k}) = \frac{V_\text{cell}}{\sum_{\boldsymbol{x}} w_\hi^2(\boldsymbol{x})}\lvert\tilde{F}_\hi(\boldsymbol{k})\rvert^2\,,
\end{equation}
%
%
These power spectra are either \textit{spherically} averaged into bandpowers $\lvert\boldsymbol{k}\rvert\,{\equiv}\,k$ to provide 1D power spectrum results or alternatively \textit{cylindrically} averaged into 2D bandpowers defined by modes perpendicular ($k_\perp$) and parallel ($k_\parallel$) to the radial direction z, which we assume is the line-of-sight in a plane-parallel approximation, valid for the current MeerKLASS sky sizes. They are finally divided by their relevant foreground transfer functions (\autoref{eq:TF_HI} and \autoref{eq:TF_gHI}) to reconstruct signal loss for each $k$-bin.

\subsubsection{Modelling the power spectrum}

Here we outline the model of the power spectrum estimations for which we mostly follow the work in \citet{Blake:2019ddd}, with some extensions. We model the \hi\ power spectrum using the equation already introduced in \autoref{eq:PkHImodel} when discussing mock simulations. We repeat the model here for convenience;
\begin{equation}
    P_{\hi}(\boldsymbol{k})\equiv P_{\hi}(k,\mu) = \frac{\overline{T}_{\hi}^2 \left(b_\hi + f \mu^2\right)^2P_\text{m}(k)}{1 + \left(k\mu\sigma_\text{v}/H_0\right)^2}\,.
\end{equation}
Additional modelling of observational and mapping effects is also necessary since these will distort the power.  To model the MeerKAT beam and the additional smoothing from the reconvolution process we assume the field has undergone a constant smoothing by a 2D Gaussian kernel whose standard deviation is given by $R_\text{beam}\txteq{=}\sqrt{\gamma}\,R^{\max}_\text{beam}$ in comoving units. $\gamma\txteq{=}1.4$ accounts for the reconvolution to a common effective resolution and $R^{\max}_\text{beam}$ is the Gaussian approximation of the MeerKAT beam size at the deepest redshift (and lowest frequency $\nu_\text{min}$)
\begin{equation}
    R^{\max}_\text{beam} = \chi(z_{\max})\,\sigma_\text{beam}(\nu_{\min})\,,    
\end{equation}
where $\sigma_\text{beam}$ is given as before from \autoref{eq:sigbeamsize} and $\chi(z)$ is the comoving distance to redshift $z$, which in this case, is the maximum redshift for our frequency range. The perpendicular modes will then be damped according to
\begin{equation}
    \tilde{\pazocal{B}}_\text{beam}(\boldsymbol{k}) = \exp \left[\frac{-k^2\left(1-\mu^2\right) R_{\text {beam }}^2}{2}\right]\,.
\end{equation}
For this work, we find $R_\text{beam}\txteq{=}14.1\,h^{-1}\text{Mpc}$. 

Similar to the beam, there will also be distortions to the power caused by the finite resolution of the intensity maps.  We approximate the pixel size in comoving units by $s_\text{pix}\txteq{=}\chi(z_\text{eff})\Delta\theta$ where $\Delta\theta\txteq{=}0.3\,\deg$ is the pixel size used for the MeerKLASS L-band mapping. The frequency channel depth is approximated by taking the mean across all spacings between redshifts corresponding to channel boundaries i.e.\ $s_\text{chan}\txteq{=}\Sigma_i \left[\chi(z_i) - \chi(z_{i{+}1})\right]/N_\text{chan}$. These quantities are used in damping functions to model the suppressed resolution from mapping into pixels and frequency channels, the former being the far more dominant effect for the wide-pixel MeerKLASS intensity maps. These functions are given as
\begin{equation}
    \tilde{\pazocal{B}}_\text{pix}(\boldsymbol{k}) = \operatorname{sinc}\left(\frac{k\sqrt{1-\mu^2}\, s_{\text {pix}}}{2}\right)\,,
\end{equation}
and
\begin{equation}\label{eq:ChannelDamping}
    \tilde{\pazocal{B}}_{\text {chan }}(\boldsymbol{k})=\operatorname{sinc}\left(\frac{k\,\mu\, s_{\text {chan }}}{2}\right)\,,
\end{equation}
The model and damping functions are discretised onto the same 3D grid of $k$-modes as the FFT for the real data, then convolved with the survey window function. The model for the observed \hi\ auto-power spectrum therefore becomes
\begin{equation}\label{eq:PkHIobs}
    P_{\hi}^\text{obs}(\boldsymbol{k}) = \frac{W^2_\hi(\boldsymbol{k}) \circledast \left[P_{\hi}(\boldsymbol{k}) \,\pazocal{B}_\text{beam}^2(\boldsymbol{k})\,\pazocal{B}_\text{pix}^2(\boldsymbol{k})\,\pazocal{B}_\text{chan}^2(\boldsymbol{k})\right]}{\sum_{\boldsymbol{x}} w^2_\hi(\boldsymbol{x}) W^2_\hi(\boldsymbol{x})}\,,
\end{equation}
where we have normalised by the sum of the weights and window function, $W_\hi$, to obtain an amplitude that matches the weighted and windowed estimator. For the intensity mapping window function, we assume a simple binary function; 1 where we have scan coverage and 0 everywhere else. There is scope for improving this approximation and further work is needed to examine how intensity mapping surveys are impacted by any selection effects. 

A similar approach is followed for modelling the cross-power with GAMA. The cross-correlation power spectrum is given by
\begin{equation}\label{eq:PkHIg}
    P_{\hi,\mathrm{g}}(\boldsymbol{k})= \frac{\overline{T}_\hi\left[r b_\hi b_{\mathrm{g}}+b_\hi f \mu^2+b_{\mathrm{g}} f \mu^2+f^2 \mu^4\right] P_{\mathrm{m}}(k)}{1 + \left(k\mu\sigma_\text{v}/H_0\right)^2}\,.    
\end{equation}
where we introduce $r$, the correlation coefficient between the \hi\ fluctuations and the GAMA galaxy over-densities. This accounts for the different astrophysics in each tracer and non-linear clustering, which can mean the two fields are not exact multiples of each other and there is instead some stochasticity which will suppress their cross-power \citep[see e.g.][for applications in the context of line-intensity mapping]{MoradinezhadDizgah:2021dei,Sato-Polito:2022wiq,Obuljen:2022cjo}. We assume $r$ to be constant, which is sufficient for the current signal-to-noise, but in reality it will exhibit some scale dependence. The correlation coefficient only enters on the single term in \autoref{eq:PkHIg} because it only influences the biased tracer fields, not their \textit{unbiased} velocity fields which are modulated by $f$. The observational cross power is then similarly convolved with the survey window functions but only requires single factors of the damping functions $\pazocal{B}$; 
\begin{equation}\label{eq:PkHIgObs}
\begin{aligned}
    P_{\hi\text{,g}}^\text{obs} &(\boldsymbol{k}) = \\
    &\frac{W_\hi(\boldsymbol{k})W^*_\text{g}(\boldsymbol{k}) \circledast \left[P_{\hi\text{,g}}(\boldsymbol{k}) \,\pazocal{B}_\text{beam}(\boldsymbol{k})\,\pazocal{B}_\text{pix}(\boldsymbol{k})\,\pazocal{B}_\text{chan}(\boldsymbol{k})\right]}{\sum_{\boldsymbol{x}} w_\hi(\boldsymbol{x}) w_\text{g}(\boldsymbol{x}) W_\hi(\boldsymbol{x}) W_\text{g}(\boldsymbol{x}) }\,.
\end{aligned}
\end{equation}
We do not include modelling of uncertainties in the GAMA galaxy positioning, such as redshift estimation. The GAMA radial kernel equivalent of \autoref{eq:ChannelDamping} will likely have a larger impact due to spectroscopic redshift estimation errors. However, this will still be a sub-dominant contribution to the overall modelling relative to the beam and intensity map pixelisation, thus it is reasonable to neglect observational effects from the GAMA galaxies.

We perform a simple least-squares fit for an overall amplitude to our cross-power, equivalent to assuming fixed values for all parameters except for a fitted $\overline{T}_\hi$. This is deliberately simplistic to keep the model and fitting in a low-dimensional form. The small overlap sample in the cross-correlation means the statistical noise from such measurements is not yet suited to high-dimensional modelling. We are not therefore extending much from previous work \citep[e.g.][]{eBOSS:2021ebm,Cunnington:2022uzo} in this regard. Furthermore, the auto-\hi\ still has evidence of additive bias, as we will discuss, and therefore can only be trusted for upper limits in parameter inference. 

For the power amplitude fitting, we assume the parameter values outlined by \autoref{tab:fid_params}. The \hi\ bias comes from interpolated results from hydrodynamical simulations \citep{Villaescusa-Navarro:2018vsg}, which can be approximated by the polynomial
\begin{equation}
    b_\hi(z)=0.842+0.693 z-0.046 z^2\,.
\end{equation}
The galaxy bias is obtained by fitting the amplitude of the GAMA auto-power spectrum. The growth rate is obtained from $f\txteq{=}\Omega_\text{m}(z_\text{eff})^{0.545}$ for our Planck 18 cosmology. The correlation coefficient $r$ and velocity dispersion $\sigma_\text{v}$ are poorly constrained nuisance parameters tuned by eye to provide an adequate fit. We discuss model and parameter fitting further in \secref{sec:ParamConstraints} where we present results from a simultaneous MCMC fit on several parameters, demonstrating the degeneracies between them and discuss plans for breaking such degeneracies, but this is not a central focus of this work.

\begin{table}
    \begin{tabular}{|l|ccccc|}
        \hline
        \textbf{Paramete}r & $b_\hi$ & $b_\text{g}$ & $r$ & $f$ & $\sigma_\text{v}$ \\
        \textbf{Fiducial value} & 1.13  & 1.90 & 0.90 & 0.74 & 400\,km\,s$^{-1}$ \\
        \hline
    \end{tabular}
\caption{Fiducial parameters assumed in model fitting.}
\label{tab:fid_params}
\end{table}

\subsection{Error estimation with the transfer function}\label{sec:TFforErrors}

In previous analysis \citep{Cunnington:2022uzo}, an analytical approach was primarily used for estimating errors on the power spectrum measurements. However, this approach makes various assumptions, mainly that thermal noise dominates the error budget and that Gaussian statistics are valid. Since we are now reaching regions in $k$-space where thermal noise no longer dominates with MeerKLASS, we are required to relax these assumptions. The assumption of Gaussianity is arguably not solid either, given the additive systematics we know to still be present within the data. \textcolor{black}{Furthermore, an extra contribution from variance in the PCA foreground cleaning process is not accounted for in this analytical error estimation.}

A more robust approach to estimating the errors is to use the transfer function scatter across the different mock realisations $\pazocal{T}_i(\boldsymbol{k})$. This approach has been used in previous analysis \citep[e.g.][]{Anderson:2017ert} and has also been validated in simulations \citep{Cunnington:2023jpq}. \textcolor{black}{This error will estimate the contribution from sample variance, galaxy shot-noise, foreground and systematic residual, and variance from signal loss caused by eigenmode variation. 
The contribution from residual foreground and systematics to the error using this transfer function has not been robustly validated and is difficult to do so. Whilst they are present in each transfer function realisation they are not varying; only the response to them from each iteration's mock signal varies. The ideal way to robustly investigate this would be to simulate all residual systematics and foregrounds, but as we have discussed elsewhere, this is not currently possible for any \hi\ intensity mapping experiments, given their complexity and novelty. We defer this detailed investigation to future work and, for now, are content with our approach.} \textcolor{black}{As we will show, our estimated errors, if anything, appear conservative based on some analysis and are therefore not a major concern for the confidence gleaned from our conclusions.
We believe this is because it is the variance from signal loss caused by eigenmode variation that is the most dominant contribution to the error with our current data. We concluded this by comparing the variance between mock injection either before or after the foreground cleaning. When mocks are injected before foreground cleaning, the variance in the power spectrum is boosted by nearly an order of magnitude across most scales. This highlights the importance of optimising contaminant subtraction which is the focus of \citet{2024arXiv241206750C}.} 
For it to perform effectively, the mocks in the transfer function calculation must emulate the real data effectively, which we have strived to do so, as outlined in \secref{sec:Mocks}.

The transfer function for one single realisation is defined by 
\begin{equation}
    \pazocal{T}_i^{\hi,\text{g}}(\boldsymbol{k})=\frac{\pazocal{P}\left(\textbf{\textsf{M}}_\text{clean},\textbf{\textsf{M}}_\text{g}\right)}{\pazocal{P}\left(\textbf{\textsf{M}}_\hi,\textbf{\textsf{M}}_\text{g}\right)}\,,
\end{equation}
i.e.\ equivalent to \autoref{eq:TF_gHI} but without the averaging over $N_\text{mock}$. To reconstruct signal in a measured power spectrum, we take the ensemble average of \autoref{eq:TF_gHI} i.e.\ $\pazocal{T}(\boldsymbol{k})\txteq{=}\langle\pazocal{T}_i(\boldsymbol{k})\rangle_{N_\text{mock}}$, and divide the estimated power by this; $\hat{P}_\text{rec}(\boldsymbol{k})\txteq{=}\hat{P}(\boldsymbol{k})/\pazocal{T}(\boldsymbol{k})$. Note we are dropping the subscript \hi,g for brevity and also because the formalism we present here is identically applicable to the \hi-auto. The assumption made when using the transfer function for error estimation is that the scatter in the reconstructed power comes from the transfer function distribution $\pazocal{T}(\boldsymbol{k})\txteq{+}\delta\pazocal{T}_i(\boldsymbol{k})$ contributed by each $i$th mock. We can isolate the scatter component from the array of calculated transfer functions by simply subtracting the ensemble mean for each $i$th mock
\begin{equation}
    \delta\pazocal{T}_i(\boldsymbol{k}) = \pazocal{T}_i(\boldsymbol{k}) - \pazocal{T}(\boldsymbol{k})\,.
\end{equation}
Assuming any errors add in quadrature, error propagation suggests
\begin{equation}
    \frac{\delta \hat{P}^\text{rec}_i(\boldsymbol{k})}{\hat{P}_\text{rec}(\boldsymbol{k})} = \sqrt{\left(\frac{\delta \hat{P}_i(\boldsymbol{k})}{\hat{P}(\boldsymbol{k})}\right)^2 + \left(\frac{\delta \pazocal{T}_i(\boldsymbol{k})}{\pazocal{T}(\boldsymbol{k})}\right)^2}\,,
\end{equation}
however, we assume all quantifiable error lies in the transfer function distribution, hence $\delta \hat{P}_i\txteq{=}0$ and so
\begin{equation}
    \delta \hat{P}^\text{rec}_i(\boldsymbol{k}) = \hat{P}_\text{rec}(\boldsymbol{k})\frac{\delta \pazocal{T}_i(\boldsymbol{k})}{\pazocal{T}(\boldsymbol{k})}\,.
\end{equation}
%
Our setup thus provides $N_\text{mock}$ realisations of $\delta \hat{P}_i^\text{rec}$ and the covariance can be computed from this for any $k$-bin.
We reiterate that this formalism is valid whichever way the $\boldsymbol{k}$-modes are binned e.g.\ for a 2D cylindrical averaged or 1D spherically averaged power. In this work, our error bars in any 1D power spectrum estimate represent the limits of the 68th percentile regions of the distribution of $\delta \hat{P}^\text{rec}_i(k)$. If the uncertainties are Gaussian symmetric, this is equivalent to $1\,\sigma$ error bars, however, as we will later show, this is not perfectly the case for all $k$-modes. 

\subsection{Spherical averaging $k$-cuts}\label{sec:kcuts}

\begin{figure*}
    \centering
    \includegraphics[width=1\linewidth]{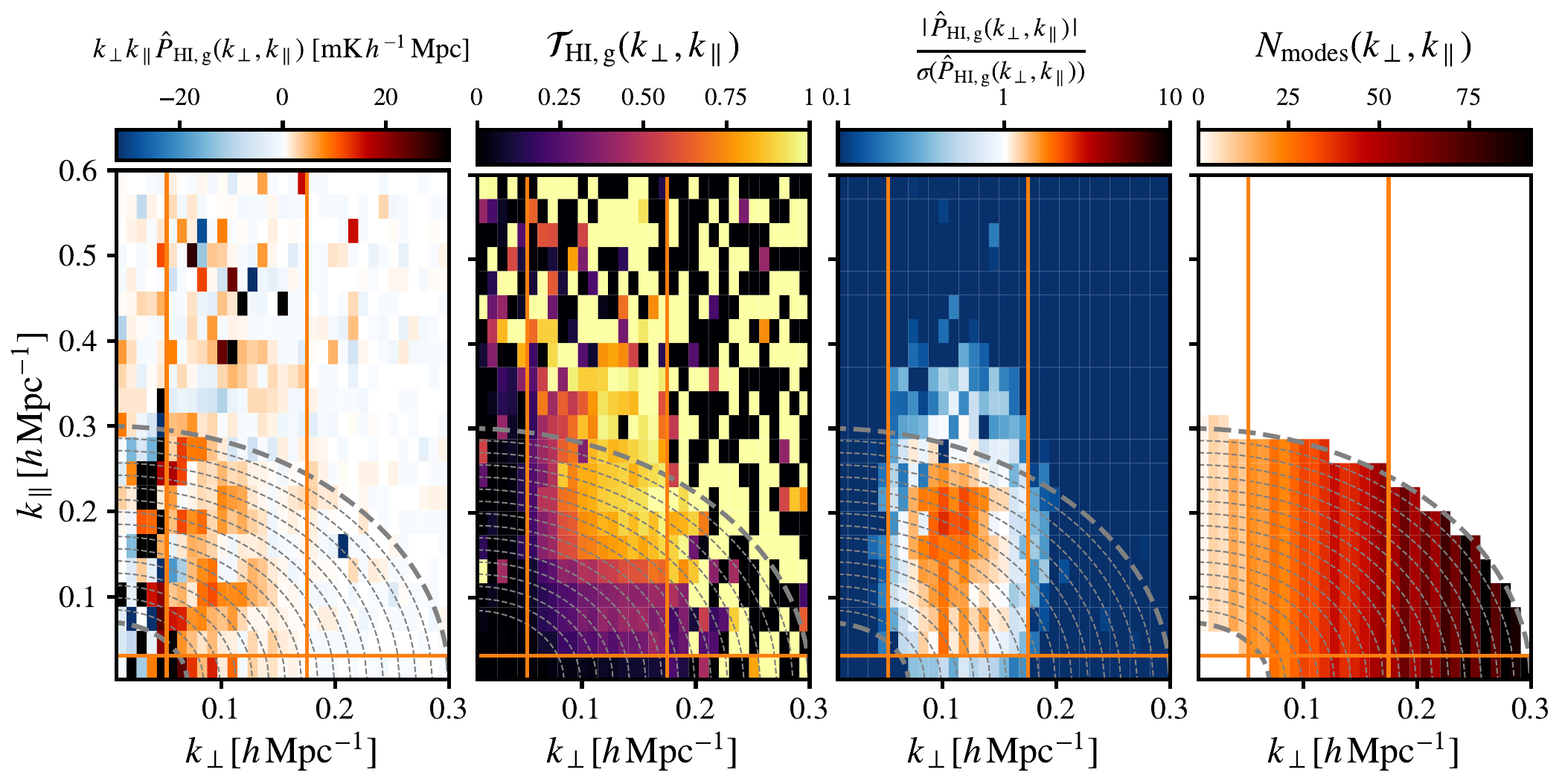}
    \caption{Cylindrical ($k_\perp,k_\parallel$) analysis of power, signal-loss and signal-to-noise. The first panel from the left shows the cross power between the MeerKLASS deep field and GAMA galaxies. The second panel is the transfer function used to reconstruct the power in $\hat{P}_{\hi,\text{g}}$. The third panel shows the signal-to-noise by dividing the cross power by the error estimation discussed in the text. The final fourth panel counts the number of discrete $\boldsymbol{k}$ wavenumber in the Fourier grid, demonstrating the concentration of available modes. The gray contours mark the boundaries of the $k$-bins used in spherically averaged power $P(k)$. The orange lines indicate regions used for $k$-cuts discussed in the text.}
    \label{fig:2DTF_SNR}
\end{figure*}

Before presenting spherically averaged power spectra from the MeerKLASS data, we first present results from the cross-correlation with GAMA galaxies in cylindrical-($k_\perp,k_\parallel)$ space. This is shown in \autoref{fig:2DTF_SNR} and motivates a discussion on where in $k$-space we have the most signal-to-noise so we can make optimal spherically averaged $\hat{P}(k)$ measurements. The dashed gray lines on all panels represent the $|\boldsymbol{k}|$-bin boundaries we use for the spherically averaged power spectra, restricting this to $k\txteq{<}0.3\hMpc$ to avoid highly non-linear cosmology and increasingly poor signal-to-noise. The orange solid lines represent boundaries for potential $k$-cuts which we will discuss.

The first panel of \autoref{fig:2DTF_SNR} shows the cross-power between GAMA G23 galaxies and the MeerKLASS deep field, cleaned by removing $\Nfg\txteq{=}10$ PCA modes. The $k_\perp k_\parallel$ scaling is for demonstrative purposes so the high power at low-$k$ does not saturate the plot. At $k\txteq{\gtrsim}0.15\hMpc$ the large MeerKAT beam begins to have an impact and will damp \hi\ fluctuations, which explains why the cross-power signal tends to zero at high-$k_\perp$. Except for the extreme low-$k$ regions, the rest of the $k$-space reveals a reasonably consistent positive cross-correlation signal, providing strong evidence of consistent clustering between the MeerKLASS maps and GAMA galaxies. As discussed in the previous \secref{sec:TFsignalloss}, the signal loss in this power has been corrected with the foreground transfer function. The second panel shows the corresponding transfer function for the $\Nfg\txteq{=}10$ clean. Low values of $\pazocal{T}_{\hi,\text{g}}$ signify high signal loss. This is evidence that nearly all signal is lost for the lowest-$k_\parallel$ modes which is expected since these modes should be highly degenerate with the removed long radial modes of the continuum foregrounds. The high signal loss at low-$k_\perp$ is interesting and not seen in previous MeerKLASS work \citep{Cunnington:2022uzo} with a similar method, albeit with less robust mocks. This is likely caused by the small GAMA footprint (see \autoref{fig:FGmap_and_counts}) which means the large-ranging modes at these low-$k$ are not adequately resolved and deliver little signal which is easily coupled to foreground modes and stripped away in the clean. This can be explored further in follow-up work with wider surveys where we expect this not to be a problem.

Due to the noisy power at low-$k$ and damped power at high-$k_\perp$ in \autoref{fig:2DTF_SNR}, we can choose to cut these regions of $k$-space from our spherically averaged power to obtain a more robust result. The orange solid lines represent chosen boundaries, where modes beyond these boundaries are excluded in the averaging of each $k$-bin (shown by the dashed gray lines). Providing the modelling of any power spectra also incorporates the $k$-cuts, then it will not bias any results. The $k$-cuts approach is further motivated by considering the signal-to-noise of the cross-power which we show in the third panel of \autoref{fig:2DTF_SNR}. Here the errors $\sigma(\hat{P}_{\hi,\text{g}})$ are computed using the transfer function by taking the standard deviation from the $\delta \hat{P}_i^\text{rec}(k_\perp,k_\parallel)$ distribution (see previous \secref{sec:TFforErrors}). This result provides a strikingly clear window where signal-to-noise is optimal; the orange to red regions in the third panel. The orange $k$-cut boundary lines have been chosen to target this region. One limitation of this approach is that it excludes the regions of $k$-space with the most modes, as shown by the fourth and final panel. We will present comparisons between the spherically averaged $P(k)$ with and without these $k$-cuts in the upcoming \secref{sec:GAMAcross}. 

\subsection{\hi\ auto-correlation power spectrum}

\begin{figure}
    \centering
    \includegraphics[width=1\linewidth]{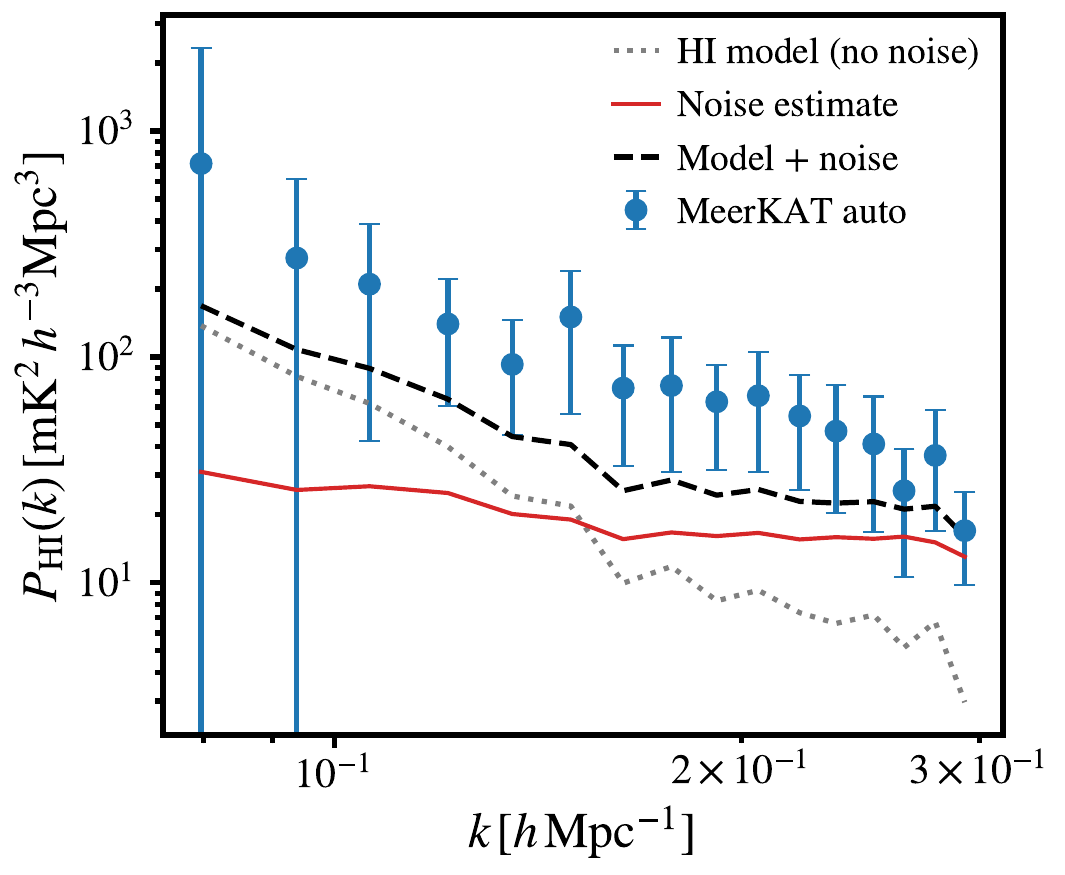}
    \caption{The \hi\ auto-power spectrum for the MeerKLASS L-band deep-field foreground cleaned by removing $\Nfg\txteq{=}10$ PCA modes. The red solid line is the noise estimate from measuring the average power of 10 Gaussian random fields with rms given by $\sigma_\text{N}$ (\autoref{eq:sigma_N}). The grey dotted line is the observational \hi\ power spectrum model (\autoref{eq:PkHIobs}) and the black dashed line is the combination of this \hi\ model plus the thermal noise estimate.}
    \label{fig:HIauto_power}
\end{figure}

We next examine results for the auto-correlation \hi\ power spectrum from the MeerKLASS L-band deep field. Previous single-dish intensity mapping results have been thermal noise-dominated across all scales. We are now crossing an important threshold with MeerKLASS where the \hi\ fluctuations should be dominating the noise at large scales. We demonstrate this in \autoref{fig:HIauto_power}, where we show the estimated \hi\ power spectrum from the MeerKLASS deep field (blue data points) and overlay an estimate of the thermal noise (red solid line). As outlined in \secref{sec:HIIMmocks}, this noise is modelled by generating Gaussian random fields with the expected rms given by \autoref{eq:sigma_N} which utilises the observational hit map. Despite being the \hi\ auto-power, we still restrict the measurement to the GAMA-field footprint, for consistency with the cross-correlation results. We tested using the full field but found no improvement in the auto-power. We leave exploration of the full-field to follow-up work which focuses on the \hi-auto power spectrum. The results in \autoref{fig:HIauto_power} include the scale $k$-cuts discussed in the previous section.

\autoref{fig:HIauto_power} also shows the expected \hi\ power (grey dotted line) from \autoref{eq:PkHIobs}. For this auto-\hi\ model, we do not fit the amplitude to avoid additive bias from residual systematics over inflating the model estimate. Instead we use the scale-independent $\overline{T}_\hi$ fit from the upcoming cross-power measurement. This model shows how the cosmological \hi\ power surpasses the thermal noise on large scales ($k\txteq{\lesssim}0.15\hMpc$). The thermal noise will be present even in the absence of clustering, thus the total expected power can be reasonably approximated as a simple addition of the \hi\ power, plus the thermal noise (shown by the black dashed line).

The small discrepancy between the deep-field power estimate and the model represents the lowest upper limit from a \hi\ auto-correlation on these large scales. We find the mean residual factor between model and data to be just $2.36$ across all $k$. Whilst the model and data are close in agreement, inferred parameter values for the bias and mean \hi\ temperature from fitting the amplitude of the auto power yields inconsistent values with the GAMA cross-correlation (discussed next). This is therefore evidence of low-level residual systematics that are additively biasing the power. In this evaluation, we inherently assume that the transfer function reconstructs any signal loss in the Gaussian thermal noise. If signal loss in the Gaussian noise is underestimated, the residual between the model and data may be higher than we report here. We therefore intend to robustly test this assumption when pursuing a more dedicated study of the \hi\ auto-correlation.

In follow-up work, we will focus on improving this upper limit from the \hi\ auto power which can be done in two ways: \textit{Firstly}, by implementing some corrections to remove low-level systematic contributions such as the residual RFI structures discussed in \secref{sec:zebra}. \textit{Secondly}, we can split the observations into sub-sets i.e.\ combinations of different maps from different scan times and/or dishes. This is useful for isolating dish- and time-dependent systematics and removing their biasing contributions to the power. Since each scan and dish is observing the same patch of sky, we would ideally expect all maps to agree, thus any differences will represent thermal noise, or systematics such as residual RFI, varying polarisation leakage etc. We do not pursue these ideas in this paper since they warrant their own detailed investigation.

\subsection{Cross-correlation power with GAMA}\label{sec:GAMAcross}

\begin{figure}
    \centering
    \includegraphics[width=1\linewidth]{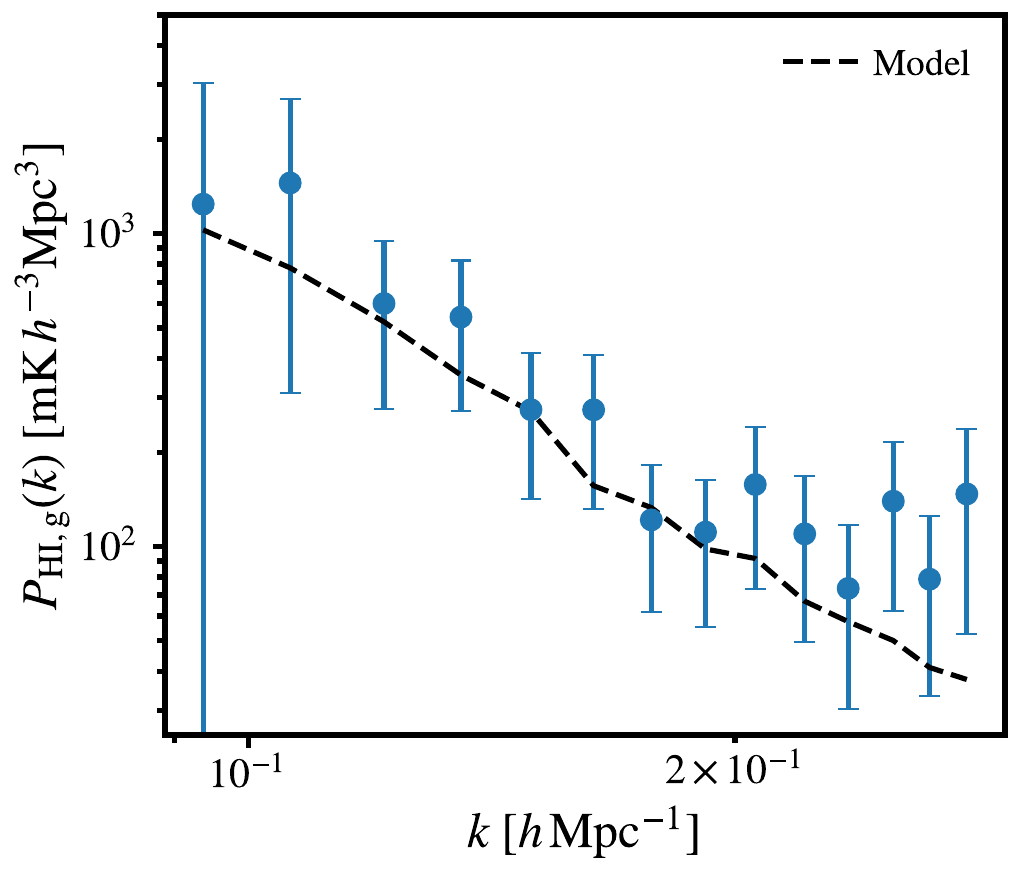}
    \caption{The cross-correlation power spectrum between GAMA galaxies and the  MeerKLASS L-band deep field, foreground cleaned by removing $\Nfg\txteq{=}10$ PCA modes. The black dashed line represents a model given by \autoref{eq:PkHIgObs} with a fitted scale-independent amplitude.}
    \label{fig:PkGAMAcross}
\end{figure}

We now focus on the cross-correlation with the GAMA galaxies for the remainder of the paper. In \autoref{fig:PkGAMAcross} we present the cross-correlation power spectrum between the GAMA galaxies and the MeerKLASS deep field. All points are positive so none are lost to the log scaling, indicating a strong, positive correlation. Generally, a good agreement with the model is obtained. We performed a null test on the result by substituting the GAMA mock galaxies (from \secref{sec:Galmocks}) in place of the real galaxies. The mocks should not correlate with the intensity maps, and indeed, the resulting cross-power converges to zero once averaged over a sufficient number of mocks. The fitted amplitude of the model is equivalent to an inferred parameter of $\overline{T}_\hi\txteq{=}0.166\,\text{mK}$, with all other values in \autoref{tab:fid_params} assumed. We discuss this further in \appref{sec:ParamConstraints}, but do not focus on parameter inference in detail due to the large statistical noise owing to the small overlap footprint between MeerKLASS and GAMA.

The error bars for the cross-power in \autoref{fig:PkGAMAcross} are obtained using the scatter from the foreground transfer function as discussed in \secref{sec:TFforErrors}. The error size appears to be over-predicted based on the scatter of the data points, which have a reasonably tight agreement with the model relative to what the error bars reflect. This is supported by examining the reduced $\chi^2$ which we find to be $\chi^2_\text{dof}\txteq{\equiv}\chi^2{/}N_\text{dof}\txteq{=}0.42$. A $\chi^2_\text{dof}\txteq{<}1$ is conventionally indicative of overinflated errors. Using the reduced $\chi^2$ however is not a reliable test where there are non-linear contributions, as is likely the case for these observations that are still prone to residual systematics. In these cases, the number of degrees of freedom cannot be robustly determined \citep{Andrae:2010gh}. For our case we simply assume $N_\text{dof}\txteq{=}13$, for our 14 $k$-bins minus a single fitted degree of freedom, representing the amplitude fit. Furthermore, noise within the data causes high uncertainty in $\chi^2$ estimates, which despite the relatively low noise levels we are now reaching, may still be a concern. Alternatives to $\chi^2$ have been proposed, and specifically for 21cm applications 
\citep[e.g.][]{Tauscher:2018uxi}, but these have not been pursued in this work. We only use the $\chi^2$ as an approximate guide, mainly for relative comparisons between different cases.

\begin{figure}
    \centering
    \includegraphics[width=1\linewidth]{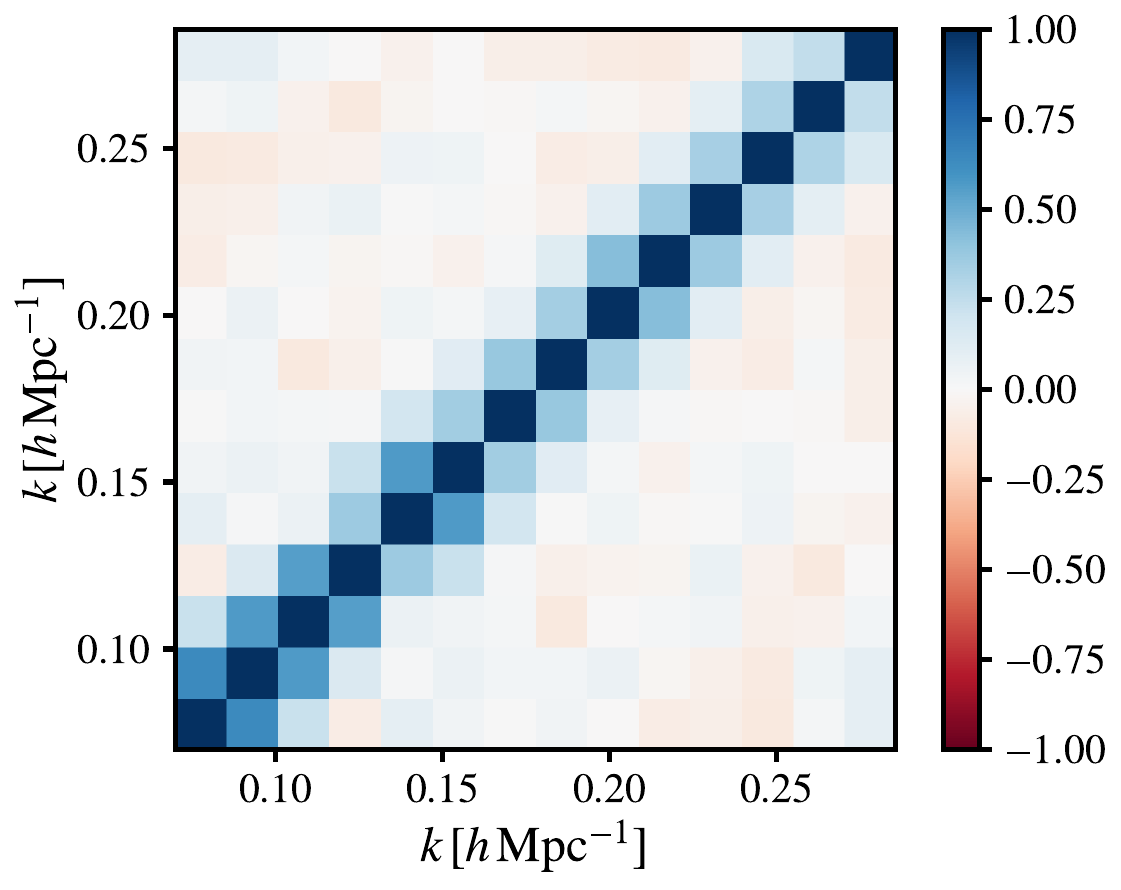}
    \caption{$k$-bin correlation matrix for the cross-power between the MeerKLASS deep field and GAMA galaxies, shown in \autoref{fig:PkGAMAcross}.}
    \label{fig:covariance_kbin}
\end{figure}

For the estimation of $\chi^2$, we include the full contribution from off-diagonals in the $k$-bin covariance $\textbf{\textsf{C}}$, which can be trivially estimated for our error estimation method, simply the covariance over $\hat{P}_i^\text{rec}(k)\txteq{=}\hat{P}_\text{rec}(k)\txteq{+}\delta \hat{P}^\text{rec}_i(k)$ (see \secref{sec:TFforErrors}). The $\chi^2$ is then estimated from
%
%
\begin{equation}
    \chi^2=(\hat{\boldsymbol{p}}-\boldsymbol{p}_{\rm mod})^\text{T} \, \textbf{\textsf{C}}^{-1} \, (\hat{\boldsymbol{p}}-\boldsymbol{p}_{\rm mod})\,,
\end{equation}
where $\hat{\boldsymbol{p}}$ is the data vector and $\boldsymbol{p}_\text{mod}$ the corresponding model. We present the normalised covariance matrix (or correlation matrix) in \autoref{fig:covariance_kbin}, showing evidence that the $k$-bins are slightly correlated, especially at low-$k$. This could be contributing to the large error bars, hence why we include the \textit{full} covariance in the $\chi^2$ estimation.

\begin{figure*}
    \centering
    \includegraphics[width=1\linewidth]{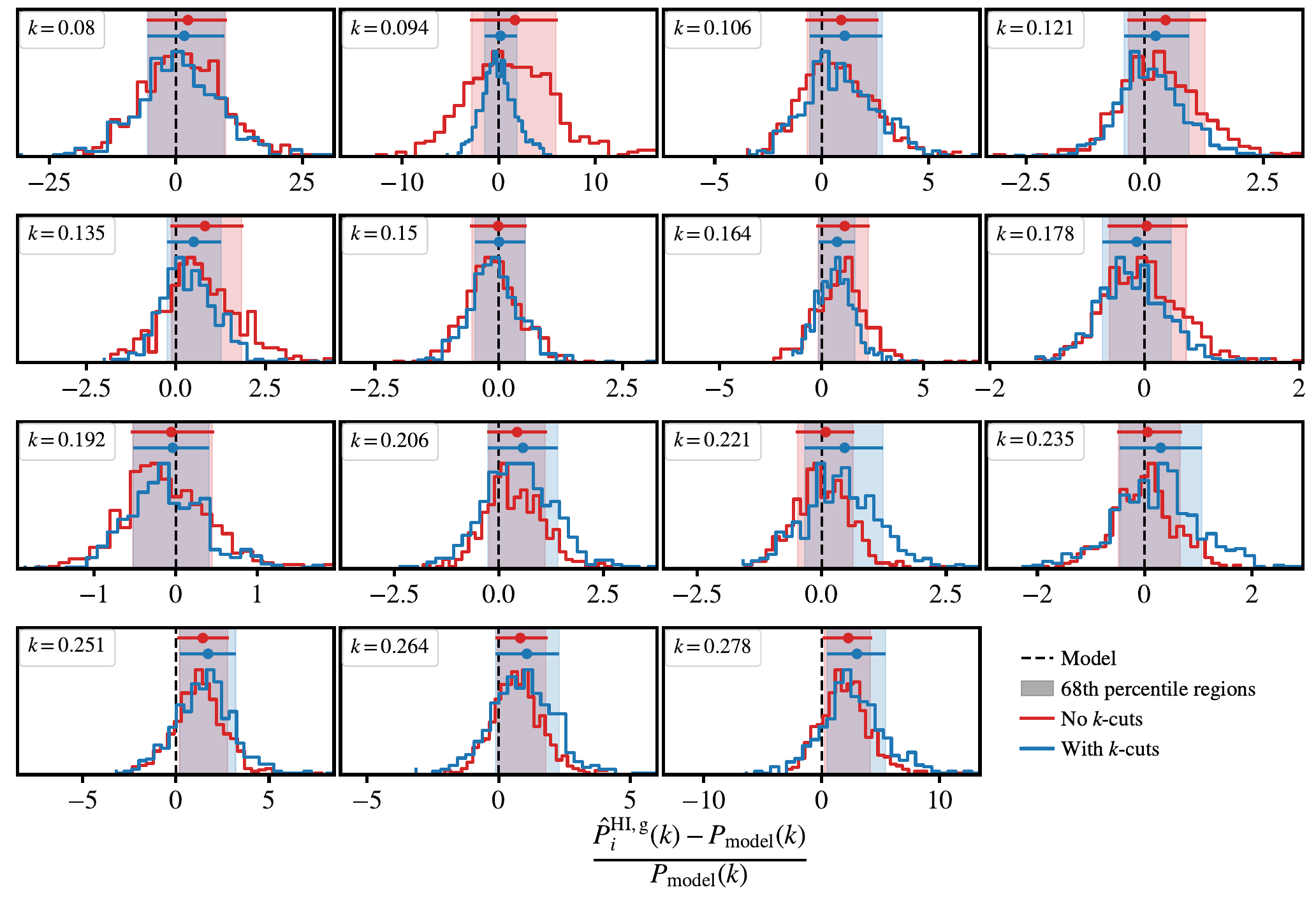}
    \caption{
    Estimated power distribution for each $k$-bin from the cross-correlation between MeerKLASS intensity maps and GAMA. This is computed using the power distribution in the 500 mocks used in the foreground transfer function.
    $k$-bin value is given in the top-left corner of each panel. The power distributions are centred on the fitted model at each $k$ so a zero median in the distribution represents perfect accuracy. Shaded regions show the 68th percentiles for the distributions which are used as the limits for the $1\,\sigma$ error bars in the power spectra plots.}
    \label{fig:kbin_histrograms}
\end{figure*}

To investigate the errors from the cross-power further, we plot the full distributions for the reconstructed power spectra $\hat{P}_i^\text{rec}$ from each $i$th mock in the foreground transfer function computation. \autoref{fig:kbin_histrograms} shows this for each $k$-bin, including an additional low-$k$ bin which we otherwise exclude due to the large error on these scales. We centre the distributions on the model and normalise by it, to aid comparison with the case of no $k$-cuts (red results), which includes the areas of $k$-space with low signal-to-noise (discussed in \secref{sec:kcuts}). The plot shows how the distributions have some non-Gaussian features, which could also impact error estimation, along with the correlated $k$-bins shown by the covariance. We find the average kurtosis across all $k$-bins for the $k$-cut case to be $-0.476$, indicating the distribution is \textit{flatter} than a normal distribution with more contribution in the tails. This could also widen the central percentiles and thus increase the error bar size. 

\autoref{fig:kbin_histrograms} shows a mild improvement in performance for the case with $k$-cuts. The distributions around the model are less broad at low-$k$, which is expected since this was one of the regions where signal-to-noise was poorest (see third panel of \autoref{fig:2DTF_SNR}). In terms of the Gaussianity of the distributions, we find a better kurtosis score for the case with $k$-cuts, suggesting the profiles are more Gaussian. In contrast, we find the results \textit{without} $k$-cuts to be less skewed, although this is marginal and in general, the skewness of the distributions is not concerning. Any improvement coming from the $k$-cuts is small and we also find little difference in detection significance, as we will explore shortly. As the volume of data grows beyond this small overlap sample, the $k$-cut results may prove more optimal since we will no longer be cutting modes from an already sample variance-limited data set. This is something that future data sets with MeerKLASS will reveal.

\textcolor{black}{Perfect Gaussian distributions in \autoref{fig:kbin_histrograms} is an unrealistic expectation, even in the high-precision limit. This is explored in \citet{Wilensky:2022sfh}, albeit for 21cm epoch of reionisation analyses. They highlight how in cross-multiplication estimators such as our galaxy cross-correlation, implicit assumptions made in the central limit theorem have consequences for the Gaussianity of error distributions. However, our distributions in \autoref{fig:kbin_histrograms} serve mainly as a guide
where approximate Gaussianity is sought to demonstrate reasonable robustness in our data and error estimation procedure. This is supported by \citet{Wilensky:2022sfh}, which concludes that extreme violations of the central limit theorem are required to break the approximate Gaussianity in error distributions. This will require careful consideration in future precision cosmology involving parameter inference. However, for current pathfinding applications, the Gaussianity and noise within the distributions of \autoref{fig:kbin_histrograms} provide deeper context into error estimation, without overly stretching central limit theorem assumptions.}

\subsubsection{Varying the foreground clean}

\begin{figure}
    \centering
    \includegraphics[width=1\linewidth]{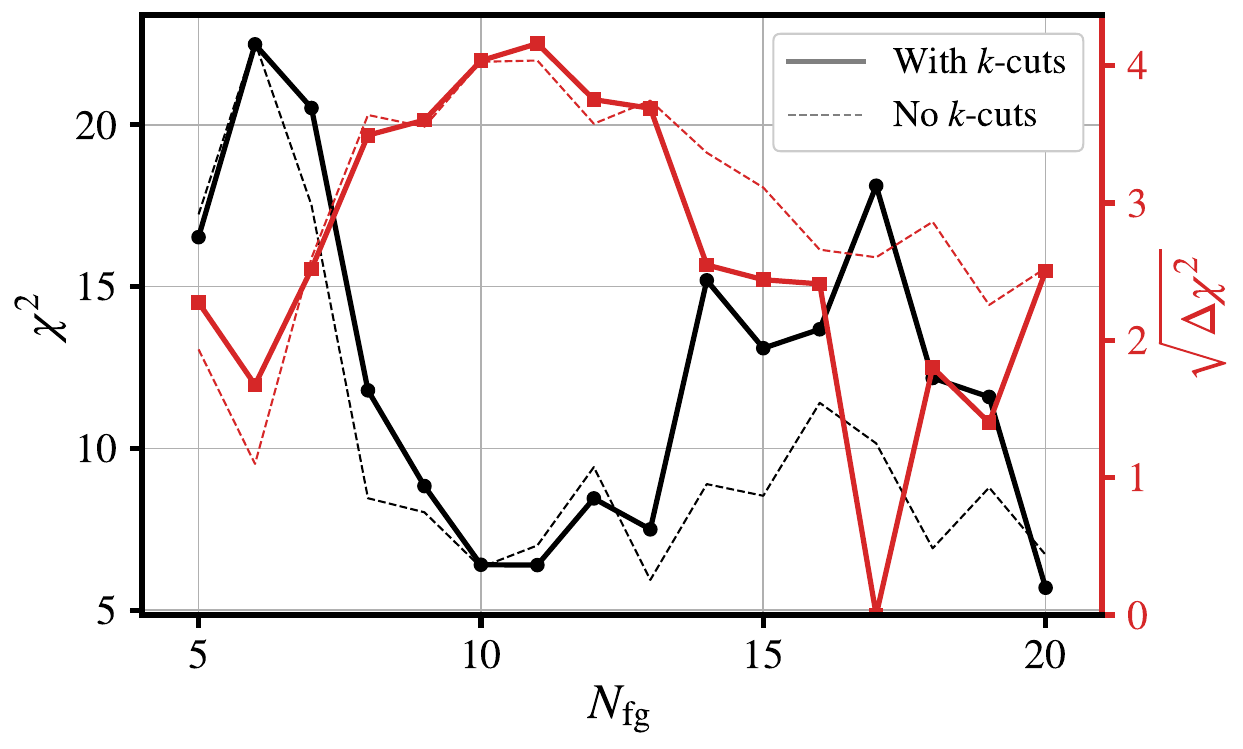}
    \caption{Variation of cross-correlation results with changing $\Nfg$ i.e.\ the number of PCA modes removed in the foreground clean. The left (black) axis shows the $\chi^2$, the lower values generally represent a better fit between cross-power data and model. The right (red) axis shows the cross-correlation power spectrum detection significance. Solid and thin-dashed lines represent the case with and without $k$-cuts respectively.}
    \label{fig:NfgsVariation}
\end{figure}

So far we have only considered the chosen case of removing $\Nfg\txteq{=}10$ PCA modes for the foreground clean. \autoref{fig:NfgsVariation} shows how the $\chi^2$ (which we continue to use as a rough goodness-of-fit metric) varies with $\Nfg$. As seen in previous work \citep{Cunnington:2022uzo}, too low an $\Nfg$ does not sufficiently clean the foregrounds, causing a poor fit. Going too high with $\Nfg$ also diminishes the fit. Ideally, we would expect high $\Nfg$ to cause more signal loss which should be reconstructed by the transfer function, giving unbiased results and thus not diminishing the fit. A higher uncertainty could be expected with greater signal loss, but this should be accounted for by the increasing error bars i.e.\ the $\chi^2$ should stabilise and not start to rise again at high $\Nfg$. On further scrutiny, we found that it is a single data point at $k\txteq{\sim}0.2\hMpc$ with very little power distorting the fits at higher $\Nfg$. This is evidence of low-level residual systematics that become more influential as signal is stripped away in the more aggressive foreground clean. Therefore, with future results where systematics are more controlled, this is expected to be less of a problem. For the case with no $k$-cuts (thin dashed lines), there is a plateau in $\chi^2$ at higher $\Nfg$, as we argued should be the expected situation. This is because the no $k$-cut power spectra were less impacted by the $k\txteq{\sim}0.2\hMpc$ anomaly at high $\Nfg$ which is potentially suggesting that the $k$-cuts are overly aggressive for this scale. With larger overlapping fields, the more accessible $k$-bins will naturally mitigate this discrepency. The no $k$-cut results in \autoref{fig:NfgsVariation} are therefore encouraging that a model and data fit should reach a plateau for high $\Nfg$ as expected with the current analysis pipeline. 

The right hand side (red) $y$-axis in \autoref{fig:NfgsVariation}, also shows the detection significance defined by $\sqrt{\Delta\chi^2}\txteq{\equiv}\sqrt{\chi^2 - \chi_\text{null}^2}$, where $\chi^2_\text{null}$ is the result's $\chi^2$ with a null model with zero power. The cross-correlation with GAMA is returning a modest ${>}\,4\sigma$ detection for the most optimal choices of foreground clean, lower than previous MeerKLASS detections \citep{Cunnington:2022uzo}. This is however still impressive, given that galaxies only overlap with approximately a quarter of the full MeerKLASS deep field. The detection significance is also restricted by the more robust approach to error estimation, which as we have already discussed, provides conservative estimates potentially inflated by residual systematics and correlated $k$-bins. To demonstrate this point, we artificially construct a scenario where a $\chi^2_\text{dof}\txteq{\sim}1$ is obtained for the $\Nfg\txteq{=}10$ case. This is achieved by assuming a perfectly diagonal covariance and
decreasing the diagonal errors by $40\%$. This boosts the detection significance to $9.8\,\sigma$, verifying that it is the conservative approach to errors lowering the detection significance, rather than the data-model residual. This is of course only a toy demonstration but it validates the argument that if large errors, which are suspected to be driven by systematics and $k$-bin correlations, can be contained, then detection robustness should improve. This is expected in future work where we will pursue dedicated studies into systematic mitigation. Some mode coupling is always likely, driven at least by non-linear clustering. However, we are confident that improvements can be made in covariance. MeerKLASS observations ongoing now are covering wider areas with more overlap with Stage-IV optical galaxy surveys. These observations are also using MeerKAT's UHF-band, which will allow a deeper usable redshift range, extending the $0.39\txteq{<}z\txteq{<}0.46$ used in this work to approximately $0.4\txteq{<}z\txteq{<}1.4$. This will naturally improve the large statistical noise due to the small volume overlap in this work.


\section{Stacking} \label{sec:stack}
In this section, we explore the possibility of a stacking detection of the \hi\ signal in the MeerKLASS L-band deep-field intensity map. The technique of spectral line stacking \citep{2001Sci...293.1800Z} can be used to detect the \hi\ content of galaxies below noise limit. In combination with spectroscopic galaxy surveys, spectral line stacking probes the average emission line profile of selected galaxy samples. Complimentary to the intensity mapping technique, the stacking measurement of the emission line profile probes the \hi\ mass function of galaxy samples (e.g.\ \citealt{2022ApJ...940L..10B}), the relation of \hi\ content with its host environment (e.g.\ \citealt{2020ApJ...894...92G,2023ApJ...950L..18B,2023MNRAS.523.2693D,2023MNRAS.518.4646R}), and the baryonic Tully-Fisher relation (e.g.\ \citealt{2016MNRAS.455.3136M}). Interferometric observations of MeerKAT using the MeerKAT International GigaHertz Tiered Extragalactic Exploration survey (MIGHTEE; \citealt{2016mks..confE...6J,2021A&A...646A..35M}) have been used to produce \hi\ measurements with regard to the scaling relation \citep{2022ApJ...935L..13S} with galaxy properties and the dependency on the large-scale structure environment \citep{2024MNRAS.529.4192S}. See also \cite{2021MNRAS.508.1195P,2022MNRAS.513.2168T,2023MNRAS.525..256P,2023MNRAS.522.5308P} for direct detections of \hi\ galaxies in the MIGHTEE survey.

For intensity mapping experiments, stacking analysis has been performed on the \hi\ intensity maps using the Parkes telescope \citep{Anderson:2017ert,2019MNRAS.489..385T,2020MNRAS.498.5916T} and the CHIME telescope \citep{CHIME:2022kvg}, as well as on CO intensity maps from the CO Mapping Array Project \citep{2024ApJ...965....7D} and Ly$\alpha$ emission of [O III]-selected galaxies from the Hobby-Eberly Telescope Dark Energy Experiment \citep{2022ApJ...929...90L,2022ApJ...934L..26L}.

MeerKLASS, on the other hand, faces unique challenges. The $\txteq{\sim}1\,$deg instrument power beam introduces heavy contamination from nearby sources. For the GAMA galaxy sample used, the number density of the sources is $\txteq{\sim} 35\,{\rm deg^{-2}}$ when integrated along the frequency sub-band. Therefore, when stacking each source, the stacking procedure will inevitably include emissions from more than one GAMA galaxy. The velocity width of the \hi\ sources further scatter the 21\,cm emission into nearby channels. The double-counting of sources from the beam creates a plateau of excess signal as shown in \autoref{fig:mockstack}, and can be resolved by PCA cleaning, which we discuss in \hyperref[apdx:stack]{Appendix \ref{apdx:stack}}. However, the residual after PCA cleaning also incorporates systematic effects. The chromatic structure of the systematic effects produces oscillations in the stacked profile, with their amplitudes comparable to the signal. For this work, we are only interested in using stacking as part of the validation test. As discussed later, we obtain a detection of a stacked signal in both angular and spectral space, which can be validated by performing a null test with random galaxy positions. The existence of a stacked signal further demonstrates that we have reached the depth needed for the possible detection of the \hi\ power spectrum using the MeerKLASS L-band intensity maps. We leave a more detailed study into refining the stacking technique for probing \hi\ science to follow-up work.

\subsection{The stacking procedure of the intensity maps}
We use the PCA-cleaned map described in \hyperref[sec:FGcleaning]{Section \ref{sec:FGcleaning}} and the positions of the GAMA galaxies to perform a 3D stacking in $(\delta\alpha,\delta\phi,\delta\nu)$ space so that
\begin{equation}
    I(\delta\alpha,\delta\phi,\delta\nu) = \frac{\sum_i I(\alpha_i{+}\delta\alpha,\,\phi_i{+}\delta\phi,\,\nu_i{+}\delta\nu) w_i}{\sum_i w_i},
\end{equation}
where $(\alpha_i,\phi_i)$ are the right ascension and declination of the map pixel in which the $i^{\rm th}$ galaxy resides, $I$ is the flux density, and $w_i\txteq{=}w_{\hi}(\alpha_i{+}\delta\alpha,\phi_i{+}\delta\phi,\nu_i{+}\delta\nu)$ is the reconvolved weight of each pixel defined in \hyperref[subsec:reconvolution]{Section \ref{subsec:reconvolution}}. We choose the angular and the spectral bins to be consistent with the map resolution with $\Delta \alpha\txteq{=}\Delta \phi\txteq{=}0.3\,$deg. The frequency resolution corresponds to an effective velocity resolution of $\Delta v\txteq{=}62.83\,$km\,s$^{-1}$ at the centre frequency of the sub-band $997.37\,$MHz. Note that no Doppler correction has been performed on the data. For our analysis, the intensity maps are coherently averaged in frequencies for PCA cleaning and therefore can not be averaged in velocity channels.

In order to calculate the statistical noise for the stacked cube, we perform the same stacking routine on random galaxy positions, similar to the shuffling null test in \secref{sec:GAMAcross}. The positions are randomly sampled uniformly within the region of the GAMA galaxy sample. The redshifts are randomly sampled following the redshift distribution of the survey. The random process is repeated for 100 realisations to calculate the mean and the standard deviation of the reference cube. The reference cube is used for performing the null test, i.e. checking that the stacked emission signal in the reference cube is consistent with zero.

After obtaining the stacked cubes, we create the angular and spectral stacking by summing over the cubes along the radial and transverse directions. To control the systematics generated by the foreground cleaning, we limit the voxels that go into the summation. For angular stacking, only voxels that are within 150\,km\,s$^{-1}$ of the central channel are considered, corresponding to the typical width of the HI emission line profile. For spectral stacking, only the voxels whose centre positions are within the $1\,$deg area of the centre voxel in each channel are included, corresponding to the size of the primary beam. The same summation is performed on the random realisation to calculate the mean and the standard deviation of the reference image and spectrum. We note that there is signal loss due to an insufficient number of voxels included. However, as the random realisations follow the same procedure, it does not affect the validation of the detection which is our focus in this work. 

\begin{figure}
    \centering
    \includegraphics[width=1\linewidth]{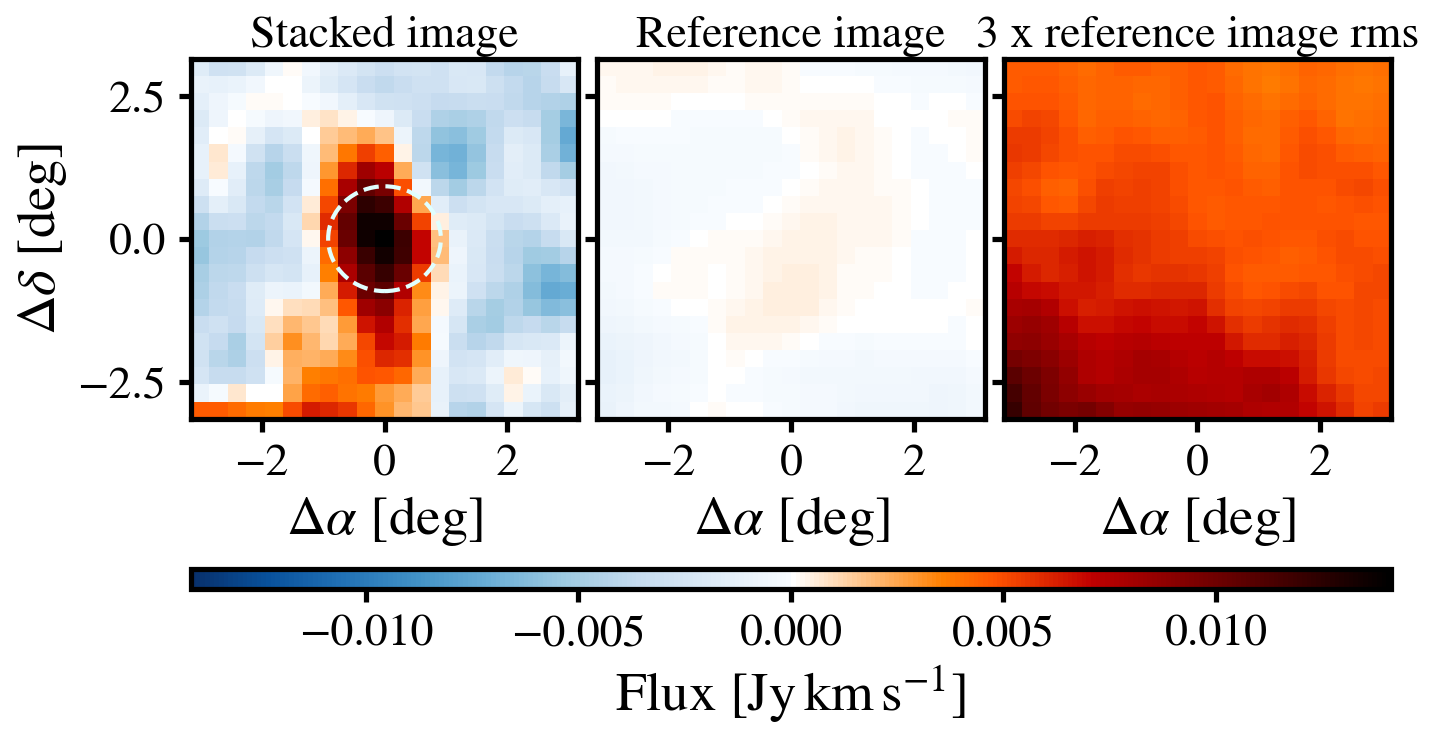}
    \caption{Left panel: The angular stacking of the GAMA galaxies on the \hi\ intensity map. The dashed circle denotes the FWHM of the beam. Central panel: The reference null detection image from 100 realisations of random galaxy positions. Right panel: The standard deviation of the reference image multiplied by 3 to visualise the 99.7\% confidence interval.}
    \label{fig:ang_stack}
\end{figure}

\subsection{The detection of stacked \hi\ signal}

In \autoref{fig:ang_stack}, we show the results for the angular stacking of the GAMA galaxies on the \hi\ intensity map. The stacked map shows a clear detection at the centre of the image, with the centre pixel achieving a detection with a statistical significance of $8.35\,\sigma$. The stacked \hi\ emission follows the structure of the beam. 
The extended structure of emission outside the beam FWHM is a result of the relatively large thermal noise fluctuation and contamination from nearby sources in the stacking procedure as discussed above. We note that, as shown in \autoref{fig:FGmap_and_counts}, the lower part of the GAMA galaxy sample region sits near the edge of the observed intensity maps. As a result, the lower half of the stacked image has a higher level of statistical fluctuations as shown in the right panel of \autoref{fig:ang_stack}.
Furthermore, the reference image is consistent with a null detection, demonstrating that the stacked signal does not arise from systematic effects in the map.

\begin{figure*}
    \centering
    \includegraphics[width=0.8\linewidth]{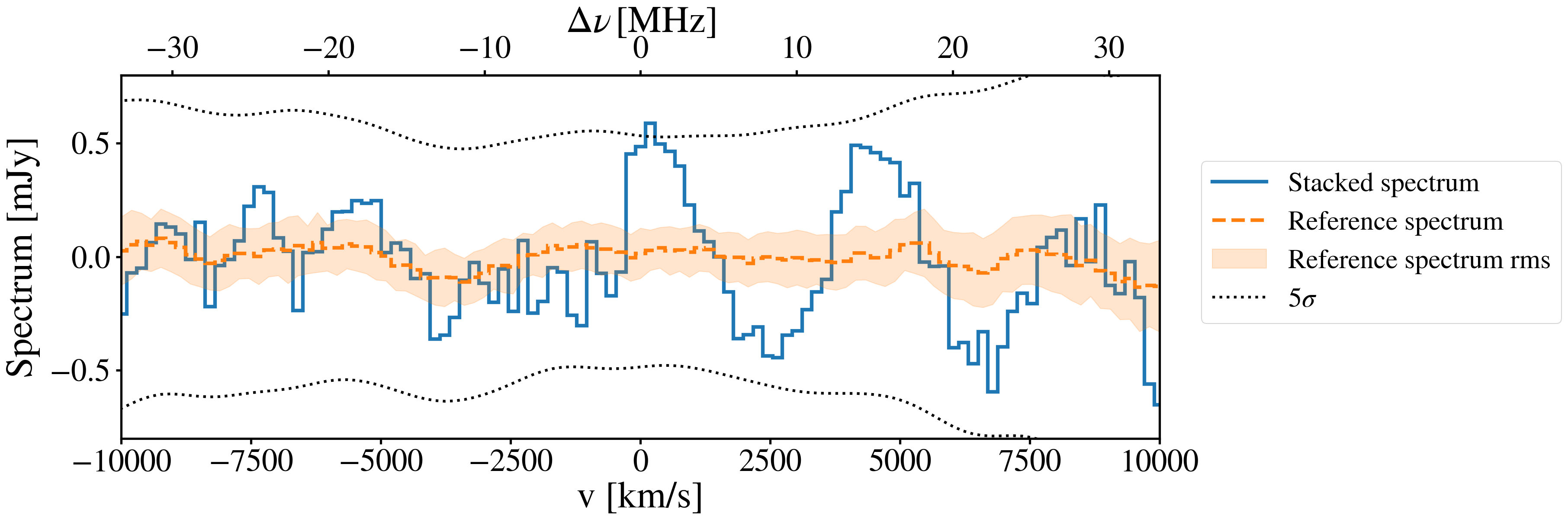}
    \caption{The stacked spectrum of the GAMA galaxies on the \hi\ intensity map. The blue solid line marks the measured stacked spectrum. The orange dashed line denotes the reference spectrum from 100 random realisations, and the shaded region denotes the standard deviation of the realisations. The black dotted line denotes the $5\,\sigma$ limit of the reference spectrum rms and is smoothed with a Gaussian kernel of 300\,km/s for better illustration.}
    \label{fig:spectral_stack}
\end{figure*}

In \autoref{fig:spectral_stack}, we show the stacked spectrum along the line-of-sight. 
The peak of the stacked spectrum shows a detection with a significance of $6.30\,\sigma$. The detection is validated by the reference spectrum which is consistent with a null detection. However, as seen in the reference spectrum, there is a systematic noise component with fluctuations on large frequency intervals. The systematics can be most visibly seen at the peaks and troughs of the stacked signal at $v \txteq{\sim} 2500,\,4500,\,$ and $7000\,$km/s. We have identified the source of this fluctuation as the residual systematics in the map after PCA cleaning, which is demonstrated in \hyperref[apdx:stack]{Appendix \ref{apdx:stack}}.
As stated previously, we are only interested in validating the detection against the null test and leave the suppressing of the systematic effects for future work.

The stacking results show a clear detection of \hi\ signal in the positions of the GAMA galaxies. The detection of an emission line signal is statistically significant against the null test.
At the same time, the residual systematics in the data, while being sub-dominant comparing to the \hi\ signal, induce structures in the stacked spectrum. We intend to follow up on the findings in this work and develop analysis techniques to extract the underlying emission line signal in the future, as well as using the stacked signal to understand the residual systematics.

\section{Conclusions}\label{sec:conclusion}

In this work we have presented L-band observations from the MeerKLASS intensity mapping campaign. The survey covered a single patch of approximately $200\,\deg^2$, repeatedly scanned 41 times (27 scans remained after aggressive RFI flagging), which we have referred to as the MeerKLASS L-band deep field. The observations used in our analysis came from only a small section of the full L-band ($971.2\txteq{<}\nu\txteq{<}1023.6$ at redshift $0.39\txteq{<}z\txteq{<}0.46$) to avoid RFI which we find is more present in this band relative to the lower frequency UHF-band ($580\txteq{<}\nu\txteq{<}1000$) which all future MeerKLASS surveys will be conducted in. We detailed the calibration strategy used on these observations which included a new approach of self-calibration (\secref{sec:SelfCal}), exploiting the data's higher signal-to-noise to devise a less model-dependent calibration. The improvement in data quality relative to previous MeerKLASS pilot surveys is evident (see e.g.\ \autoref{fig:LoS_spectra}) and encouraging for future MeerKLASS surveys gathering ever-larger data sets.

We demonstrated the enhanced signal-to-noise from the calibrated deep-field intensity maps by presenting the auto-correlation power spectrum (\autoref{fig:HIauto_power}) which revealed the lowest upper-limit from a \hi\ power spectrum at these cosmological scales $k\txteq{<}0.3\hMpc$. This remains a factor of ${\sim}\,2$ above where we would expect the \hi\ power to lie. We will pursue this limit further in follow-up work where we will explore a systematic removal strategy to reduce the quantity of flagged data plus the cross-correlation of divided data sets to mitigate the additive bias from time-dependent residual systematics. 

We also presented results from the cross-correlation power spectrum between the MeerKLASS deep-field and 2269 overlapping GAMA galaxies with spectroscopic redshifts. Despite the small field of overlap (${\sim}25\%$ of the deep field patch), we still obtained a clear ${>}\,4\,\sigma$ detection of the cross-power spectrum (\autoref{fig:PkGAMAcross}). Furthermore, we presented a detection of \hi\ emission from stacking the MeerKLASS maps onto the positions of the GAMA galaxies. This yielded a $8.35\,\sigma$ and $6.30\,\sigma$ detection from angular and spectral stacking respectively, the first stacking detections with MeerKAT in single-dish mode (see \secref{sec:stack}). 

We implemented several novel strategies into the power spectrum analysis pipeline, including a selective procedure to target the region of $k$-space with the highest signal-to-noise. The third panel of \autoref{fig:2DTF_SNR} demonstrated the clear window of Fourier modes where power dominates over the error. By implementing some minima and maxima cuts on $k_\perp$ and $k_\parallel$ we were able to exclude regions of $k$-space dominated by error. This provides a more robust result since error-dominated modes will not be included in final spherical averaging. Comparison of the error profiles for each $k$-bin (\autoref{fig:kbin_histrograms}) demonstrated a mild improvement in terms of Gaussianity when using the $k$-cuts, however, the sacrifice of many modes contained at high-$k$ could be restricting this improvement. We plan to investigate this strategy in future work where larger data sets will reveal the optimal approach.

We formalised our approach to covariance estimation for the power spectra measurements which utlises the foreground transfer function distribution from each mock realisation. This distribution will be determined by all foreseeable sources of error; residual systematics, thermal noise fluctuations, galaxy shot-noise, cosmic variance and any uncertainty in signal-loss correction. This should therefore be a robust method for error estimation. However, we found the errors in the spherically averaged cross-power spectrum appear to be large relative to their scatter about the model. This was supported by a $\chi^2$ analysis. Further inspection revealed that the $k$-bins are mildly correlated and the error profiles for each $k$-bin are not perfectly Gaussian, both of which could be inflating the errors. We hypothesise that this should be naturally mitigated with future MeerKLASS analysis where systematics can be more reliably controlled. 

This work also presented validation tests of our improved analysis pipeline. This was achieved by generating a suite of simulations that emulate the real data products. The mock data was then put through the same pipeline as the real data where we could validate its performance. Results in \autoref{fig:PksFromMocks} showed good agreement with the expected model based on the known inputs to generate the mocks, thus validating processes such as Cartesian field regridding and power spectrum estimation. Measured power from the mocks in \autoref{fig:PksFromMocks_wFG} which included the foreground cleaning also showed good consistency with real data.  


The MeerKLASS campaign is ongoing and will continue to be so in the pre-SKAO era. Further observations have already begun in the telescope's UHF-band. These lower frequencies should be less prone to RFI, meaning far fewer channels will be flagged. Furthermore, the lower frequency and wider wave-band mean a much larger portion of redshift space will be covered, providing intensity maps of unprecedented volume.

\section*{Acknowledgements}

The authors would like to acknowledge the ongoing contributions to the MeerKLASS collaboration from members Suman Chatterjee, Karin Fornazier, Tamirat Gogo, Wenkai Hu, Piyanat Kittiwisit and Sifiso Mahlalela which provided valuable insights for this paper. We would also like to thank Jacob Burba, whose UK SKA Regional Centre project involved a full run-through of our post-calibration analysis pipeline, providing useful validation. 

JW acknowledges support from the National SKA Program of China (No. 2020SKA0110100).
SCu is supported by a UK Research and Innovation Future Leaders Fellowship grant [MR/V026437/1]. 
This result is part of a project that has received funding from the European Research Council (ERC) under the European Union's Horizon 2020 research and innovation programme (Grant agreement No. 948764; PB).
JLB acknowledges funding from the Ramón y Cajal Grant RYC2021-033191-I, financed by MCIN/AEI/10.13039/501100011033 and by
the European Union “NextGenerationEU”/PRTR, as well as the project UC-LIME (PID2022-140670NA-I00), financed by MCIN/AEI/ 10.13039/501100011033/FEDER, UE.
SCa acknowledges support from the Italian Ministry of University and Research, PRIN 2022 `EXSKALIBUR -- Euclid-Cross-SKA: Likelihood Inference Building for Universe Research', from the Italian Ministry of Foreign Affairs and International
Cooperation (grant no.\ ZA23GR03), and from the European Union -- Next Generation EU. 
IPC is supported by the European Union within the Next Generation EU programme [PNRR-4-2-1.2 project No. SOE\textunderscore0000136, RadioGaGa].
JF acknowledges support of Funda\c{c}\~{a}o para a Ci\^{e}ncia e a Tecnologia through the Investigador FCT Contract No. 2020.02633.CEECIND/CP1631/CT0002, the FCT project PTDC/FIS-AST/0054/2021, and the research grants UIDB/04434/2020 and UIDP/04434/2020. 
AP is a UK Research and Innovation Future Leaders Fellow [grant MR/X005399/1]. ZC's research is supported by a UK Research and Innovation Future Leaders Fellowship [grant MR/X005399/1].
MGS acknowledges support from the South African Radio Astronomy Observatory and National Research Foundation (Grant No. 84156).
SCa acknowledges support from the Italian Ministry of University and Research (\textsc{mur}), PRIN 2022 `EXSKALIBUR – Euclid-Cross-SKA: Likelihood Inference Building for Universe's Research', Grant No.\ 20222BBYB9, CUP D53D2300252 0006, from the Italian Ministry of Foreign Affairs and International
Cooperation (\textsc{maeci}), Grant No.\ ZA23GR03, and from the European Union -- Next Generation EU.
LW is a UK Research and Innovation Future Leaders Fellow [grant MR/V026437/1]. 

The MeerKAT telescope is operated by the South African Radio Astronomy Observatory, which is a facility of the National Research Foundation, an agency of the Department of Science and Innovation. We acknowledge the use of the Ilifu cloud computing facility, through the Inter-University Institute for Data Intensive Astronomy (IDIA).

GAMA is a joint European-Australasian project based around a spectroscopic campaign using the Anglo-Australian Telescope. The GAMA input catalogue is based on data taken from the Sloan Digital Sky Survey and the UKIRT Infrared Deep Sky Survey. Complementary imaging of the GAMA regions is being obtained by a number of independent survey programmes including GALEX MIS, VST KiDS, VISTA VIKING, WISE, Herschel-ATLAS, GMRT and ASKAP providing UV to radio coverage. GAMA is funded by the STFC (UK), the ARC (Australia), the AAO, and the participating institutions. The GAMA website is \href{https://www.gama-survey.org/}{https://www.gama-survey.org/}.

For the purpose of open access, the author has applied a Creative Commons Attribution (CC BY) licence to any Author Accepted Manuscript version arising from this submission. 

\section*{Data Availability}

The data underlying this article will be shared upon reasonable request to the corresponding author. Access to the raw data used in the analysis is public (for access information please contact archive@ska.ac.za).



\bibliographystyle{mnras}
\bibliography{bib} 

\newcommand{\noop}[1]{}
\begin{thebibliography}{}
\makeatletter
\relax
\def\mn@urlcharsother{\let\do\@makeother \do\$\do\&\do\#\do\^\do\_\do\%\do\~}
\def\mn@doi{\begingroup\mn@urlcharsother \@ifnextchar [ {\mn@doi@}
  {\mn@doi@[]}}
\def\mn@doi@[#1]#2{\def\@tempa{#1}\ifx\@tempa\@empty \href
  {http://dx.doi.org/#2} {doi:#2}\else \href {http://dx.doi.org/#2} {#1}\fi
  \endgroup}
\def\mn@eprint#1#2{\mn@eprint@#1:#2::\@nil}
\def\mn@eprint@arXiv#1{\href {http://arxiv.org/abs/#1} {{\tt arXiv:#1}}}
\def\mn@eprint@dblp#1{\href {http://dblp.uni-trier.de/rec/bibtex/#1.xml}
  {dblp:#1}}
\def\mn@eprint@#1:#2:#3:#4\@nil{\def\@tempa {#1}\def\@tempb {#2}\def\@tempc
  {#3}\ifx \@tempc \@empty \let \@tempc \@tempb \let \@tempb \@tempa \fi \ifx
  \@tempb \@empty \def\@tempb {arXiv}\fi \@ifundefined
  {mn@eprint@\@tempb}{\@tempb:\@tempc}{\expandafter \expandafter \csname
  mn@eprint@\@tempb\endcsname \expandafter{\@tempc}}}

\bibitem[\protect\citeauthoryear{Agrawal, Makiya, Chiang, Jeong, Saito  \&
  Komatsu}{Agrawal et~al.}{2017}]{Agrawal:2017khv}
Agrawal A.,  Makiya R.,  Chiang C.-T.,  Jeong D.,  Saito S.,   Komatsu E.,
  2017, \mn@doi [JCAP] {10.1088/1475-7516/2017/10/003}, 10, 003

\bibitem[\protect\citeauthoryear{Alonso, Bull, Ferreira  \& Santos}{Alonso
  et~al.}{2015}]{Alonso:2014dhk}
Alonso D.,  Bull P.,  Ferreira P.~G.,   Santos M.~G.,  2015, \mn@doi [Mon. Not.
  Roy. Astron. Soc.] {10.1093/mnras/stu2474}, 447, 400

\bibitem[\protect\citeauthoryear{Anderson et~al.}{Anderson
  et~al.}{2018}]{Anderson:2017ert}
Anderson C.~J.,  et~al., 2018, \mn@doi [Mon. Not. Roy. Astron. Soc.]
  {10.1093/mnras/sty346}, 476, 3382

\bibitem[\protect\citeauthoryear{Andrae, Schulze-Hartung  \& Melchior}{Andrae
  et~al.}{2010}]{Andrae:2010gh}
Andrae R.,  Schulze-Hartung T.,   Melchior P.,  2010, \texttt{arXiv}, {}
  (\mn@eprint {arXiv} {1012.3754})

\bibitem[\protect\citeauthoryear{Ansari et~al.,}{Ansari
  et~al.}{2012}]{Ansari:2011bv}
Ansari R.,  et~al., 2012, \mn@doi [Astron. Astrophys.]
  {10.1051/0004-6361/201117837}, 540, A129

\bibitem[\protect\citeauthoryear{{Asad} et~al.,}{{Asad}
  et~al.}{2021}]{2021MNRAS.502.2970A}
{Asad} K.~M.~B.,  et~al., 2021, \mn@doi [\mnras] {10.1093/mnras/stab104}, \href
  {https://ui.adsabs.harvard.edu/abs/2021MNRAS.502.2970A} {502, 2970}

\bibitem[\protect\citeauthoryear{{Astropy Collaboration} et~al.,}{{Astropy
  Collaboration} et~al.}{2022}]{Astropy}
{Astropy Collaboration} et~al., 2022, \mn@doi [\apj]
  {10.3847/1538-4357/ac7c74}, \href
  {https://ui.adsabs.harvard.edu/abs/2022ApJ...935..167A} {935, 167}

\bibitem[\protect\citeauthoryear{Battye, Davies  \& Weller}{Battye
  et~al.}{2004}]{Battye:2004re}
Battye R.~A.,  Davies R.~D.,   Weller J.,  2004, \mn@doi [MNRAS]
  {10.1111/j.1365-2966.2004.08416.x}, 355, 1339

\bibitem[\protect\citeauthoryear{Battye et~al.,}{Battye
  et~al.}{2012}]{Battye:2012fd}
Battye R.~A.,  et~al., 2012, \texttt{arXiv}, {} (\mn@eprint {arXiv}
  {1209.1041})

\bibitem[\protect\citeauthoryear{{Bera}, {Kanekar}, {Chengalur}  \&
  {Bagla}}{{Bera} et~al.}{2022}]{2022ApJ...940L..10B}
{Bera} A.,  {Kanekar} N.,  {Chengalur} J.~N.,   {Bagla} J.~S.,  2022, \mn@doi
  [\apjl] {10.3847/2041-8213/ac9d32}, \href
  {https://ui.adsabs.harvard.edu/abs/2022ApJ...940L..10B} {940, L10}

\bibitem[\protect\citeauthoryear{{Bera}, {Kanekar}, {Chengalur}  \&
  {Bagla}}{{Bera} et~al.}{2023}]{2023ApJ...950L..18B}
{Bera} A.,  {Kanekar} N.,  {Chengalur} J.~N.,   {Bagla} J.~S.,  2023, \mn@doi
  [\apjl] {10.3847/2041-8213/acd0b3}, \href
  {https://ui.adsabs.harvard.edu/abs/2023ApJ...950L..18B} {950, L18}

\bibitem[\protect\citeauthoryear{Bernal, Breysse, Gil-Mar\'\i{}n  \&
  Kovetz}{Bernal et~al.}{2019}]{Bernal:2019jdo}
Bernal J.~L.,  Breysse P.~C.,  Gil-Mar\'\i{}n H.,   Kovetz E.~D.,  2019,
  \mn@doi [Phys. Rev. D] {10.1103/PhysRevD.100.123522}, 100, 123522

\bibitem[\protect\citeauthoryear{Beutler et~al.,}{Beutler
  et~al.}{2011}]{Beutler:2011hx}
Beutler F.,  et~al., 2011, \mn@doi [Mon. Not. Roy. Astron. Soc.]
  {10.1111/j.1365-2966.2011.19250.x}, 416, 3017

\bibitem[\protect\citeauthoryear{Bharadwaj, Nath, Nath  \& Sethi}{Bharadwaj
  et~al.}{2001}]{Bharadwaj:2000av}
Bharadwaj S.,  Nath B.,  Nath B.~B.,   Sethi S.~K.,  2001, \mn@doi [J.
  Astrophys. Astron.] {10.1007/BF02933588}, 22, 21

\bibitem[\protect\citeauthoryear{Blake}{Blake}{2019}]{Blake:2019ddd}
Blake C.,  2019, \mn@doi [Mon. Not. Roy. Astron. Soc.] {10.1093/mnras/stz2145},
  489, 153

\bibitem[\protect\citeauthoryear{{CHIME Collaboration} et~al.}{{CHIME
  Collaboration} et~al.}{2022}]{CHIME:2022dwe}
{CHIME Collaboration} et~al., 2022, \mn@doi [Astrophys. J. Supp.]
  {10.3847/1538-4365/ac6fd9}, 261, 29

\bibitem[\protect\citeauthoryear{{CHIME Collaboration} et~al.}{{CHIME
  Collaboration} et~al.}{2023}]{CHIME:2022kvg}
{CHIME Collaboration} et~al., 2023, \mn@doi [Astrophys. J.]
  {10.3847/1538-4357/acb13f}, 947, 16

\bibitem[\protect\citeauthoryear{Carucci, Irfan  \& Bobin}{Carucci
  et~al.}{2020}]{Carucci:2020enz}
Carucci I.~P.,  Irfan M.~O.,   Bobin J.,  2020, \mn@doi [Mon. Not. Roy. Astron.
  Soc.] {10.1093/mnras/staa2854}, 499, 304

\bibitem[\protect\citeauthoryear{{Carucci} et~al.,}{{Carucci}
  et~al.}{2024}]{2024arXiv241206750C}
{Carucci} I.~P.,  et~al., 2024, \texttt{arXiv}, {} (\mn@eprint {arXiv}
  {2412.06750})

\bibitem[\protect\citeauthoryear{Chang, Pen, Peterson  \& McDonald}{Chang
  et~al.}{2008}]{Chang:2007xk}
Chang T.-C.,  Pen U.-L.,  Peterson J.~B.,   McDonald P.,  2008, \mn@doi [Phys.
  Rev. Lett.] {10.1103/PhysRevLett.100.091303}, 100, 091303

\bibitem[\protect\citeauthoryear{Chen}{Chen}{2012}]{Chen:2012xu}
Chen X.,  2012, \mn@doi [Int. J. Mod. Phys. Conf. Ser.]
  {10.1142/S2010194512006459}, 12, 256

\bibitem[\protect\citeauthoryear{{Coles} \& {Jones}}{{Coles} \&
  {Jones}}{1991}]{ColesLognormal1991}
{Coles} P.,  {Jones} B.,  1991, \mn@doi [Mon. Not. Roy. Astron. Soc.]
  {10.1093/mnras/248.1.1}, \href
  {https://ui.adsabs.harvard.edu/abs/1991MNRAS.248....1C} {248, 1}

\bibitem[\protect\citeauthoryear{Cunnington}{Cunnington}{2022}]{Cunnington:2022ryj}
Cunnington S.,  2022, \mn@doi [Mon. Not. Roy. Astron. Soc.]
  {10.1093/mnras/stac576}, 512, 2408

\bibitem[\protect\citeauthoryear{Cunnington \& Wolz}{Cunnington \&
  Wolz}{2024}]{Cunnington:2023aou}
Cunnington S.,  Wolz L.,  2024, \mn@doi [Mon. Not. Roy. Astron. Soc.]
  {10.1093/mnras/stae333}, 528, 5586

\bibitem[\protect\citeauthoryear{Cunnington, Pourtsidou, Soares, Blake  \&
  Bacon}{Cunnington et~al.}{2020}]{Cunnington:2020mnn}
Cunnington S.,  Pourtsidou A.,  Soares P.~S.,  Blake C.,   Bacon D.,  2020,
  \mn@doi [Mon. Not. Roy. Astron. Soc.] {10.1093/mnras/staa1524}, 496, 415

\bibitem[\protect\citeauthoryear{Cunnington, Irfan, Carucci, Pourtsidou  \&
  Bobin}{Cunnington et~al.}{2021}]{Cunnington:2020njn}
Cunnington S.,  Irfan M.~O.,  Carucci I.~P.,  Pourtsidou A.,   Bobin J.,  2021,
  \mn@doi [Mon. Not. Roy. Astron. Soc.] {10.1093/mnras/stab856}, 504, 208

\bibitem[\protect\citeauthoryear{Cunnington et~al.}{Cunnington
  et~al.}{2022}]{Cunnington:2022uzo}
Cunnington S.,  et~al., 2022, \mn@doi [Mon. Not. Roy. Astron. Soc.]
  {10.1093/mnras/stac3060}, 518, 6262

\bibitem[\protect\citeauthoryear{Cunnington et~al.}{Cunnington
  et~al.}{2023}]{Cunnington:2023jpq}
Cunnington S.,  et~al., 2023, \mn@doi [Mon. Not. Roy. Astron. Soc.]
  {10.1093/mnras/stad1567}, 523, 2453

\bibitem[\protect\citeauthoryear{{DES Collaboration}}{{DES
  Collaboration}}{2022}]{DES:2021wwk}
{DES Collaboration} 2022, \mn@doi [Phys. Rev. D] {10.1103/PhysRevD.105.023520},
  105, 023520

\bibitem[\protect\citeauthoryear{{DESI Collaboration} et~al.}{{DESI
  Collaboration} et~al.}{2024}]{DESI:2024mwx}
{DESI Collaboration} et~al., 2024, \texttt{arXiv}, {} (\mn@eprint {arXiv}
  {2404.03002})

\bibitem[\protect\citeauthoryear{{Dev} et~al.,}{{Dev}
  et~al.}{2023}]{2023MNRAS.523.2693D}
{Dev} A.,  et~al., 2023, \mn@doi [\mnras] {10.1093/mnras/stad1575}, \href
  {https://ui.adsabs.harvard.edu/abs/2023MNRAS.523.2693D} {523, 2693}

\bibitem[\protect\citeauthoryear{Drinkwater et~al.}{Drinkwater
  et~al.}{2010}]{Drinkwater:2009sd}
Drinkwater M.~J.,  et~al., 2010, \mn@doi [Mon. Not. Roy. Astron. Soc.]
  {10.1111/j.1365-2966.2009.15754.x}, 401, 1429

\bibitem[\protect\citeauthoryear{Driver et~al.}{Driver
  et~al.}{2009}]{Driver:2009hm}
Driver S.~P.,  et~al., 2009, \mn@doi [Astron. Geophys.]
  {10.1111/j.1468-4004.2009.50512.x}, 50, 5.12

\bibitem[\protect\citeauthoryear{Driver et~al.}{Driver
  et~al.}{2011}]{Driver:2010zb}
Driver S.~P.,  et~al., 2011, \mn@doi [Mon. Not. Roy. Astron. Soc.]
  {10.1111/j.1365-2966.2010.18188.x}, 413, 971

\bibitem[\protect\citeauthoryear{Driver et~al.}{Driver
  et~al.}{2022}]{Driver:2022vyh}
Driver S.~P.,  et~al., 2022, \mn@doi [Mon. Not. Roy. Astron. Soc.]
  {10.1093/mnras/stac472}, 513, 439

\bibitem[\protect\citeauthoryear{{Dunne} et~al.,}{{Dunne}
  et~al.}{2024}]{2024ApJ...965....7D}
{Dunne} D.~A.,  et~al., 2024, \mn@doi [\apj] {10.3847/1538-4357/ad2dfc}, \href
  {https://ui.adsabs.harvard.edu/abs/2024ApJ...965....7D} {965, 7}

\bibitem[\protect\citeauthoryear{Engelbrecht et~al.}{Engelbrecht
  et~al.}{2024}]{Engelbrecht:2024eoc}
Engelbrecht B.,  et~al., 2024, \texttt{arXiv}, {} (\mn@eprint {arXiv}
  {2404.17908})

\bibitem[\protect\citeauthoryear{Feldman, Kaiser  \& Peacock}{Feldman
  et~al.}{1994}]{Feldman:1993ky}
Feldman H.~A.,  Kaiser N.,   Peacock J.~A.,  1994, \mn@doi [Astrophys. J.]
  {10.1086/174036}, 426, 23

\bibitem[\protect\citeauthoryear{{Fixsen}}{{Fixsen}}{2009}]{2009ApJ...707..916F}
{Fixsen} D.~J.,  2009, \mn@doi [\apj] {10.1088/0004-637X/707/2/916}, \href
  {https://ui.adsabs.harvard.edu/abs/2009ApJ...707..916F} {707, 916}

\bibitem[\protect\citeauthoryear{Foreman-Mackey, Hogg, Lang  \&
  Goodman}{Foreman-Mackey et~al.}{2013}]{Foreman-Mackey:2012any}
Foreman-Mackey D.,  Hogg D.~W.,  Lang D.,   Goodman J.,  2013, \mn@doi [Publ.
  Astron. Soc. Pac.] {10.1086/670067}, 125, 306

\bibitem[\protect\citeauthoryear{Foreman, Obuljen  \& Simonovi\'c}{Foreman
  et~al.}{2024}]{Foreman:2024kzw}
Foreman S.,  Obuljen A.,   Simonovi\'c M.,  2024, \texttt{arXiv}, {}
  (\mn@eprint {arXiv} {2405.18559})

\bibitem[\protect\citeauthoryear{{Guo}, {Jones}, {Haynes}  \& {Fu}}{{Guo}
  et~al.}{2020}]{2020ApJ...894...92G}
{Guo} H.,  {Jones} M.~G.,  {Haynes} M.~P.,   {Fu} J.,  2020, \mn@doi [\apj]
  {10.3847/1538-4357/ab886f}, \href
  {https://ui.adsabs.harvard.edu/abs/2020ApJ...894...92G} {894, 92}

\bibitem[\protect\citeauthoryear{Harper \& Dickinson}{Harper \&
  Dickinson}{2018}]{Harper:2018ncl}
Harper S.,  Dickinson C.,  2018, \mn@doi [Mon. Not. Roy. Astron. Soc.]
  {10.1093/mnras/sty1495}, 479, 2024

\bibitem[\protect\citeauthoryear{{Haslam}, {Klein}, {Salter}, {Stoffel},
  {Wilson}, {Cleary}, {Cooke}  \& {Thomasson}}{{Haslam}
  et~al.}{1981}]{1981A&A...100..209H}
{Haslam} C.~G.~T.,  {Klein} U.,  {Salter} C.~J.,  {Stoffel} H.,  {Wilson}
  W.~E.,  {Cleary} M.~N.,  {Cooke} D.~J.,   {Thomasson} P.,  1981, \aap, \href
  {https://ui.adsabs.harvard.edu/abs/1981A&A...100..209H} {100, 209}

\bibitem[\protect\citeauthoryear{{Haslam}, {Salter}, {Stoffel}  \&
  {Wilson}}{{Haslam} et~al.}{1982}]{1982A&AS...47....1H}
{Haslam} C.~G.~T.,  {Salter} C.~J.,  {Stoffel} H.,   {Wilson} W.~E.,  1982,
  \aaps, \href {https://ui.adsabs.harvard.edu/abs/1982A&AS...47....1H} {47, 1}

\bibitem[\protect\citeauthoryear{Heymans et~al.}{Heymans
  et~al.}{2021}]{Heymans:2020gsg}
Heymans C.,  et~al., 2021, \mn@doi [Astron. Astrophys.]
  {10.1051/0004-6361/202039063}, 646, A140

\bibitem[\protect\citeauthoryear{{Hurley-Walker} et~al.,}{{Hurley-Walker}
  et~al.}{2022}]{2022PASA...39...35H}
{Hurley-Walker} N.,  et~al., 2022, \mn@doi [\pasa] {10.1017/pasa.2022.17},
  \href {https://ui.adsabs.harvard.edu/abs/2022PASA...39...35H} {39, e035}

\bibitem[\protect\citeauthoryear{Irfan \& Bull}{Irfan \&
  Bull}{2021}]{Irfan:2021bci}
Irfan M.~O.,  Bull P.,  2021, \mn@doi [Mon. Not. Roy. Astron. Soc.]
  {10.1093/mnras/stab2855}, 508, 3551

\bibitem[\protect\citeauthoryear{Irfan et~al.,}{Irfan
  et~al.}{2021}]{Irfan:2021xuk}
Irfan M.~O.,  et~al., 2021, \mn@doi [Mon. Not. Roy. Astron. Soc.]
  {10.1093/mnras/stab3346}, 509, 4923

\bibitem[\protect\citeauthoryear{{Jarvis} et~al.,}{{Jarvis}
  et~al.}{2016}]{2016mks..confE...6J}
{Jarvis} M.,  et~al., 2016, in MeerKAT Science: On the Pathway to the SKA. p.~6
  (\mn@eprint {arXiv} {1709.01901}), \mn@doi{10.22323/1.277.0006}

\bibitem[\protect\citeauthoryear{Jing}{Jing}{2005}]{Jing:2004fq}
Jing Y.~P.,  2005, \mn@doi [Astrophys. J.] {10.1086/427087}, 620, 559

\bibitem[\protect\citeauthoryear{{Jones}, {Haynes}, {Giovanelli}  \&
  {Moorman}}{{Jones} et~al.}{2018}]{2018MNRAS.477....2J}
{Jones} M.~G.,  {Haynes} M.~P.,  {Giovanelli} R.,   {Moorman} C.,  2018,
  \mn@doi [\mnras] {10.1093/mnras/sty521}, \href
  {https://ui.adsabs.harvard.edu/abs/2018MNRAS.477....2J} {477, 2}

\bibitem[\protect\citeauthoryear{Lian, Xu, Zhu  \& Hu}{Lian
  et~al.}{2020}]{Lian:2020jgd}
Lian X.,  Xu H.,  Zhu Z.,   Hu D.,  2020, \mn@doi [Mon. Not. Roy. Astron. Soc.]
  {10.1093/mnras/staa1179}, 496, 1232

\bibitem[\protect\citeauthoryear{{Liske} et~al.,}{{Liske}
  et~al.}{2015}]{LiskeGAMA15}
{Liske} J.,  et~al., 2015, \mn@doi [\mnras] {10.1093/mnras/stv1436}, \href
  {https://ui.adsabs.harvard.edu/abs/2015MNRAS.452.2087L} {452, 2087}

\bibitem[\protect\citeauthoryear{{Lujan Niemeyer} et~al.,}{{Lujan Niemeyer}
  et~al.}{2022a}]{2022ApJ...929...90L}
{Lujan Niemeyer} M.,  et~al., 2022a, \mn@doi [\apj] {10.3847/1538-4357/ac5cb8},
  \href {https://ui.adsabs.harvard.edu/abs/2022ApJ...929...90L} {929, 90}

\bibitem[\protect\citeauthoryear{{Lujan Niemeyer} et~al.,}{{Lujan Niemeyer}
  et~al.}{2022b}]{2022ApJ...934L..26L}
{Lujan Niemeyer} M.,  et~al., 2022b, \mn@doi [\apjl]
  {10.3847/2041-8213/ac82e5}, \href
  {https://ui.adsabs.harvard.edu/abs/2022ApJ...934L..26L} {934, L26}

\bibitem[\protect\citeauthoryear{{Maddox} et~al.,}{{Maddox}
  et~al.}{2021}]{2021A&A...646A..35M}
{Maddox} N.,  et~al., 2021, \mn@doi [\aap] {10.1051/0004-6361/202039655}, \href
  {https://ui.adsabs.harvard.edu/abs/2021A&A...646A..35M} {646, A35}

\bibitem[\protect\citeauthoryear{Masui et~al.}{Masui
  et~al.}{2013}]{Masui:2012zc}
Masui K.~W.,  et~al., 2013, \mn@doi [Astrophys. J. Lett.]
  {10.1088/2041-8205/763/1/L20}, 763, L20

\bibitem[\protect\citeauthoryear{Matshawule, Spinelli, Santos  \&
  Ngobese}{Matshawule et~al.}{2021}]{Matshawule:2020fjz}
Matshawule S.~D.,  Spinelli M.,  Santos M.~G.,   Ngobese S.,  2021, \mn@doi
  [Mon. Not. Roy. Astron. Soc.] {10.1093/mnras/stab1688}, 506, 5075

\bibitem[\protect\citeauthoryear{{Meyer}, {Meyer}, {Obreschkow}  \&
  {Staveley-Smith}}{{Meyer} et~al.}{2016}]{2016MNRAS.455.3136M}
{Meyer} S.~A.,  {Meyer} M.,  {Obreschkow} D.,   {Staveley-Smith} L.,  2016,
  \mn@doi [\mnras] {10.1093/mnras/stv2458}, \href
  {https://ui.adsabs.harvard.edu/abs/2016MNRAS.455.3136M} {455, 3136}

\bibitem[\protect\citeauthoryear{Moradinezhad~Dizgah, Nikakhtar, Keating  \&
  Castorina}{Moradinezhad~Dizgah et~al.}{2022}]{MoradinezhadDizgah:2021dei}
Moradinezhad~Dizgah A.,  Nikakhtar F.,  Keating G.~K.,   Castorina E.,  2022,
  \mn@doi [JCAP] {10.1088/1475-7516/2022/02/026}, 02, 026

\bibitem[\protect\citeauthoryear{{Mozdzen}, {Mahesh}, {Monsalve}, {Rogers}  \&
  {Bowman}}{{Mozdzen} et~al.}{2019}]{2019MNRAS.483.4411M}
{Mozdzen} T.~J.,  {Mahesh} N.,  {Monsalve} R.~A.,  {Rogers} A.~E.~E.,
  {Bowman} J.~D.,  2019, \mn@doi [\mnras] {10.1093/mnras/sty3410}, \href
  {https://ui.adsabs.harvard.edu/abs/2019MNRAS.483.4411M} {483, 4411}

\bibitem[\protect\citeauthoryear{Murray}{Murray}{2018}]{Murray2018}
Murray S.~G.,  2018, \mn@doi [Journal of Open Source Software]
  {10.21105/joss.00850}, 3, 850

\bibitem[\protect\citeauthoryear{{Nan} et~al.,}{{Nan}
  et~al.}{2011}]{2011IJMPD..20..989N}
{Nan} R.,  et~al., 2011, \mn@doi [International Journal of Modern Physics D]
  {10.1142/S0218271811019335}, \href
  {https://ui.adsabs.harvard.edu/abs/2011IJMPD..20..989N} {20, 989}

\bibitem[\protect\citeauthoryear{Newburgh et~al.}{Newburgh
  et~al.}{2016}]{Newburgh:2016mwi}
Newburgh L.~B.,  et~al., 2016, \mn@doi [Proc. SPIE Int. Soc. Opt. Eng.]
  {10.1117/12.2234286}, 9906, 99065X

\bibitem[\protect\citeauthoryear{Niemeyer, Bernal  \& Komatsu}{Niemeyer
  et~al.}{2023}]{Niemeyer:2023yeu}
Niemeyer M.~L.,  Bernal J.~L.,   Komatsu E.,  2023, \mn@doi [Astrophys. J.]
  {10.3847/1538-4357/acfef4}, 958, 4

\bibitem[\protect\citeauthoryear{Obuljen, Simonovi\'c, Schneider  \&
  Feldmann}{Obuljen et~al.}{2023}]{Obuljen:2022cjo}
Obuljen A.,  Simonovi\'c M.,  Schneider A.,   Feldmann R.,  2023, \mn@doi
  [Phys. Rev. D] {10.1103/PhysRevD.108.083528}, 108, 083528

\bibitem[\protect\citeauthoryear{Olivari, Remazeilles  \& Dickinson}{Olivari
  et~al.}{2016}]{Olivari:2015tka}
Olivari L.~C.,  Remazeilles M.,   Dickinson C.,  2016, \mn@doi [Mon. Not. Roy.
  Astron. Soc.] {10.1093/mnras/stv2884}, 456, 2749

\bibitem[\protect\citeauthoryear{{PUMA Collaboration} et~al.}{{PUMA
  Collaboration} et~al.}{2019}]{PUMA:2019jwd}
{PUMA Collaboration} et~al., 2019, \texttt{arXiv}, {} (\mn@eprint {arXiv}
  {1907.12559})

\bibitem[\protect\citeauthoryear{Pal et~al.}{Pal et~al.}{2022}]{Pal:2022xfm}
Pal S.,  et~al., 2022, \mn@doi [Mon. Not. Roy. Astron. Soc.]
  {10.1093/mnras/stac2419}, 516, 2851

\bibitem[\protect\citeauthoryear{{Pan} et~al.,}{{Pan}
  et~al.}{2023}]{2023MNRAS.525..256P}
{Pan} H.,  et~al., 2023, \mn@doi [\mnras] {10.1093/mnras/stad2343}, \href
  {https://ui.adsabs.harvard.edu/abs/2023MNRAS.525..256P} {525, 256}

\bibitem[\protect\citeauthoryear{Paul, Santos, Chen  \& Wolz}{Paul
  et~al.}{2023}]{Paul:2023yrr}
Paul S.,  Santos M.~G.,  Chen Z.,   Wolz L.,  2023, \texttt{arXiv}, {}
  (\mn@eprint {arXiv} {2301.11943})

\bibitem[\protect\citeauthoryear{{Ponomareva} et~al.,}{{Ponomareva}
  et~al.}{2021}]{2021MNRAS.508.1195P}
{Ponomareva} A.~A.,  et~al., 2021, \mn@doi [\mnras] {10.1093/mnras/stab2654},
  \href {https://ui.adsabs.harvard.edu/abs/2021MNRAS.508.1195P} {508, 1195}

\bibitem[\protect\citeauthoryear{{Ponomareva} et~al.,}{{Ponomareva}
  et~al.}{2023}]{2023MNRAS.522.5308P}
{Ponomareva} A.~A.,  et~al., 2023, \mn@doi [\mnras] {10.1093/mnras/stad1249},
  \href {https://ui.adsabs.harvard.edu/abs/2023MNRAS.522.5308P} {522, 5308}

\bibitem[\protect\citeauthoryear{{Remazeilles}, {Dickinson}, {Banday},
  {Bigot-Sazy}  \& {Ghosh}}{{Remazeilles} et~al.}{2015}]{2015MNRAS.451.4311R}
{Remazeilles} M.,  {Dickinson} C.,  {Banday} A.~J.,  {Bigot-Sazy} M.~A.,
  {Ghosh} T.,  2015, \mn@doi [\mnras] {10.1093/mnras/stv1274}, \href
  {https://ui.adsabs.harvard.edu/abs/2015MNRAS.451.4311R} {451, 4311}

\bibitem[\protect\citeauthoryear{{Rhee} et~al.,}{{Rhee}
  et~al.}{2023}]{2023MNRAS.518.4646R}
{Rhee} J.,  et~al., 2023, \mn@doi [\mnras] {10.1093/mnras/stac3065}, \href
  {https://ui.adsabs.harvard.edu/abs/2023MNRAS.518.4646R} {518, 4646}

\bibitem[\protect\citeauthoryear{{SKA Cosmology SWG} et~al.}{{SKA Cosmology
  SWG} et~al.}{2020}]{Bacon:2018dui}
{SKA Cosmology SWG} et~al., 2020, \mn@doi [Publ. Astron. Soc. Austral.]
  {10.1017/pasa.2019.51}, 37, e007

\bibitem[\protect\citeauthoryear{Santos, Cooray  \& Knox}{Santos
  et~al.}{2005}]{Santos:2004ju}
Santos M.~G.,  Cooray A.,   Knox L.,  2005, \mn@doi [Astrophys. J.]
  {10.1086/429857}, 625, 575

\bibitem[\protect\citeauthoryear{Santos et~al.}{Santos
  et~al.}{2017}]{MeerKLASS:2017vgf}
Santos M.~G.,  et~al., 2017, in {MeerKAT Science}: {On the Pathway to the SKA}.
   (\mn@eprint {arXiv} {1709.06099})

\bibitem[\protect\citeauthoryear{Sato-Polito, Kokron  \& Bernal}{Sato-Polito
  et~al.}{2023}]{Sato-Polito:2022wiq}
Sato-Polito G.,  Kokron N.,   Bernal J.~L.,  2023, \mn@doi [Mon. Not. Roy.
  Astron. Soc.] {10.1093/mnras/stad2498}, 526, 5883

\bibitem[\protect\citeauthoryear{Shaw, Sigurdson, Sitwell, Stebbins  \&
  Pen}{Shaw et~al.}{2015}]{Shaw:2014khi}
Shaw J.~R.,  Sigurdson K.,  Sitwell M.,  Stebbins A.,   Pen U.-L.,  2015,
  \mn@doi [Phys. Rev. D] {10.1103/PhysRevD.91.083514}, 91, 083514

\bibitem[\protect\citeauthoryear{Shin, Kim, Pichon, Jeong  \& Park}{Shin
  et~al.}{2017}]{Shin:2017cwu}
Shin J.,  Kim J.,  Pichon C.,  Jeong D.,   Park C.,  2017, \mn@doi [Astrophys.
  J.] {10.3847/1538-4357/aa74b9}, 843, 73

\bibitem[\protect\citeauthoryear{{Sinigaglia} et~al.,}{{Sinigaglia}
  et~al.}{2022}]{2022ApJ...935L..13S}
{Sinigaglia} F.,  et~al., 2022, \mn@doi [\apjl] {10.3847/2041-8213/ac85ae},
  \href {https://ui.adsabs.harvard.edu/abs/2022ApJ...935L..13S} {935, L13}

\bibitem[\protect\citeauthoryear{{Sinigaglia} et~al.,}{{Sinigaglia}
  et~al.}{2024}]{2024MNRAS.529.4192S}
{Sinigaglia} F.,  et~al., 2024, \mn@doi [\mnras] {10.1093/mnras/stae713}, \href
  {https://ui.adsabs.harvard.edu/abs/2024MNRAS.529.4192S} {529, 4192}

\bibitem[\protect\citeauthoryear{Soares, Cunnington, Pourtsidou  \&
  Blake}{Soares et~al.}{2021}]{Soares:2020zaq}
Soares P.~S.,  Cunnington S.,  Pourtsidou A.,   Blake C.,  2021, \mn@doi [Mon.
  Not. Roy. Astron. Soc.] {10.1093/mnras/stab027}, 502, 2549

\bibitem[\protect\citeauthoryear{Soares, Watkinson, Cunnington  \&
  Pourtsidou}{Soares et~al.}{2022}]{Soares:2021ohm}
Soares P.~S.,  Watkinson C.~A.,  Cunnington S.,   Pourtsidou A.,  2022, \mn@doi
  [Mon. Not. Roy. Astron. Soc.] {10.1093/mnras/stab2594}, 510, 5872

\bibitem[\protect\citeauthoryear{{Spinelli}, {Bernardi}, {Garsden},
  {Greenhill}, {Fialkov}, {Dowell}  \& {Price}}{{Spinelli}
  et~al.}{2021}]{2021MNRAS.505.1575S}
{Spinelli} M.,  {Bernardi} G.,  {Garsden} H.,  {Greenhill} L.~J.,  {Fialkov}
  A.,  {Dowell} J.,   {Price} D.~C.,  2021, \mn@doi [\mnras]
  {10.1093/mnras/stab1363}, \href
  {https://ui.adsabs.harvard.edu/abs/2021MNRAS.505.1575S} {505, 1575}

\bibitem[\protect\citeauthoryear{{Spinelli}, {Carucci}, {Cunnington}, {Harper},
  {Irfan}, {Fonseca}, {Pourtsidou}  \& {Wolz}}{{Spinelli}
  et~al.}{2022}]{2022MNRAS.509.2048S}
{Spinelli} M.,  {Carucci} I.~P.,  {Cunnington} S.,  {Harper} S.~E.,  {Irfan}
  M.~O.,  {Fonseca} J.,  {Pourtsidou} A.,   {Wolz} L.,  2022, \mn@doi [\mnras]
  {10.1093/mnras/stab3064}, \href
  {https://ui.adsabs.harvard.edu/abs/2022MNRAS.509.2048S} {509, 2048}

\bibitem[\protect\citeauthoryear{Switzer et~al.}{Switzer
  et~al.}{2013}]{Switzer:2013ewa}
Switzer E.~R.,  et~al., 2013, \mn@doi [Mon. Not. Roy. Astron. Soc.]
  {10.1093/mnrasl/slt074}, 434, L46

\bibitem[\protect\citeauthoryear{Switzer, Chang, Masui, Pen  \& Voytek}{Switzer
  et~al.}{2015}]{Switzer:2015ria}
Switzer E.~R.,  Chang T.-C.,  Masui K.~W.,  Pen U.-L.,   Voytek T.~C.,  2015,
  \mn@doi [Astrophys. J.] {10.1088/0004-637X/815/1/51}, 815, 51

\bibitem[\protect\citeauthoryear{Tauscher, Rapetti  \& Burns}{Tauscher
  et~al.}{2018}]{Tauscher:2018uxi}
Tauscher K.,  Rapetti D.,   Burns J.~O.,  2018, \mn@doi [JCAP]
  {10.1088/1475-7516/2018/12/015}, 12, 015

\bibitem[\protect\citeauthoryear{{Thorne}, {Dunkley}, {Alonso}  \&
  {N{\ae}ss}}{{Thorne} et~al.}{2017}]{2017MNRAS.469.2821T}
{Thorne} B.,  {Dunkley} J.,  {Alonso} D.,   {N{\ae}ss} S.,  2017, \mn@doi
  [\mnras] {10.1093/mnras/stx949}, \href
  {https://ui.adsabs.harvard.edu/abs/2017MNRAS.469.2821T} {469, 2821}

\bibitem[\protect\citeauthoryear{{Tramonte} \& {Ma}}{{Tramonte} \&
  {Ma}}{2020}]{2020MNRAS.498.5916T}
{Tramonte} D.,  {Ma} Y.-Z.,  2020, \mn@doi [\mnras] {10.1093/mnras/staa2727},
  \href {https://ui.adsabs.harvard.edu/abs/2020MNRAS.498.5916T} {498, 5916}

\bibitem[\protect\citeauthoryear{{Tramonte}, {Ma}, {Li}  \&
  {Staveley-Smith}}{{Tramonte} et~al.}{2019}]{2019MNRAS.489..385T}
{Tramonte} D.,  {Ma} Y.-Z.,  {Li} Y.-C.,   {Staveley-Smith} L.,  2019, \mn@doi
  [\mnras] {10.1093/mnras/stz2146}, \href
  {https://ui.adsabs.harvard.edu/abs/2019MNRAS.489..385T} {489, 385}

\bibitem[\protect\citeauthoryear{{Tudorache} et~al.,}{{Tudorache}
  et~al.}{2022}]{2022MNRAS.513.2168T}
{Tudorache} M.~N.,  et~al., 2022, \mn@doi [\mnras] {10.1093/mnras/stac996},
  \href {https://ui.adsabs.harvard.edu/abs/2022MNRAS.513.2168T} {513, 2168}

\bibitem[\protect\citeauthoryear{Vanderlinde et~al.}{Vanderlinde
  et~al.}{2019}]{Vanderlinde:2019tjt}
Vanderlinde K.,  et~al., 2019, \texttt{arXiv}, {} (\mn@eprint {arXiv}
  {1911.01777})

\bibitem[\protect\citeauthoryear{Villaescusa-Navarro
  et~al.}{Villaescusa-Navarro et~al.}{2018}]{Villaescusa-Navarro:2018vsg}
Villaescusa-Navarro F.,  et~al., 2018, \mn@doi [Astrophys. J.]
  {10.3847/1538-4357/aadba0}, 866, 135

\bibitem[\protect\citeauthoryear{Wang et~al.}{Wang et~al.}{2021}]{Wang:2020lkn}
Wang J.,  et~al., 2021, \mn@doi [Mon. Not. Roy. Astron. Soc.]
  {10.1093/mnras/stab1365}, 505, 3698

\bibitem[\protect\citeauthoryear{{Westmeier}, {Jurek}, {Obreschkow},
  {Koribalski}  \& {Staveley-Smith}}{{Westmeier}
  et~al.}{2014}]{2014MNRAS.438.1176W}
{Westmeier} T.,  {Jurek} R.,  {Obreschkow} D.,  {Koribalski} B.~S.,
  {Staveley-Smith} L.,  2014, \mn@doi [\mnras] {10.1093/mnras/stt2266}, \href
  {https://ui.adsabs.harvard.edu/abs/2014MNRAS.438.1176W} {438, 1176}

\bibitem[\protect\citeauthoryear{Wilensky, Brown  \& Hazelton}{Wilensky
  et~al.}{2023}]{Wilensky:2022sfh}
Wilensky M.~J.,  Brown J.,   Hazelton B.~J.,  2023, \mn@doi [Mon. Not. Roy.
  Astron. Soc.] {10.1093/mnras/stad863}, 521, 5191

\bibitem[\protect\citeauthoryear{Wolz, Abdalla, Blake, Shaw, Chapman  \&
  Rawlings}{Wolz et~al.}{2014}]{Wolz:2013wna}
Wolz L.,  Abdalla F.~B.,  Blake C.,  Shaw J.~R.,  Chapman E.,   Rawlings S.,
  2014, \mn@doi [Mon. Not. Roy. Astron. Soc.] {10.1093/mnras/stu792}, 441, 3271

\bibitem[\protect\citeauthoryear{Wolz et~al.}{Wolz et~al.}{2017}]{Wolz:2015lwa}
Wolz L.,  et~al., 2017, \mn@doi [Mon. Not. Roy. Astron. Soc.]
  {10.1093/mnras/stw2556}, 464, 4938

\bibitem[\protect\citeauthoryear{Wolz et~al.}{Wolz
  et~al.}{2022}]{eBOSS:2021ebm}
Wolz L.,  et~al., 2022, \mn@doi [Mon. Not. Roy. Astron. Soc.]
  {10.1093/mnras/stab3621}, 510, 3495

\bibitem[\protect\citeauthoryear{Wyithe, Loeb  \& Geil}{Wyithe
  et~al.}{2008}]{Wyithe:2007rq}
Wyithe S.,  Loeb A.,   Geil P.,  2008, \mn@doi [MNRAS]
  {10.1111/j.1365-2966.2007.12631.x}, 383, 1195

\bibitem[\protect\citeauthoryear{{Zwaan}, {van Dokkum}  \& {Verheijen}}{{Zwaan}
  et~al.}{2001}]{2001Sci...293.1800Z}
{Zwaan} M.~A.,  {van Dokkum} P.~G.,   {Verheijen} M.~A.~W.,  2001, \mn@doi
  [Science] {10.1126/science.1063034}, \href
  {https://ui.adsabs.harvard.edu/abs/2001Sci...293.1800Z} {293, 1800}

\bibitem[\protect\citeauthoryear{{eBOSS Collaboration}}{{eBOSS
  Collaboration}}{2021}]{eBOSS:2020yzd}
{eBOSS Collaboration} 2021, \mn@doi [Phys. Rev. D]
  {10.1103/PhysRevD.103.083533}, 103, 083533

\makeatother
\end{thebibliography}




\appendix

\section{Observation details}

\autoref{tab:block} provides the details for the MeerKLASS L-band deep field observations, for all 41 attempted observation blocks.

\begin{table*}
\centering
 \caption{Information for all 41 observation blocks i.e.\ each of the scans for the MeerKLASS L-band deep field patch. The last column marks if the block was selected for the final dataset. An absent tick means the block was flagged, primarily due to RFI contamination.} 
 \label{tab:block}
 \begin{tabular}{ccccccccc}
  \hline
  \hline
Block ID  & Short name & Observation start time & Sunset & az range & el & Calibrator& Motion of field & Calibrated \\
(Timestamp) &(in this paper) &(UTC time)&(UTC time)&($^{\circ})$&($^{\circ})$&&&\\
\hline
obs1630519596 & {\tt obs210901} & 2021-09-01 18:06:53 & 09-01 16:14:50 & [98.0, 111.9] & 42.3 & PKS 1934-638 & Rising & \checkmark\\ 
obs1631379874 & {\tt obs210911a} & 2021-09-11 17:04:51 & 09-11 16:20:28 & [100.1, 113.2] & 37.8 & PKS 1934-638 & Rising &\\ 
obs1631387336 & {\tt obs210911b} & 2021-09-11 19:09:13 & 09-11 16:20:28 & [86.6, 108.8] & 63.0 & PKS 1934-638 & Rising & \checkmark\\ 
obs1631552188 & {\tt obs210913a} & 2021-09-13 16:56:49 & 09-13 16:21:35 & [100.1, 113.2] & 37.7 & PKS 1934-638 & Rising &\\ 
obs1631559762 & {\tt obs210913b} & 2021-09-13 19:02:58 & 09-13 16:21:35 & [86.4, 108.8] & 63.3 & PKS 1934-638 & Rising & \checkmark\\ 
obs1631659886 & {\tt obs210914} & 2021-09-14 22:52:47 & 09-14 16:22:08 & [-108.8, -85.8] & 64.1 & Pictor A & Setting &\\ 
obs1631667564 & {\tt obs210915a} & 2021-09-15 00:59:40 & 09-14 16:22:08 & [-113.1, -99.9] & 38.1 & Pictor A & Setting & \checkmark\\ 
obs1631724508 & {\tt obs210915b} & 2021-09-15 16:48:49 & 09-15 16:22:42 & [100.3, 113.4] & 37.2 & PKS 1934-638 & Rising& \\ 
obs1631732038 & {\tt obs210915c} & 2021-09-15 18:54:17 & 09-15 16:22:42 & [86.9, 108.7] & 62.6 & PKS 1934-638 & Rising & \checkmark\\ 
obs1631810671 & {\tt obs210916a} & 2021-09-16 16:44:51 & 09-16 16:23:16 & [100.2, 113.3] & 37.5 & PKS 1934-638 & Rising & \checkmark\\ 
obs1631818149 & {\tt obs210916b} & 2021-09-16 18:49:27 & 09-16 16:23:16 & [86.8, 108.8] & 62.7 & PKS 1934-638 & Rising & \checkmark\\ 
obs1631982988 & {\tt obs210918a} & 2021-09-18 16:36:47 & 09-18 16:24:23 & [100.2, 113.3] & 37.4 & PKS 1934-638 & Rising & \checkmark\\ 
obs1631990463 & {\tt obs210918b} & 2021-09-18 18:41:19 & 09-18 16:24:23 & [86.9, 108.8] & 62.6 & PKS 1934-638 & Rising & \checkmark\\ 
obs1632069690 & {\tt obs210919a} & 2021-09-19 16:41:48 & 09-19 16:24:57 & [99.5, 112.9] & 38.9 & PKS 1934-638 & Rising & \checkmark\\ 
obs1632077222 & {\tt obs210919b} & 2021-09-19 18:47:18 & 09-19 16:24:57 & [85.6, 108.8] & 64.4 & PKS 1934-638 & Rising & \checkmark\\ 
obs1632184922 & {\tt obs210921} & 2021-09-21 00:44:34 & 09-20 16:25:31 & [-113.7, -100.7] & 36.3 & Pictor A & Setting & \checkmark\\ 
obs1632505883 & {\tt obs210924} & 2021-09-24 17:51:44 & 09-24 16:27:48 & [90.4, 109.0] & 57.0 & PKS 1934-638 & Rising &\\ 
obs1632760885 & {\tt obs210927} & 2021-09-27 16:41:43 & 09-27 16:29:33 & [96.6, 111.1] & 45.2 & PKS 1934-638 & Rising & \checkmark\\ 
obs1633365980 & {\tt obs211004} & 2021-10-04 16:46:39 & 10-04 16:33:46 & [93.1, 109.7] & 52.1 & PKS 1934-638 & Rising & \checkmark\\ 
obs1633970780 & {\tt obs211011} & 2021-10-11 16:46:38 & 10-11 16:38:15 & [89.9, 108.9] & 58.0 & PKS 1934-638 & Rising &\\ 
obs1634252028 & {\tt obs211014} & 2021-10-14 22:55:17 & 10-14 16:40:15 & [-113.0, -99.8] & 38.5 & Pictor A & Setting & \checkmark\\ 
obs1634402485 & {\tt obs211016} & 2021-10-16 16:41:45 & 10-16 16:41:37 & [88.0, 108.7] & 61.0 & PKS 1934-638 & Rising & \checkmark\\ 
obs1634748682 & {\tt obs211020} & 2021-10-20 16:51:50 & 10-20 16:44:25 & [84.1, 109.1] & 66.2 & PKS 1934-638 & Rising &\\ 
obs1634835083 & {\tt obs211021} & 2021-10-21 16:51:43 & 10-21 16:45:08 & [83.4, 109.2] & 67.0 & PKS 1934-638 & Rising &\\ 
obs1637346562 & {\tt obs211119a} & 2021-11-19 18:31:59 & 11-19 17:08:34 & [-109.0, -84.6] & 65.7 & Pictor A & Setting & \checkmark\\ 
obs1637354605 & {\tt obs211119b} & 2021-11-19 20:44:44 & 11-19 17:08:34 & [-113.4, -100.4] & 37.2 & Pictor A & Setting & \checkmark\\ 
obs1637691677 & {\tt obs211123a} & 2021-11-23 18:22:58 & 11-23 17:11:59 & [-108.8, -86.4] & 63.2 & Pictor A & Setting & \checkmark\\ 
obs1637699408 & {\tt obs211123b} & 2021-11-23 20:31:33 & 11-23 17:11:59 & [-113.8, -101.0] & 35.9 & Pictor A & Setting & \checkmark\\ 
obs1638130295 & {\tt obs211128} & 2021-11-28 20:13:37 & 11-28 17:16:12 & [-113.9, -101.1] & 35.5 & Pictor A & Setting & \checkmark\\ 
obs1638294319 & {\tt obs211130a} & 2021-11-30 17:47:13 & 11-30 17:17:51 & [-108.9, -84.9] & 65.3 & Pictor A & Setting & \checkmark\\ 
obs1638301944 & {\tt obs211130b} & 2021-11-30 19:53:44 & 11-30 17:17:51 & [-113.1, -100.0] & 38.1 & Pictor A & Setting & \checkmark\\ 
obs1638386189 & {\tt obs211201} & 2021-12-01 19:18:31 & 12-01 17:18:39 & [-111.2, -96.8] & 44.7 & Pictor A & Setting & \checkmark\\ 
obs1638639082 & {\tt obs211204a} & 2021-12-04 17:33:15 & 12-04 17:21:03 & [-108.9, -85.4] & 64.6 & Pictor A & Setting& \checkmark \\ 
obs1638647186 & {\tt obs211204b} & 2021-12-04 19:47:48 & 12-04 17:21:03 & [-113.9, -101.1] & 35.7 & Pictor A & Setting &\\ 
obs1638898468 & {\tt obs211207} & 2021-12-07 17:36:20 & 12-07 17:23:20 & [-108.7, -87.7] & 61.4 & Pictor A & Setting &\\ 
obs1639157507 & {\tt obs211210} & 2021-12-10 17:33:41 & 12-10 17:25:31 & [-108.8, -88.8] & 59.7 & Pictor A & Setting & \checkmark\\ 
obs1639331184 & {\tt obs211212} & 2021-12-12 17:48:20 & 12-12 17:26:53 & [-109.3, -91.8] & 54.6 & Pictor A & Setting &\\ 
obs1639935088 & {\tt obs211219} & 2021-12-19 17:34:00 & 12-19 17:31:08 & [-109.5, -92.8] & 52.7 & Pictor A & Setting &\\ 
obs1640540184 & {\tt obs211226} & 2021-12-26 17:39:00 & 12-26 17:34:21 & [-111.0, -96.3] & 45.8 & Pictor A & Setting &\\ 
obs1640712986 & {\tt obs211228} & 2021-12-28 17:38:01 & 12-28 17:35:03 & [-111.6, -97.5] & 43.2 & Pictor A & Setting &\\ 
obs1640799689 & {\tt obs211229} & 2021-12-29 17:43:29 & 12-29 17:35:22 & [-112.2, -98.5] & 41.2 & Pictor A & Setting & \checkmark\\ 

\hline
\end{tabular}
\end{table*}

\section{PCA foreground cleaning}

\subsection{Impact of reconvolution}\label{sec:ResmoothingAppendix}

As discussed in the main text, the reconvolution of the maps before foreground cleaning (outlined in \secref{subsec:reconvolution}), slightly improves the foreground clean performance. This is not necessarily just because it homogenises the effective resolution of the beam, as is the primary aim of this technique. The smoothing will also suppress systematics in the data. Residual contributions, for example from RFI, will distort the true amplitude of the sky map. If these contributions vary with frequency, which will be the case for most systematics, then this will create small perturbations to the spectra along each pixel. Despite smoothing the maps in the \textit{angular} direction only, this will still suppress perturbations along the line of sight, hence providing slightly smoother spectra, which may be linked to the improved foreground clean. 

\begin{figure}
    \centering
    \includegraphics[width=1\linewidth]{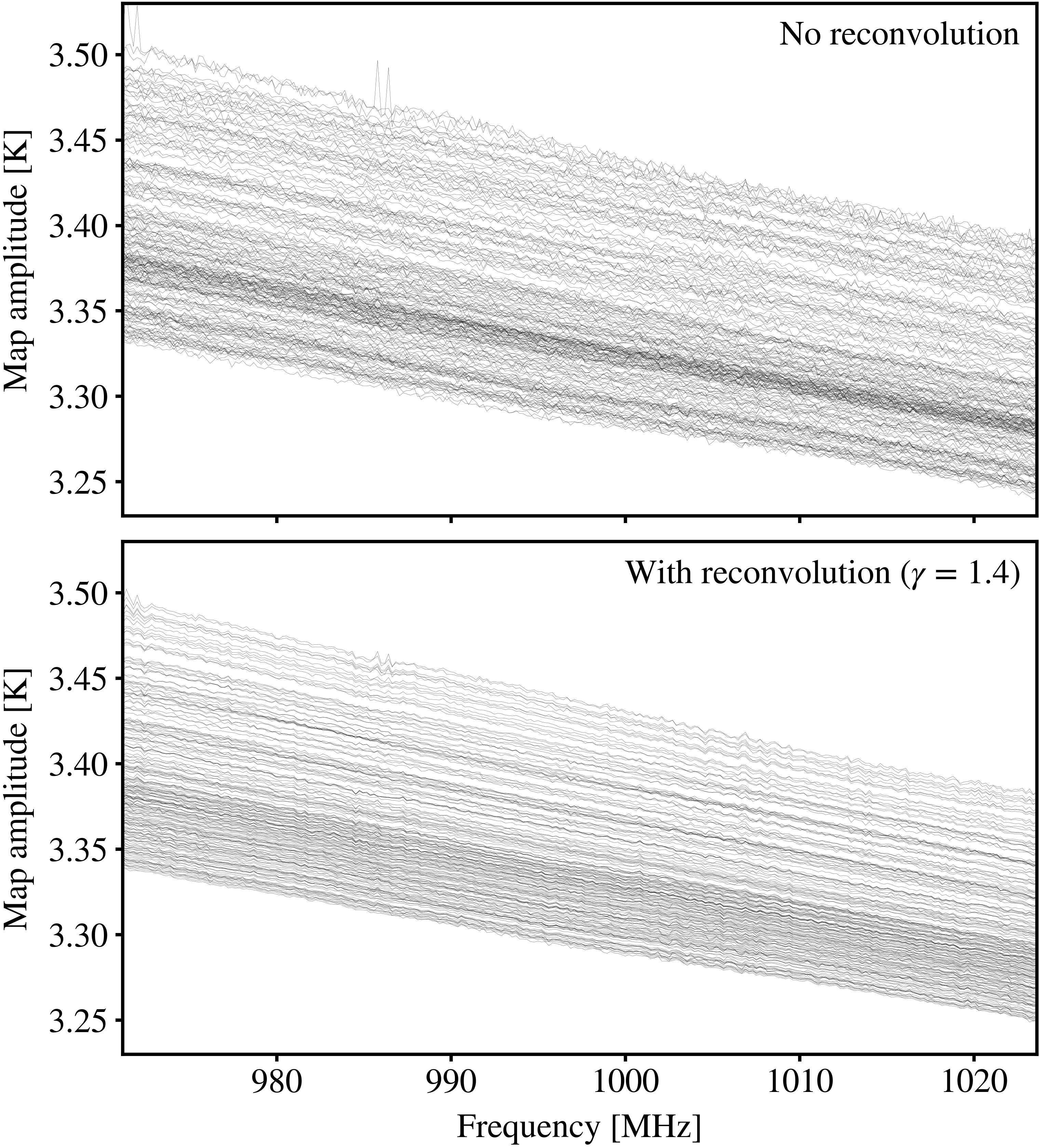}
    \caption{The impact of reconvolution on the smoothness of the sky map spectra. Each plotted line is the amplitude along frequency, which, for these pre-foreground clean sky maps, should trace the dominant synchrotron. The top panel is with no reconvolution. The bottom panel shows the smoothing effect from the reconvolution.}
    \label{fig:resmoothing_maps}
\end{figure}

We demonstrate this in \autoref{fig:resmoothing_maps} which plots the amplitudes of each pixel along frequency, as previously shown in \autoref{fig:LoS_spectra} (here we also trim the edges of the map, as outlined in \secref{sec:maptrim}). The top panel shows the spectra without reconvolution. A clear improvement in the smoothness of the spectra is evident when reconvolution is used (bottom panel). 

Future work is planned where a more realistic MeerKAT beam will be implemented into our modelling. In this work, we will also investigate this technique of reconvolution and aim to conclude if the smoothing to a common resolution is the source of any improvement or if it is just entirely from the suppression of systematics, reducing the perturbations they cause in the spectra which aids the PCA process.

\subsection{Weighted frequency covariance estimation}\label{sec:FGweights}

Here we explore the small impact that opting to use weights in the covariance estimation has on the PCA cleaning process. The weights used are constant along the frequency direction, so not to add structure to the modes and diminish the PCA performance. Furthermore, after trimming the map edges, the time stamps have a more uniform coverage (see \autoref{fig:ResmoothMap_and_weights}). This means the weights do not vary considerably across our whole field, so we would expect only minor changes by including them. 

\begin{figure}
    \centering
    \includegraphics[width=1\linewidth]{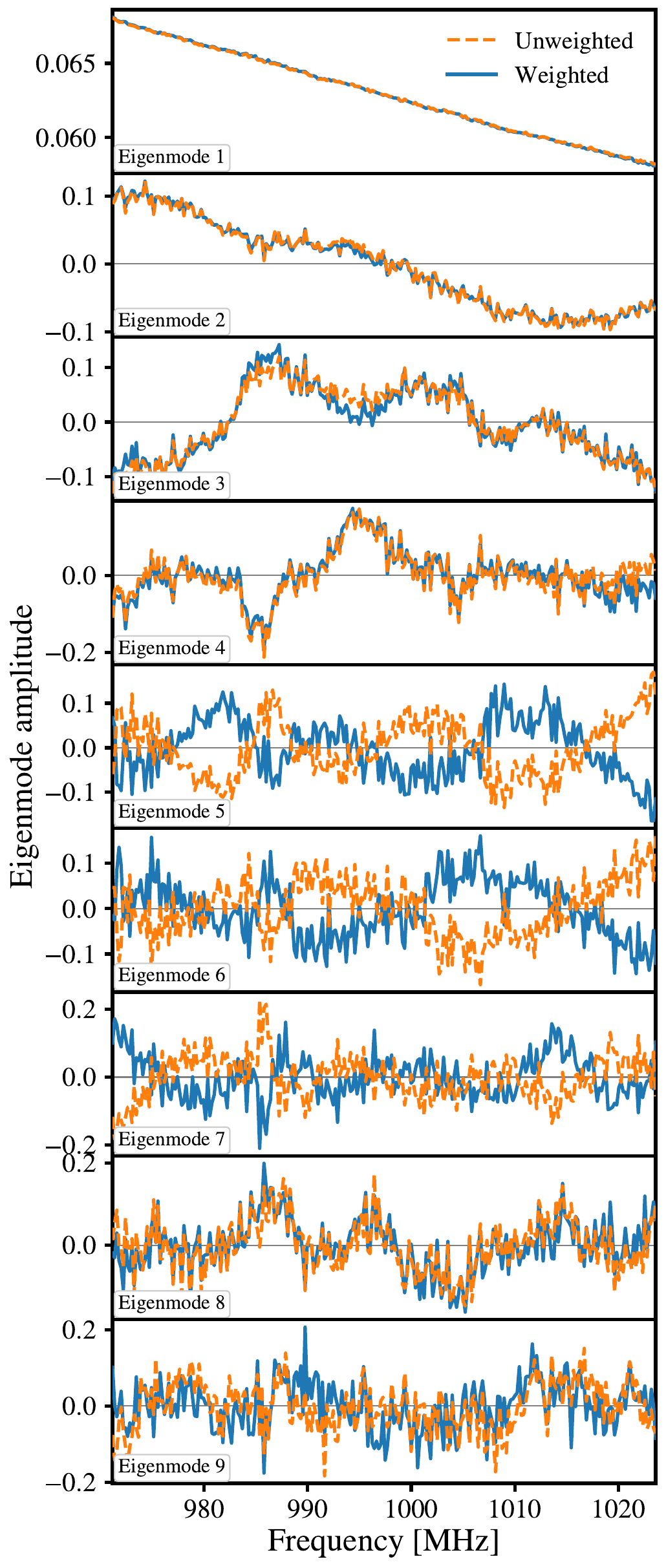}
    \caption{The first 9 eigenvectors (descending order of \textit{dominance} from top to bottom) extracted from the frequency covariance of the MeerKLASS deep field intensity maps. Eigenmode 1 shows the overwhelming synchrotron signal, a smooth slope with a spectral index power law. The different lines show the contrast between an unweighted and weighted PCA clean.}
    \label{fig:eigenmodes}
\end{figure}

\autoref{fig:eigenmodes} shows the first 9 eigenmodes from $\textbf{\textsf{U}}$, extracted from the diagonalisation of the covariance matrix computed using \autoref{eq:weightedcov}. We show the difference between the weighted case (blue solid) and the unweighted (orange dashed), equivalent to omitting the $\textbf{\textsf{w}}$ factor in the covariance calculation. As expected, the differences are marginal, especially when considering that in some modes the discrepancy is mainly an arbitrary change of sign. However, some small differences are still noticeable, particularly for the higher eigenmodes where noise will become more of a component which is effectively what we are weighting for. 

We found little evidence of improved performance in the final results whether weights were used or not in the foreground clean. The difference between the cleaned maps was at the sub-$1\mu$K level, and testing the process on the mocks showed no clear performance gain. However, we included them in our final results for completeness. Furthermore, the weighted formalism may be useful in future work focussing on \hi\ auto-correlations, where cross-correlating sub-sets of maps from different dishes or time blocks will be pursued. Here, different map weights can be used in a combined cross-covariance computation; then asymmetric singular vectors can be extracted from this, as in the Singular Value Decomposition (SVD) approach of \citet{Switzer:2013ewa}.

\begin{figure}
    \centering
    \includegraphics[width=1\linewidth]{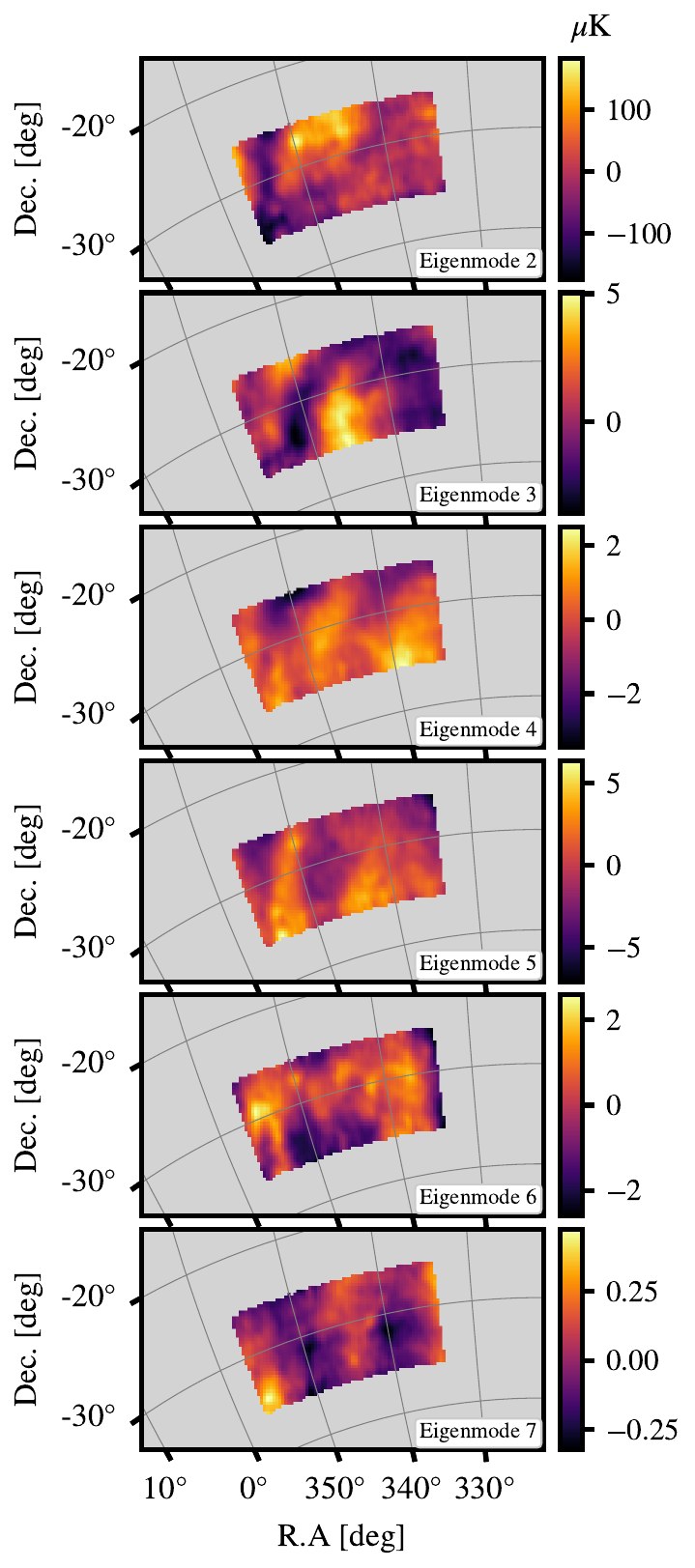}
    \caption{The first seven weighted eigenmodes from \autoref{fig:eigenmodes}, projected onto the map space, thus showing the map content removed in each PCA mode in the foreground cleaning. We do not show Eigenmode 1 as this is dominated by the synchrotron and is indistinguishable from the sky map of \autoref{fig:FGmap_and_counts}. Each map is averaged along frequency.}
    \label{fig:ProjectedModeMaps}
\end{figure}

To accompany the eigenmode plot of \autoref{fig:eigenmodes}, we also show the projected map from the first 7 eigenmodes in \autoref{fig:ProjectedModeMaps} for the weighted PCA case. We omit the first eigenmode since it is indistinguishable from the synchrotron sky maps we have shown previously. These show at map level, each component that is being removed in the PCA clean. For example, a $N_\text{fg}\txteq{=}7$ clean would remove all these components. No obvious systematic structure is apparent which could otherwise be targetted for removal by other means, rather than relying on an aggressive PCA clean. However, as we discussed in \secref{sec:zebra}, when cleans are performed individually on smaller sub-sets of time blocks, constant declination stripes become apparent in some of the components which we will aim to correct for in calibration in future work.

\subsection{Power index of the transfer function for signal resconstruction in auto- and cross-correlations}\label{sec:TF_signalloss_autoVcross}

The transfer function defined in \autoref{eq:TF_HI} is applied to the measured power spectrum to reconstruct the loss of information due to foreground cleaning. A priori, intuition would lead to applying the transfer function as defined for the cross-power spectrum of \hi\ and galaxies, and the \textit{square} of the transfer function for the \hi\ auto-power spectrum, as signal loss occurs \textit{twice} in that calculation~\citep{Switzer:2013ewa}. However, \citet{Cunnington:2023jpq} found empirically that squaring the transfer function over-corrected the signal loss and that the transfer function should be applied equally to cross- and auto-power spectra for correct modelling. In this appendix, we provide simplified analytic arguments to support such a claim. 

Let us assume the idealised scenario where we can decompose the cosmological fluctuations $\delta T_\hi(\pmb{k})\equiv r(\pmb{k})+l(\pmb{k})$, where $r$ and $l$ correspond to the residual and loss modes of the cosmological signal, respectively. Under this separation, the transfer function can be understood as
\begin{equation}
    \pazocal{T}_\hi(\pmb{k}) = \frac{\left\langle r_{\rm m}(\pmb{k})\left[r_{\rm m}(\pmb{k}')+l_{\rm m}(\pmb{k}')\right]^*\right\rangle}{\left\langle \left[r_{\rm m}(\pmb{k})+l_{\rm m}(\pmb{k})\right]\left[r_{\rm m}(\pmb{k}')+l_{\rm m}(\pmb{k}')\right]^*\right\rangle}\,,
\end{equation}
where the subscript `m' refers to quantities obtained from mock injection. 
Due to the component separation process in PCA,
$r$ and $l$ are a combination of independent components, and are therefore uncorrelated. The power spectrum of their sum is
\begin{equation}
\begin{split}
    \left\langle \left[r_{\rm m}(\pmb{k})+l_{\rm m}(\pmb{k})\vphantom{^*}\right]\right. & \left. \left[r_{\rm m}(\pmb{k}')+l_{\rm m}(\pmb{k}')\right]^*\right\rangle = \\ = \, & \frac{1}{(2\pi)^3}\delta_D(\pmb{k}-\pmb{k}')\left[P_{\rm m}^{(r)}(k)+P_{\rm m}^{(l)}(k)\right] \\= \, & \frac{1}{(2\pi)^3}\delta_D(\pmb{k}-\pmb{k}')P_{\rm m}^{\hi}(k)\,.
\end{split}
\end{equation}
Note that the equation above applies equally to the actual observed cosmological signal. Therefore, the transfer function can be now understood as
\begin{equation}
    \pazocal{T}_\hi(\pmb{k}) \sim \frac{P_{\rm m}^{(r)}(k)}{P_{\rm m}^{(r)}(k)+P_{\rm m}^{(l)}(k)}\,.
\end{equation}
Let us now consider the application of this conceptual transfer function to the power spectra measured from the clean maps. In this case, we have
\begin{equation}
    \frac{\left\langle r(\pmb{k})\,r^*(\pmb{k}')\right\rangle}{\pazocal{T}_\hi(\pmb{k})} \sim P^{(r)}(k)\left[\frac{ P_{\rm m}^{(r)}(k) + P_{\rm m}^{(l)}(k)}{P_{\rm m}^{(r)}(k)}\right]
\end{equation}
for the auto-power spectrum, and
\begin{equation}
    \frac{\langle r(\pmb{k})g^*(\pmb{k}')\rangle}{\pazocal{T}_\hi(\pmb{k})} \sim P^{(rg)}(k)\left[\frac{P_{\rm m}^{(r)}(k) + P_{\rm m}^{(l)}(k)}{P_{\rm m}^{(r)}(k)}\right]
\end{equation}
for the cross-power spectrum. Assuming $P^{(r)}\txteq{\sim}P_\text{m}^{(r)}$, it is clear that a single power of the transfer function (i.e.\ $\pazocal{T}^1$) correctly reconstructs the auto-power to the desired $P^{(r)}\txteq{+}P^{(l)}$ level. For the cross-power spectrum, let us simplify the situation and assume linear theory, for which $\delta_\text{g} \txteq{\propto} \delta T_\hi\txteq{\sim} r\txteq{+}l$. Then, $P^{(rg)}\txteq{\propto} P^{(r)}$; if we denote the proportionality factor generically as $\pazocal{A}$, we have 
\begin{equation}
    \frac{\langle r(\pmb{k})g^*(\pmb{k}')\rangle}{\pazocal{T}_\hi(\pmb{k})} \sim \pazocal{A}P^{(r)}(k)\left[1+\frac{P_{\rm m}^{(l)}(k)}{P_{\rm m}^{(r)}(k)}\right]\,, 
\end{equation}
which similarly shows that a single power of the transfer function must also be applied to the cross-power spectrum.

\section{Parameter inference from the cross-power}\label{sec:ParamConstraints}

We have not dedicated a lot of attention to the modelling of the power spectra in this work. This is mostly due to the large errors in the cross-power due to the small sample overlap and the small additive bias still present in the auto-\hi\ power spectrum. This makes high dimensional models unnecessary and we opted instead for a simple amplitude fit of a fiducial model with some assumptions on parameters (see \autoref{tab:fid_params}). Some of these parameters are poorly constrained however, so here we explore what happens if we extend the model and include an MCMC fit of four parameters; $\boldsymbol{\varphi}\txteq{=}\{\Omega_\hi,b_\hi,b_\text{g},\sigma_\text{v}\}$. $\Omega_\hi$ is the dimensionless \hi\ density parameter and can be directly linked to the mean \hi\ temperature through
\begin{equation}
    \overline{T}_\hi(z)=180 \Omega_\hi(z) \, h \frac{(1+z)^2}{\sqrt{\Omega_{\mathrm{m}}(1+z)^3+\Omega_{\Lambda}}} \, \mathrm{mK} \, \text {, }
\end{equation}
where $\Omega_\text{m}$ and $\Omega_\Lambda$ are the density parameters for matter and the cosmological constant respectively.

We assume wide and flat priors for the $\boldsymbol{\varphi}$ parameters given by 
\begin{itemize}
	\item $0<10^3\Omega_\hi<2 $
	\item $0.5<b_\hi<2.0$
	\item $0<\sigma_\text{v}<600\,[\text{km}\,\text{s}^{-1}]$
	\item $1.8<b_\text{g}<2.0$
\end{itemize}
A tighter prior is assumed for the GAMA galaxy bias since we fit the galaxy auto-correlation power spectrum, finding a best-fit bias value of $b_\text{g}\txteq{\sim}1.9$. The remaining priors are sufficiently wide, even for these poorly constrained parameters. We stress that this is still an overly simplistic approach, but our aim here is a demonstration of current modelling capabilities, not rigorous parameter inference.

\begin{figure}
    \centering
    \includegraphics[width=1\linewidth]{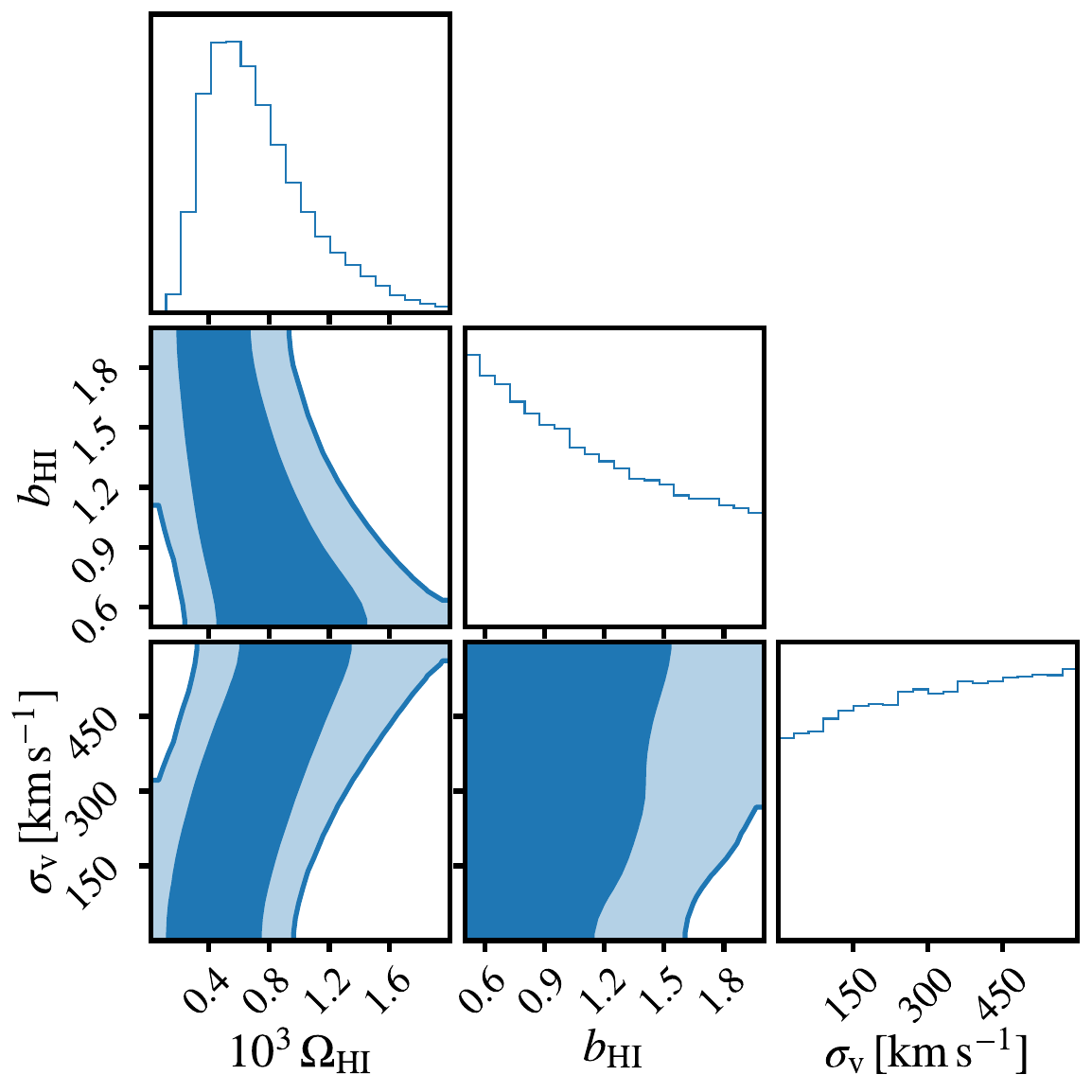}
    \caption{Parameter constraints from MCMC for $\boldsymbol{\varphi}\txteq{=}\{\Omega_\hi,b_\hi,b_\text{g},\sigma_\text{v}\}$, showing the degeneracies that exist from a monopole-only measurement. Flat priors are given in the text. We do not show $b_\text{g}$ since this is entirely prior dominated in the cross-power.}
    \label{fig:contours}
\end{figure}

\autoref{fig:contours} shows the inferred parameter constraints from the MCMC fit. We use \texttt{emcee} \citep{Foreman-Mackey:2012any} with 500 iterations of 200 walkers. We checked the chains had converged and discarded a burn-in of 40 iterations. Since it is entirely prior dominated in the cross power, we marginalise over $b_\text{g}$ and do not show its parameter contour. The nuisance parameter $\sigma_\text{v}$ is also completely prior dominated and suggests that a more refined model is needed on smaller scales, something explored in \citet{CHIME:2022kvg}. There is a constraint on $\Omega_\hi$ but this is only due to the restricted prior on $b_\hi$. Relaxing the \hi\ bias prior would inflate the $\Omega_\hi$ constraint indefinitely. 


\autoref{fig:contours} thus demonstrates the degeneracy between $\Omega_\hi$ and $b_\hi$ which has led previous efforts to merge this parameter fit in different ways \citep{Anderson:2017ert,eBOSS:2021ebm,CHIME:2022kvg,Cunnington:2022uzo}
. This degeneracy can be broken by utilising RSD which introduces an anisotropic dependence on $b_\hi$ but not on $\Omega_\hi$. We tested a measurement of the power spectrum quadrupole (a \textit{smoking-gun} of RSD) \citep{Cunnington:2020mnn,Soares:2020zaq} and found a clear non-zero measurement. However, we found that this was overwhelmingly driven by observational effects and higher signal-to-noise is required to make a distinction between this and a true RSD-induced anisotropy. Further discussion on the degeneracies between astrophysics and cosmology and the capabilities of the LIM power spectrum multipoles to constrain them can be found in \citet{Bernal:2019jdo}, where optimal parametrisations are proposed

\section{Effects of PCA in the stacked signal}\label{apdx:stack}
In this section, we use mock \hi\ galaxy signals to generate signal-only maps for stacking, and illustrate the effects of PCA in creating additional systematic noise structure.

\begin{figure}
    \centering
    \includegraphics[width=1\linewidth]{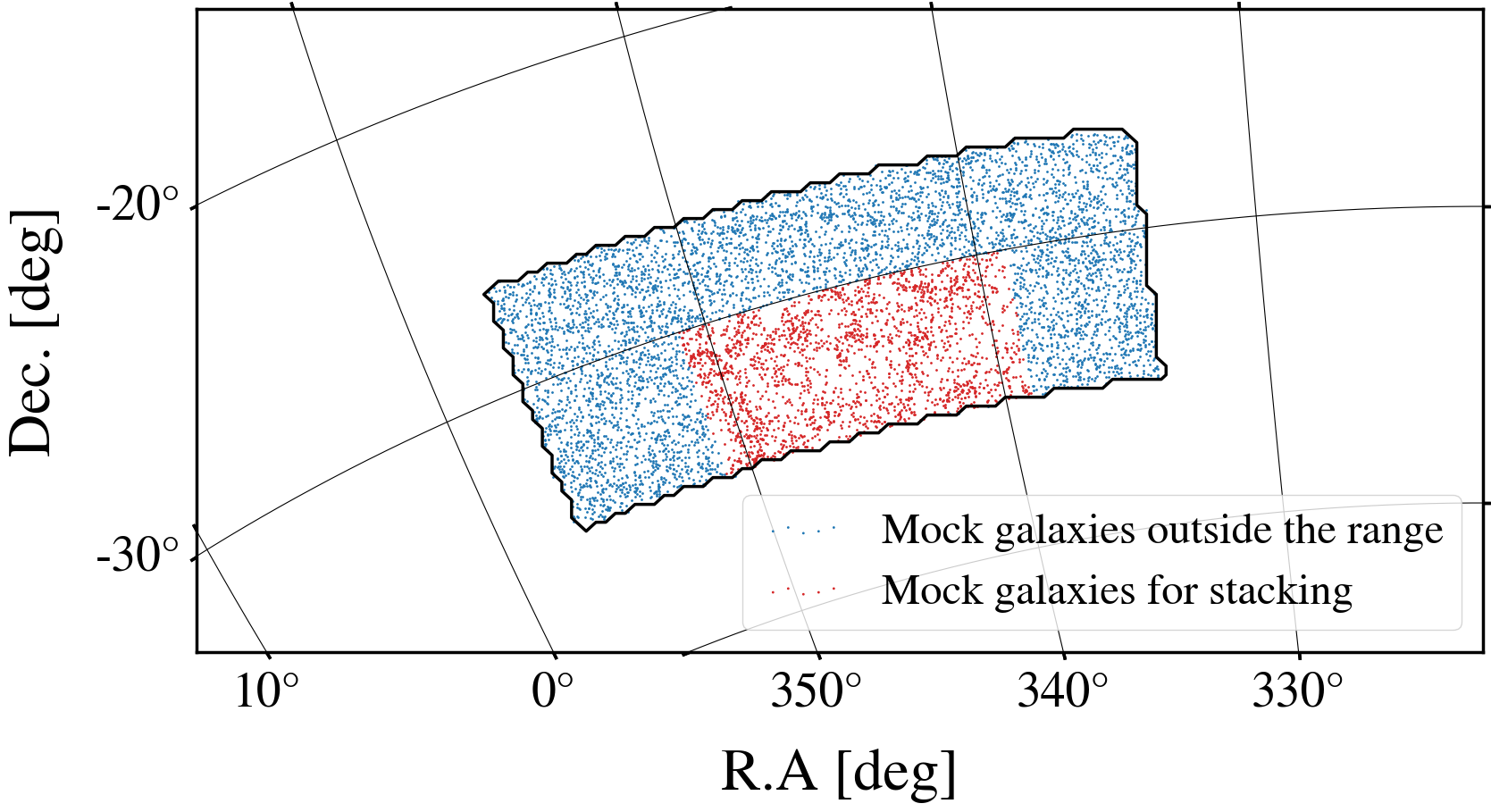}
    \caption{The positions of the mock galaxies on the sky. The blue points indicate the galaxies outside the GAMA coverage, and are not used for the mock stacking. The red points follow the same positions as the GAMA galaxies with random \hi\ mass assigned.}
    \label{fig:mockgalpos}
\end{figure}

As we are only interested in simulating the effects for stacking in this section, we use generate Poisson realisations of \hi\ sources with no clustering information as shown in \autoref{fig:mockgalpos}. For galaxies inside the survey coverage of GAMA, we use the exact same positions as the GAMA galaxies to replicate the double-counting effects. The rest of the map is then filled with galaxies at random positions assuming the same number density as the GAMA field. Random redshifts are assigned to the galaxies following the redshift kernel shown in \autoref{fig:GAMA_Nz}. Each galaxy is then assigned a \hi\ mass following the \hi\ mass function reported in \cite{2018MNRAS.477....2J}. Furthermore, we assign the velocity full width half maximum of the emission line profile $w_{50}$ to each galaxy using the Tully-Fisher relation reported in \cite{2021MNRAS.508.1195P}. The emission line profile of each galaxy is then calculated and added to the mock signal cube following the simplified busy function \citep{2014MNRAS.438.1176W}
\begin{equation}
    B_2(v) = \frac{a}{2}\times \big( {\rm erf}\big[ b(w_{50}^2-v^2)  \big] +1 \big) \times (cv^2+1),
\label{eq:busy}
\end{equation}
where erf is the error function and $a,b,c$ are parameters to be set. For each galaxy, we randomly choose $b$ logarithmically uniformly in $[10^{-3},10^{-2}]\,{\rm km^{-2}s^2}$ and $c$ in $[10^{-2},10^{0}]\,{\rm km^{-2}s^2}$. The amplitude parameter $a$ is then set by the \hi\ mass of the galaxy. We find that varying the choice of the busy function parameters does not significantly impact the results due to the relatively low velocity resolution. The signal is then convolved with the resmoothed beam described in \hyperref[subsec:reconvolution]{Section \ref{subsec:reconvolution}}.

\begin{figure}
    \centering
    \includegraphics[width=1\linewidth]{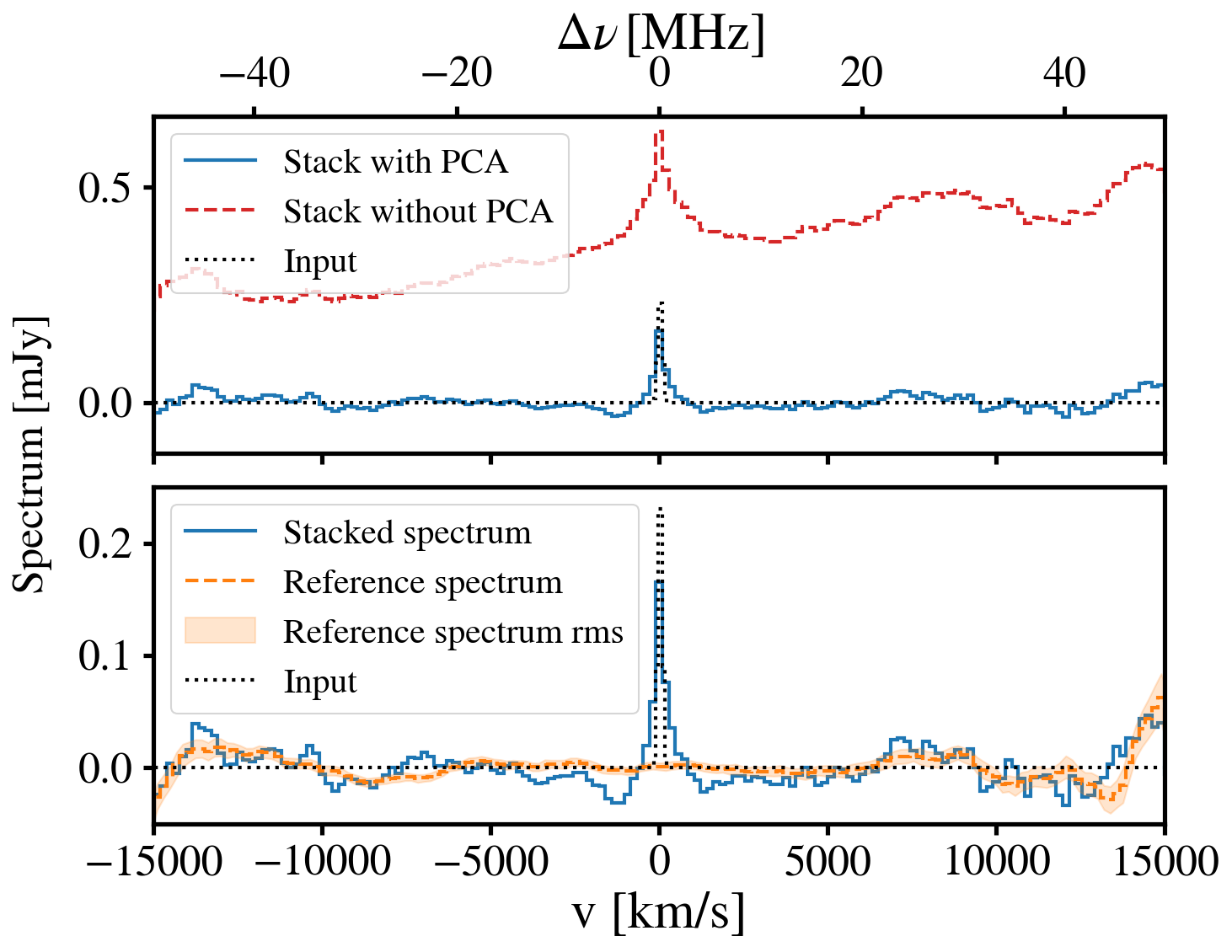}
    \caption{Top panel: The stacked spectra of the signal-only mock cube with and without PCA cleaning against the input mock signal. Bottom panel: The stacked spectrum of the signal-only mock cube with PCA cleaning, compared against the reference spectrum stacked on random positions. The dashed line shows the mean of the reference spectrum from 100 realisations and the shaded region shows the standard deviation among the realisations. }
    \label{fig:mockstack}
\end{figure}

As discussed in \hyperref[sec:stack]{Section \ref{sec:stack}}, the low angular resolution of the map induces leakage from nearby sources in the stacked profile. To illustrate this, we perform the stacking routine on the signal-only mock cube and compare against the input mock signal in \autoref{fig:mockstack}. The input signal is calculated by averaging the emission line profile of each source according to \autoref{eq:busy}. The double-counting introduces a large plateau of contamination across the frequencies. We note that the increasing slope of the plateau is due to the redshift distribution of sources. We then apply the PCA cleaning to the mock signal cube and repeat the stacking procedure. Instead of calculating the mixing matrix from the mock cube, we use the exact same mixing matrix obtained from the map for PCA and show the results in \autoref{fig:mockstack} (orange dashed line). The PCA removes the plateau to recover an emission line profile. However, it introduces several effects. There is signal loss due to the cleaning as well as the fact that we did not sum over an area much larger than the beam. Moreover, the PCA extends the width of the profile as seen in the lower panel of \autoref{fig:mockstack}. 

Apart from the effects on the recovered signal, PCA also creates a systematic noise component as seen in the stacked spectrum. To further demonstrate this, we repeat the reference spectrum calculation described in \hyperref[sec:stack]{Section \ref{sec:stack}} on the mock signal and present the results in \autoref{fig:mockstack}. 
Compared to the reference spectrum, the stacked mock signal outside the centre area of the peak shows deviations. The deviation from the reference spectrum appears to be coherent over large velocity intervals $\Delta v\txteq{\sim }5000\,$km/s. The scale of the interval is consistent with the systematic effects seen in the stacked spectrum in \autoref{fig:spectral_stack}. We emphasize that this is purely the combination of the \hi\ signal and the PCA cleaning since the mock map does not contain any other component. We further find that, when increasing the angular area for summation from $(1\,{\rm deg})^2$ to $(3\,{\rm deg})^2$, the amplitude of the coherent structure increases. 

\begin{figure}
    \centering
    \includegraphics[width=1\linewidth]{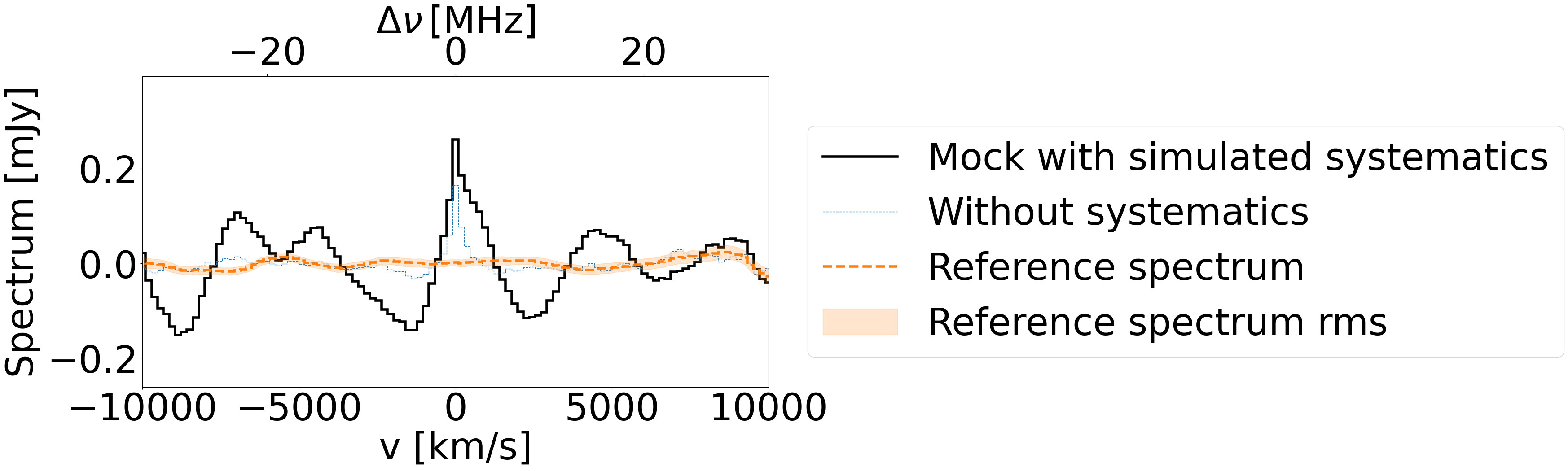}
    \caption{The stacked spectra of the signal-only mock cube with and without simulated systematics. The reference spectrum and its standard deviation, which are consistent with a null detection, is also shown for reference.}
    \label{fig:mocksystematics}
\end{figure}

The structure of the PCA cleaning is caused by the residual systematics in the data. Following the structure of the stacked spectrum in \autoref{fig:spectral_stack}, we generate random Gaussian realisations of systematic noise components and convolve the systematics to the mock signal. We then repeat the stacking procedure as well as reference spectrum calculation and show the results in \autoref{fig:mocksystematics}. As one can see, the mock signal affected by the systematics shows similar oscillating structure as seen in the data. The reference spectrum, on the other hand, shows consistentcy with a null detection, and demonstrates the fact that the detection reported in \hyperref[sec:stack]{Section \ref{sec:stack}} is a result of the detection \hi\ signal with residual systematics, and does not arise purely from contamination in the data.

We intend to simulate the mock data cube with foregrounds, noise and systematics to fully understand the impact of residual systematics on the stacked spectrum in the follow-up work.


\bsp	
\label{lastpage}
\end{document}